\documentclass[twocolumn,prb,notitlepage,preprintnumbers,amsmath,amssymb,superscriptaddress]{revtex4-1}
\usepackage{graphicx}
\usepackage{dcolumn}
\usepackage{bm}

\usepackage{amsmath}
\usepackage{subfigure}
\usepackage{hyperref}
\usepackage{soul}
\usepackage[normalem]{ulem}
\hypersetup{
    colorlinks,
    citecolor=blue,
    filecolor=black,
    linkcolor=blue,
    urlcolor=blue
}
\usepackage{tikzsymbols}
\usepackage{amssymb}
\usepackage{amsmath}
\usepackage{enumitem}
\usepackage[margin=0.75in]{geometry}
\usepackage{natbib}
\usepackage{subfigure}
\usepackage{physics}
\usepackage{lipsum}
\usepackage{graphicx}
\usepackage{mdframed}
\usepackage[skins,theorems]{tcolorbox}
\tcbset{highlight math style={enhanced,
  colframe=green,colback=white,arc=0pt,boxrule=1pt}}
\graphicspath{ {/} }
\usepackage{color}   
\usepackage{hyperref}
\usepackage{float}
\hypersetup{
    colorlinks=true, 
    linktoc=all,     
    linkcolor=blue,  
}

\newcommand{\bigzero}{\mbox{\normalfont\Large 0}}

\begin{document}


\title{Supercurrent Interference in Semiconductor Nanowire Josephson Junctions}
\author{Praveen Sriram}
\affiliation{%
Department of Electrical Engineering,
Indian Institute of Technology Bombay, Powai, Mumbai 400076, India}%
\altaffiliation[Present Address: ]{Department of Applied Physics, Stanford University, 348 Via Pueblo Mall, Stanford, CA 94305, USA}

\author{Sandesh S. Kalantre}
\affiliation{%
Department of Physics,
Indian Institute of Technology Bombay, Powai, Mumbai 400076, India }
\altaffiliation[Present Address: ]{Joint Quantum Institute, University of Maryland, College Park, MD 20742, USA}
%

\author{Kaveh Gharavi}
\affiliation{%
Institute for Quantum Computing,
University of Waterloo, Waterloo, Ontario N2L 3G1, Canada
}%
\affiliation{Department of Physics and Astronomy, University of Waterloo, Waterloo, Ontario N2L 3G1, Canada}
\author{Jonathan Baugh}
\affiliation{%
Institute for Quantum Computing,
University of Waterloo, Waterloo, Ontario N2L 3G1, Canada
}%
\affiliation{Department of Physics and Astronomy, University of Waterloo, Waterloo, Ontario N2L 3G1, Canada}
\affiliation{Department of Chemistry, University of Waterloo, Waterloo, Ontario N2L 3G1, Canada}
\author{Bhaskaran Muralidharan \textsuperscript{1}}
\email{bm@ee.iitb.ac.in}
%



\date{\today}
\begin{abstract}
Semiconductor-superconductor hybrid systems provide a promising platform for hosting unpaired Majorana fermions towards the realisation of fault-tolerant topological quantum computing. In this study, we employ the Keldysh Non-Equilibrium Green's function formalism to model quantum transport in normal-superconductor junctions. We analyze  III-V semiconductor nanowire Josephson junctions (InAs/Nb) using a three-dimensional discrete lattice model described by the Bogoliubov-de Gennes Hamiltonian in the tight-binding approximation, and compute the Andreev bound state spectrum and current-phase relations. Recent experiments [Zuo et al., Phys.\ Rev.\ Lett.\ \textbf{119},187704 (2017)] and [Gharavi et al., arXiv:1405.7455v2 (2014)] reveal critical current oscillations in these devices, and our simulations confirm these to be an interference effect of the transverse sub-bands in the nanowire. We add disorder to model coherent scattering and study its effect on the critical current oscillations, with an aim to gain a thorough understanding of the experiments. The oscillations in the disordered junction are highly sensitive to the particular realisation of the random disorder potential, and to the gate voltage. A macroscopic current measurement thus gives us information about the microscopic profile of the junction. Finally, we study dephasing in the channel by including elastic phase-breaking interactions. The oscillations thus obtained are in good qualitative agreement with the experimental data, and this signifies the essential role of phase-breaking processes in III-V semiconductor nanowire Josephson junctions.
\end{abstract}
\maketitle
%
%

\section{\label{sec:Intro}Introduction}
Semiconductor-superconductor hybrid junctions have generated significant interest over the last decade. In particular, III-V semiconductor (InAs/InSb) nanowires in proximity to an s-wave superconductor have been extensively studied as a platform for topological superconductivity\cite{Zhang,Mourik,Deng1557,Chene1701476,ADas_2012,Rokhinson_2012}. Majorana bound states (MBSs) emerge as zero energy edge excitations in a gapped bulk spectrum of the topological superconducting nanowire \cite{Majorana,Majorana2006,Kitaev_2001,BeenakkerReview2,Sau_2010,Sau_2010_2,Alicea_2010,Alicea_2012,LutchynPRL}. Signatures of MBS have been reported as a zero-bias conductance peak in tunnelling experiments \cite{Zhang,Mourik,Deng1557,Chene1701476,ADas_2012,Wimmer_2011}.  The $4\pi$ Majorana-Josephson effect has been predicted and observed in nanowire Josephson junctions tuned to the topologically non-trivial regime \cite{Kitaev_2001, Rokhinson_2012}.  
With the massive progress being made with nanowire setups, it is anticipated that the focus will shift from the detection to the demonstration of non-Abelian statistics and finally to topological quantum information processing\cite{Alicea_2011,Sarma_2015,Aasen_2016,Hyart,Karzig_2017,Plugge_2017}. The 4$\pi$ Majorana-Josephson effect forms the basis of braiding and readout schemes of a recent topological qubit proposal\cite{Stenger_2019}.
 
 Josephson junctions based on semiconductor-superconductor hybrids form the basis for microwave quantum circuity\cite{Lasrsen}, and superconducting qubits\cite{deLange,Hassler_2011}. They afford an attractive alternative for a scalable computing architecture with the possibility of an all-electric qubit control\cite{Kringhoj,Lasrsen,Casparis}. 

Several studies have focused on the structure of transverse subbands and magnetoconductance due to radial confinement in semiconductor nanowires \cite{Hernandez,Blomers,Aritra,Cayao2015} and carbon nanotubes\cite{CNT}. Recent experiments study the critical current as a function of the magnetic field and gate voltage in nanowire Josephson junctions tuned to the few-subband regime\cite{IQC,Frolov,Szombati}. For a magnetic field oriented along the nanowire axis, \citeauthor{Frolov} measured a strong suppression of the critical current at fields on the order of 100 mT in InSb weaklinks with NbTiN contacts. At higher fields, the critical current exhibited local minima (nodes). Similar results were obtained by \citeauthor{IQC} for InAs-Nb Josephson junctions. Unlike the Fraunhofer diffraction in wide planar junctions, the critical current nodes were aperiodic in the magnetic field, and highly sensitive to local fluctuations in the gate voltage. Motivated by these experiments, the object of this paper is to theoretically analyze few-mode nanowire Josephson junctions in a magnetic field oriented along the nanowire axis. 
We thus employ the Keldysh {Non-Equilibrium Green's Function formalism} (NEGF)\cite{Keldysh, datta_1995, datta_2005, DuBois, Rammer, NEGFChina} to model quasiparticle transport in the junction, and compute the evolution of the critical current as a function of the axial field and chemical potential. Based on the simulations, we attribute the observed oscillations to the interference of the transverse subbands in the nanowire. These results are crucial in the design of Majorana setups\cite{Aasen_2016,Hyart,Karzig_2017,Plugge_2017} and in interpreting experiments, particularly for those based on critical current measurements\cite{Cayao2017,San_Jose_2013,Cayao2018,SanJose}.  

Quantum transport traditionally involves excited states and the use of a variant of the Landauer-B{\"u}ttiker's scattering theory\cite{datta_1995,BEENAKKER19911,Aniket} for performing transport calculations. This essentially involves solving the Schr{\"o}dinger equation and an appropriate treatment of the boundary conditions. In a superconductor, however, zero-bias transport is essentially a ground state phenomenon supported by Cooper pairs condensed at the fermi level\cite{DeGennes,Tinkham}. \citeauthor{BTK} generalised the scattering theory approach to hybrid semiconductor-superconductor junctions by solving the Bogoliubov-de Gennes equation across the N-S interface\cite{BTK}. Beenakker applied this formalism for mesoscopic N-S junctions, thus providing a multichannel generalization of Blonder's results\cite{Beenakker}. This technique has been prevalent in the literature\cite{IQC1,Bagwell} ever since, and it forms the basis for numerous simulation packages such as Kwant \cite{Kwant}. Despite its benefits, the scattering theory approach is not very convenient in dealing with disordered junctions. While phase-coherent scattering processes can be included via random on-site potentials, it is difficult to model phase-relaxing interactions. Moreover, this formalism becomes intractable whenever a self-consistent determination of the order parameter becomes necessary. This self-consistent computation can be performed using the correlation Green's function\cite{LevyYeyatiPRB, LevyYeyatiPRL}, and various dephasing mechanisms such as electron-electron and electron-phonon interactions can be included through suitable self-energy operators in the NEGF formalism. The compatibility with phase-breaking processes is one of the main advantages of NEGF over scattering theory. Our results indicate dephasing to be essential in achieving qualitative agreement with the experiment, and this is one of the key takeaways of this paper.  

\begin{figure}[!htbp]
     \begin{center}
        \subfigure[]{%
    \label{fig:schema}
    \includegraphics[width=0.5\textwidth,keepaspectratio]{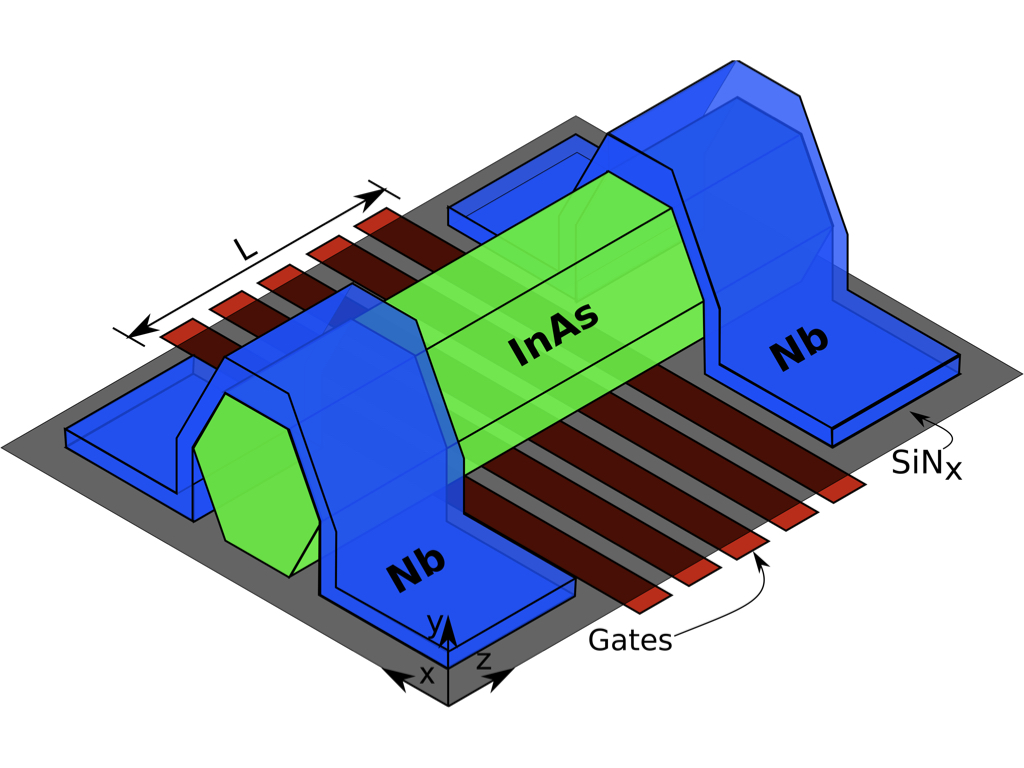}
        }\\%
        \subfigure[]{%
 	 \label{fig:tightb}  
	  \includegraphics[width=0.5\textwidth,keepaspectratio]{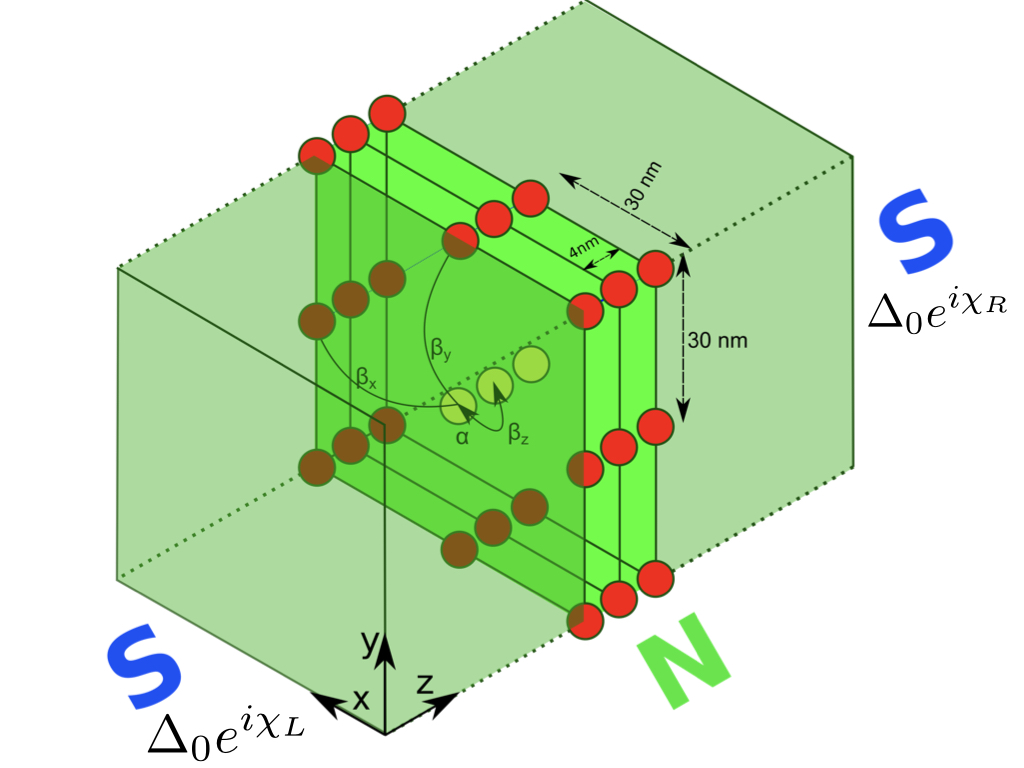}
        }
\end{center}
\caption[Schematic of the nanowire Josephson junction]{(a) Schematic of the nanowire Josephson junction. The length L of the junction is defined by the superconducting contact separation. A set of bottom gates tune the local chemical potential. Our model treats the junction as a normal (N) cuboidal cross-section (green) with flat superconducting (S) leads (blue). (b) The discrete lattice structure of our model -- highlighted in a section of the nanowire. The red spheres form the lattice sites in an effective tight-binding approximation. The length of the nanowire is controlled by the number of layers of the N-region. A potential U on the inner lattice points (yellow sites) confines the particles to surface of the nanowire. The transverse square cross-section is 60 nm wide. 
     }%
   \label{fig:SNS_3d}

            \end{figure}

This paper is organised as follows. We start with the Bogoliubov-de Gennes mean-field description of a one-dimensional nanowire Josephson junction (SNS).
In Sec.~\ref{sec:Model}, we describe the junction in a tight-binding model and outline the key aspects of the NEGF formalism. The details of this formalism have been relegated to Appendix~\ref{sec:app}. Employing the NEGF formalism, we compute the Andreev bound state (ABS) spectrum and current phase relationship (CPR) for this junction. 
Previous work almost exclusively focused on the Andreev approximation regime, which assumes the chemical potential of the nanowire ($\mu$) to be much larger than the superconducting order parameter ($\Delta_0$), i.e., $\mu \gg \Delta_0$\cite{BeenakkerReview,Bagwell,Kulik,BTK}. We go beyond this Andreev approximation limit and investigate the bound states which anti-cross at a superconducting phase difference of $\pi$ between the leads. This anti-crossing in the ABS is further analyzed in Appendix~\ref{sec:BeyondAA}.
In Sec.~\ref{sec:radialconf}, we model three-dimensional Josephson junctions in a discrete lattice tight-binding model (Fig.~\ref{fig:SNS_3d}).
The radial confinement gives rise to transverse angular momentum subbands which pick up characteristic phases in a magnetic field. Section~\ref{sec:labelling} details the procedure we follow to label these angular momentum subbands. In Sec.~\ref{sec:oscillations}, we reproduce the critical supercurrent oscillations in the presence of an axial magnetic field. Our results confirm these observed oscillations to be arising from the interference between orbital channels of the junction.
With the aim of gaining a thorough understanding of the experiments, we consider scattering processes in the nanowire and study the effect of disorder, gate voltage fluctuations, and phase-breaking processes on the critical current oscillations.

\section{\label{sec:Model} Formalism}


Superconducting correlations are induced in a proximitised semiconductor by electron-hole conversions at the interface, a process known as Andreev reflection\cite{Andreev,Andreev2}. 
Low bias transport in normal (N)-superconductor (S) junctions involves Andreev reflections at the interface. 
We first consider a one-dimensional SNS junction consisting of a semiconductor nanowire with supercondcuting contacts. We model this system using the Bogoliubov-de Gennes (BdG) mean-field Hamiltonian within the tight-binding approximation, $H = \mathcal{H}_0 + \mathcal{H}_{\text{p}}$, where
\begin{align}
\mathcal{H}_0 &= \int dz \sum_{\sigma}\psi_{\sigma}^{\dagger}(z)\left(-\frac{\hbar^2}{2m^*}\partial_z^2 + V(z) - \mu \right)\psi_{\sigma}(z)
\label{eq:model1}
\\
\mathcal{H}_p &= \int dz \psi_{\uparrow}^{\dagger}(z)\Delta(z)\psi_{\downarrow}^{\dagger}(z) + h.c.
\label{eq:model2}
\end{align}
$\mathcal{H}_0$ is the single-particle effective Hamiltonian, $\psi_{\sigma}$ is the field operator with spin index $\sigma 
\in \{\uparrow,\downarrow  \}$, $m^*$ is the electron effective mass, and $V$ models a potential energy induced in the junction. The chemical potential is defined as the energy difference between the lowest occupied subband and the Fermi energy, and is denoted by $\mu$. We assume an identical effective mass in the N and S regions thus neglecting the Fermi wave-vector mismatch at the interface. $\Delta(z)$  is the superconducting order parameter along the junction, which we assume to be constant with jump-discontinuities at the N/S interfaces
\begin{equation}
\Delta(z) = \mathcal{\vartheta}(-z)\Delta_0e^{i\chi_L} + \mathcal{\vartheta}(z-L)\Delta_0 e^{i\chi_R}
\end{equation}
where $\vartheta(x)$ is the unit step function at $x=0$, $\chi_{L,R}$ is the superconducting phase of the left and right leads respectively, and $\phi=\chi_L-\chi_R$ is the phase difference.
In the $\left[\psi_{\uparrow}^{\dagger}(r), \psi_{\downarrow}(r)\right]$ Nambu basis, we have the BdG equation 
\begin{equation}
\begin{bmatrix}
\mathcal{H}_0 & \Delta(z) \\
\Delta^*(z) & -\mathcal{H}_0^*
\end{bmatrix} 
\begin{bmatrix}
u(z) \\
v(z)
\end{bmatrix} 
=
E
\begin{bmatrix}
u(z) \\
v(z)
\end{bmatrix} 
\label{eq:BdG}
\end{equation}

\begin{figure}[!htbp]
            \centering
            \includegraphics[width=0.4\textwidth,keepaspectratio]{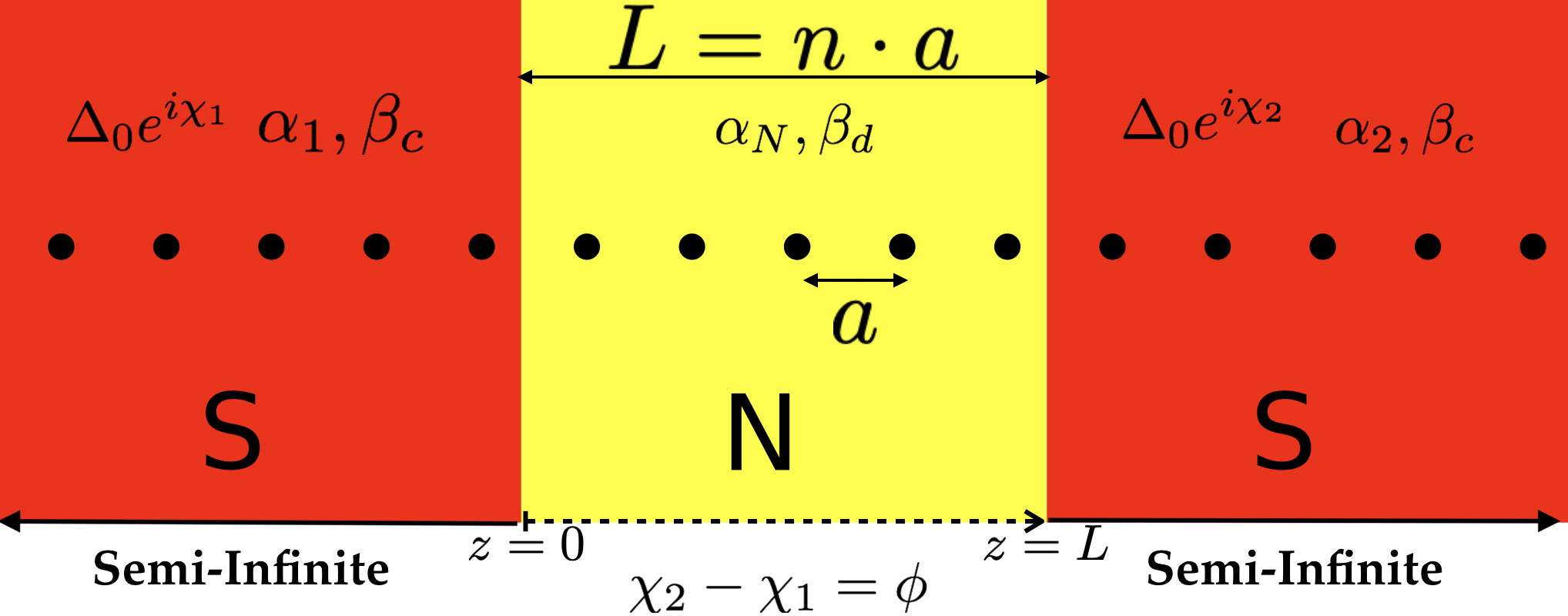}
\caption{Schematic of the SNS junction with a semi-infinite superconducting leads and an N device region. The length $L$ of the nanowire is given by the number of lattice points $(n)$ in the N-region $[L=n\cdot a]$, where $a$ is the effective lattice constant. $\alpha$ and $\beta$ are the tight-binding onsite and nearest neighbour coupling terms respectively. $\Delta_0$ and $\chi$ is the magnitude and phase of the superconducting order parameter.}
 	 \label{fig:SNS_1d}
            \end{figure}

The device is divided into three parts -- a normal semiconductor section with a length $L$ extended over $z \in [0,L]$, and semi-infinite superconducting contacts extending to $ z = \pm \infty$ on either side of the semiconductor (Fig.~\ref{fig:SNS_1d}). 

We discretise the continuum model of Eqs.~\ref{eq:model1},~\ref{eq:model2} into a lattice model with a spacing of $a$. This is shown in Fig.~\ref{fig:SNS_1d}. The superconductors are modelled as semi-infinite leads, while the number of lattice points in the normal region controls the length of the nanowire.
The on-site tight-binding parameters in the normal and superconducting regions, in the Nambu representation are \\
\begin{equation}
\alpha_N=
\begin{bmatrix}
2t-\mu & 0\\
0 & -2t+\mu
\end{bmatrix}
\end{equation}
\begin{equation}
\alpha_S=
\begin{bmatrix}
2t-\mu & \Delta\\
\Delta^{*} & -2t+\mu
\end{bmatrix}
\label{eq:alpha}
\end{equation}
\\
where, $t = {\hbar^2}/\left({2m^*a^2}\right)$ is the nearest neighbour tight binding hopping parameter. 
The hopping matrix is given by 
\begin{equation}
\beta =
 \begin{bmatrix}
-t & 0\\
0 & t
\end{bmatrix}
\label{eq:beta}
\end{equation}
{
This is not an atomic model, but an ``effective'' discrete lattice description of the junction. Parabolic dispersion relations correspond to the parameter space  $\mu \ll t$. Typical experiments are setup in the $\mu \gg \Delta_0$ (Andreev approximation) regime. Hence, the choice of the effective lattice parameter $a$ is bound by the inequalities 
\begin{equation}
t \gg \mu \gg \Delta_0
\end{equation}
}

 The device Hamiltonian is subsequently written as
 \begin{equation}
 H = \sum_i^n  c_i^{\dagger}\alpha_{N/S}c_i +  \sum_{|i-j|=1}^n  c_i^{\dagger}\beta c_j 
 \end{equation}
 where $c^{\dagger}_i$ is the creation operator of the Nambu spinor $\left[\psi_{\uparrow}^{\dagger}(r), \psi_{\downarrow}(r)\right]$ at site $i$, and $n=L/a$ is the number of sites in the device.
The Hamiltonian of the normal region can be written in the general form
 \begin{equation}
{H} = \mqty(\alpha_N & \beta & 0 & \dots & 0 \\ \beta^{\dagger} & \alpha_N & \beta & 0 & 0 \\ 0 & \beta^{\dagger}& \alpha_N & \beta  & \vdots \\ \vdots & 0 & \ddots & \ddots & \beta \\ 0 & \dots & \dots & \beta^{\dagger} & \alpha_N)\label{eq:Hamiltonian}\end{equation}
\subsection{\label{sec:SNS}Andreev bound states in SNS junctions}
Andreev reflections at the N/S interfaces give rise to Andreev bound states in the semiconductor. We use the NEGF formalism to compute these bound state energies as a function of the superconducting phase difference ($\phi$) of the leads.
The retarded Green's function in the energy domain is given by 
\begin{equation}
G^r(E) = \left(E\mathbb{I} + i \eta-\mathcal{H}-\Sigma_1^r-\Sigma_2^r\right)^{-1}
\end{equation}
where $E$ denotes the energy, $\mathbb{I}$ is the identity matrix and $\eta$ is an infinitesimal real constant. The Hamiltonian $\mathcal{H}$ is given by Eq.~\ref{eq:Hamiltonian}. The self-energy terms $\Sigma^r_{1,2}$ model the coupling of the device to the semi-infinite leads. The self-energy is not hermitian, and its anti-hermitian part is responsible for the finite lifetime of the electron in the device. This subsequently contributes to broadening the energy levels in the device. The self-energies are computed using the surface-Green's function, which requires an iterative procedure as outlined in Appendix~\ref{sec:app}. 



\begin{figure}[!htbp]
            \centering
            \includegraphics[width=0.5\textwidth,keepaspectratio]{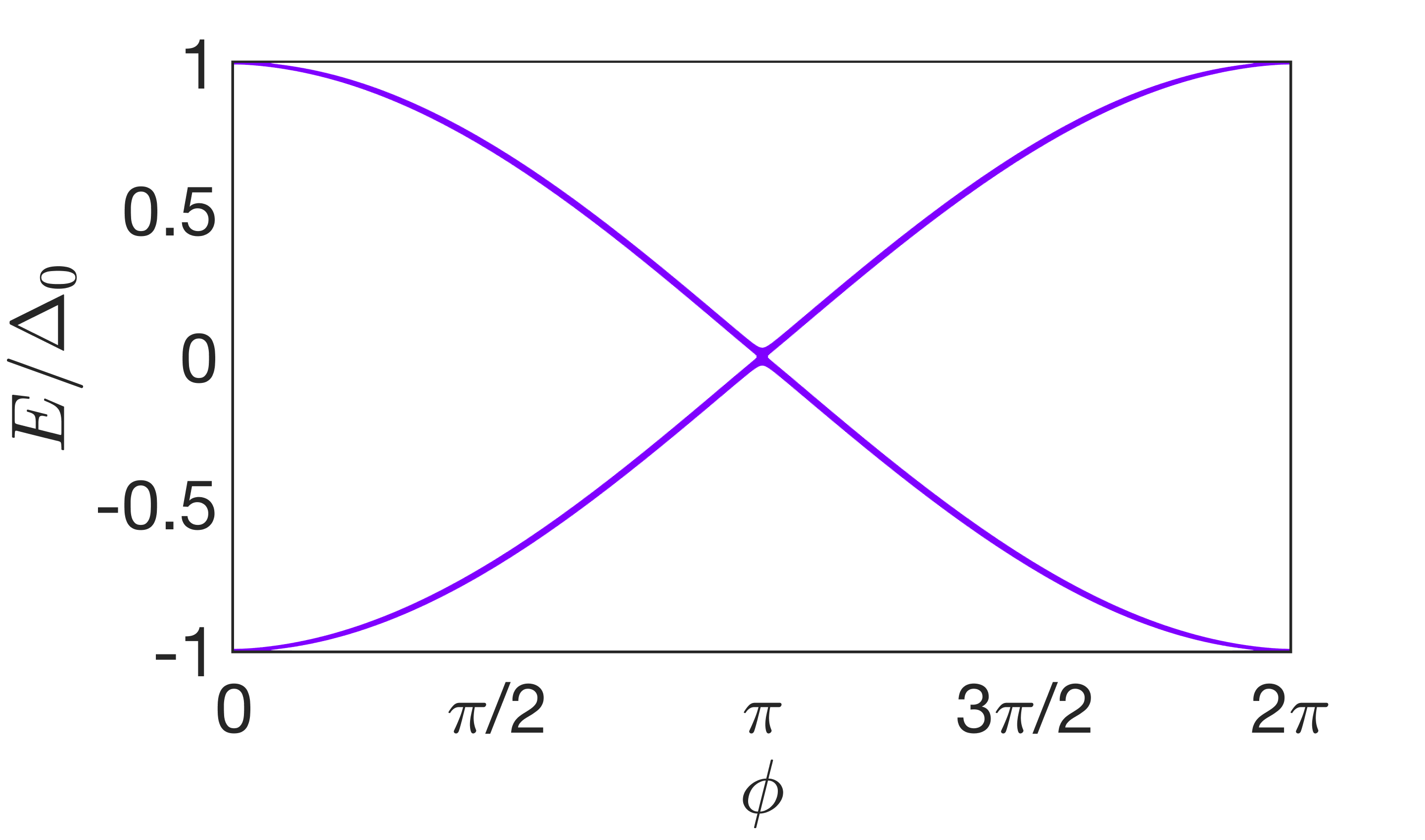}
\caption{Andreev bound state spectrum as a function of the superconductor phase difference for a clean, short 1-dimensional SNS junction. The junction is tuned into the Andreev approximation regime with $\mu = 30\Delta_0$.}
             	 \label{fig:ABS_AA}

            \end{figure}

We compute the density of states ($d$) in the nanowire as the trace of the spectral Green's function 
\begin{equation}
d(E) = \frac{1}{2\pi}\Tr \left[A(E)\right] = \frac{1}{2\pi}\Tr \left [i \left(G^r(E) - G^a(E)\right) \right ]
\label{eq:DOS}
\end{equation}
The real-valued singularities of the density of states are the Andreev bound state (ABS) energies. This is computed as a function of the phase difference ($\phi$) of the order parameter of the contacts and is shown in Fig.~\ref{fig:ABS_AA}. The parameters for this computation are consistent with the Andreev approximation\cite{Andreev,Andreev2,BeenakkerReview,Ashida} ($\mu \gg \Delta_0$). As discussed in Appendix~\ref{sec:BeyondAA}, the breakdown of this approximation is manifested as an avoided level crossing in the ABS spectrum.

\subsection{Current phase relationship}
The current-phase relationship (CPR) links the macroscopic current flow in the junction to the phase gradient of the superconducting order parameter\cite{Josephson,Spanton2017}. The traditional approach to computing the CPR involves a demarcation of the bound state and continuum currents. The bound state current involves transport in the sub-gap energy range ($\abs{E} < \Delta_0$) while the continuum current is supported by the continuous energy spectrum outside the gap. Once the ABS spectrum is computed from scattering theory, a thermodynamic relation is used to calculate the bound state current, and the transmission formalism is used for the continuum current. The total current is the sum of the bound state and continuum currents\cite{Kulik}.



By contrast, when using the NEGF formalism the current-energy density can be computed at contact $i$, as a function of the phase difference $\phi$ using the current operator\cite{datta_1995,datta_2005}

\begin{equation}
    J_i(E) = \frac{2e}{h}f(E)\text{Tr}\left[\Re\left( G^a(E)\Sigma_i^a(E)-G^r(E)\Sigma_i^r(E) \right)\tau_z\right]
\end{equation}
where $f(E) = 1/\left(\exp\left(E/k_BT\right)+1\right)$ is the Fermi-Dirac occupation probability for a given energy level and $k_B$ is the Boltzmann constant. $G^{r(a)}$ and $\Sigma_i^{r(a)}$ are the retarded (advanced) Green's function and contact $i$ self-energy respectively. To incorporate the opposite charge of electrons and holes we use the Pauli-z operator ($\tau_z$) in the particle-hole Nambu space. This current operator is reviewed in the Appendix~\ref{sec:app}. The total current at a phase difference $\phi$ is then given by 
\begin{equation}
    I(\phi) = \int_{-\infty}^{\infty}J_i(E)dE
\end{equation}

%

\begin{figure}[!htbp]
     \begin{center}
        \subfigure[]{%
    \label{fig:CPR_shortChannel}
    \includegraphics[width=0.45\textwidth,keepaspectratio]{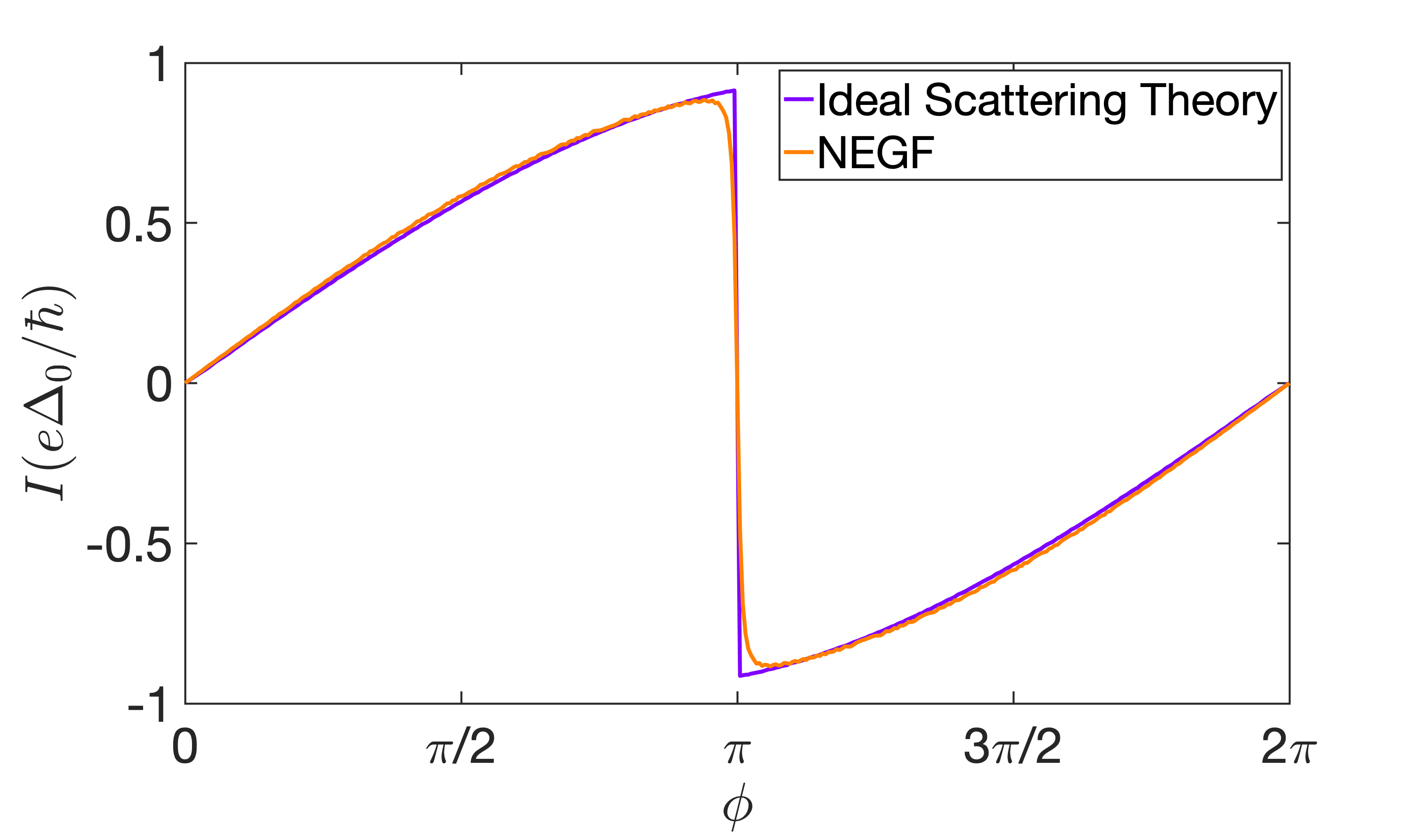}
        }\\%
        \subfigure[]{%
 	 \label{fig:CPR_longChannel}  
	  \includegraphics[width=0.45\textwidth,keepaspectratio]{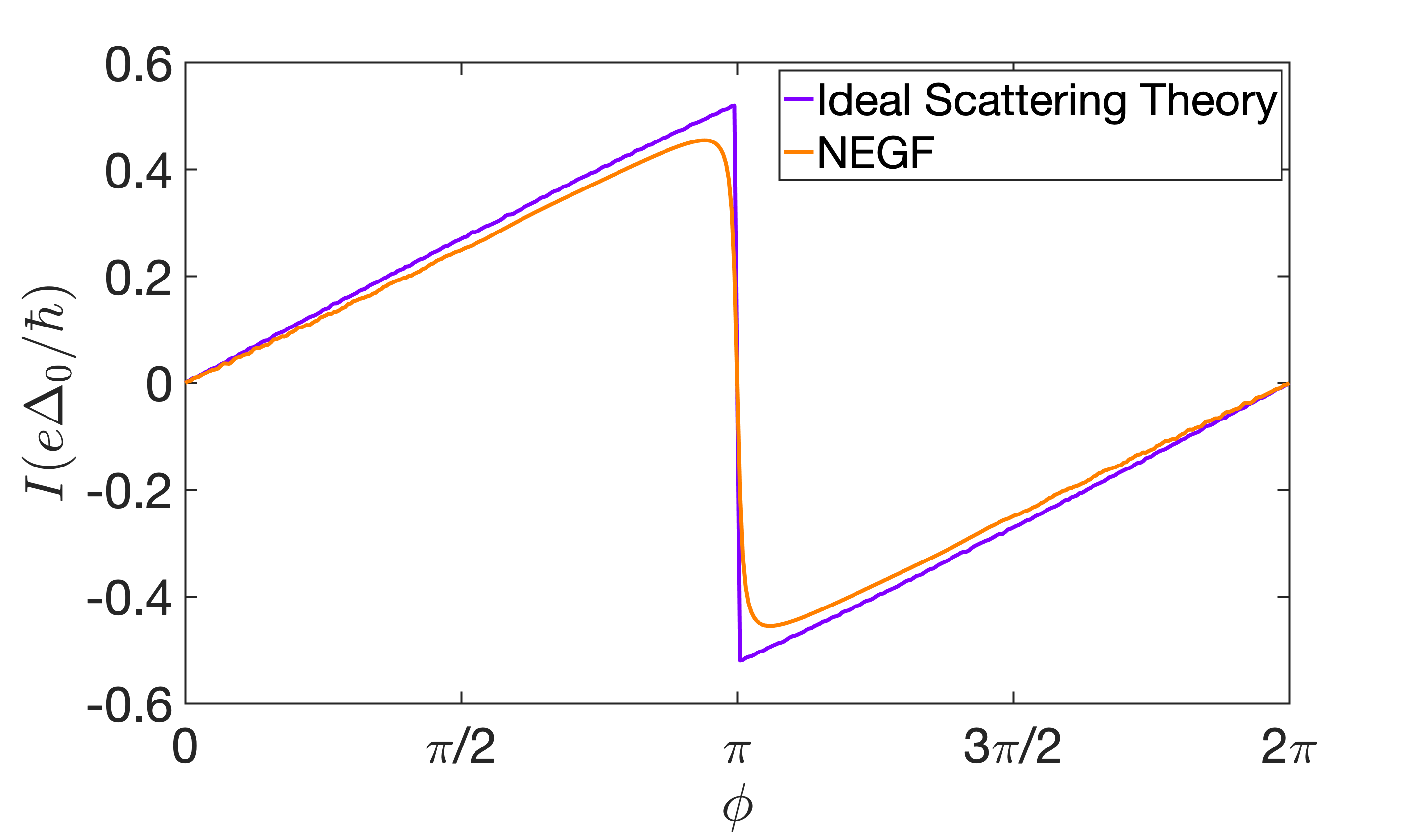}
        }
\end{center}
\caption[Current-Phase-Relation (CPR) in an SNS junction]{The total current-phase-relation in an SNS junction computed from the NEGF current operator (orange) is compared with the ideal scattering theory (purple) calculations (see text). The CPR is plotted in the (a) short channel (L=40 nm) and (b) long channel (L = 400 nm) limit, chemical potential $\mu=30\Delta_0$ and healing length $\xi_0=222$ nm. 
We recover the expected saw-tooth profile for the total current using the NEGF formalism.
     }%
   \label{fig:subfigures}

            \end{figure}

Figures~\ref{fig:CPR_shortChannel} and~\ref{fig:CPR_longChannel} compares the current phase-relations for a short ($L<\xi_0$) and long junction ($L>\xi_0$) respectively, as calculated from ideal scattering theory and NEGF. Here, $\xi_0$ is the healing length as defined in the following section (Eq.~\ref{eq:xi}). By the term ``{ideal} scattering theory'' we refer to a scattering approach which explicitly neglects normal reflections at the N/S interface in a clean junction\cite{Kulik,Bagwell,IQC1,BeenakkerReview}.
With this assumption of no normal reflections, the bound states in a clean junction cross at $\phi=\pi$.  Hence, there's a discontinuity at $\phi=\pi$ in the CPR calculated using this method. The NEGF result is expected to match scattering theory exactly in the $\mu \gg \Delta_0$ limit. 

\section{\label{sec:radialconf}Transverse subbands in Josephson Junctions}
We now consider a more realistic three-dimensional model of the junction with a magnetic field along the nanowire axis, parallel to the direction of current flow. Figure~\ref{fig:SNS_3d} illustrates a discrete lattice model of the three-dimensional nanowire. The junction is along the $z$-direction and the transverse subbands are on the $x$-$y$ plane. In III-V semiconductors (InAs, InN) the charge carriers are typically confined close to the surface due to a positive surface potential, forming a surface accumulation layer. In accordance with this, we use a shell conduction model by including a large surface confining potential $U$ 
at the core of the nanowire (yellow sites in Fig.~\ref{fig:SNS_3d}). The radial confinement and azimuthal periodicity of the nanowire gives rise to transverse subbands.

The single-electron Hamiltonian of the nanowire is 
\begin{equation}
\mathcal{H}_0 = -\mu + \frac{-\hbar^2}{2m^*}\frac{\partial^2}{\partial z^2} + \mathcal{H}_T + U
\end{equation}
where $z$ is the longitudinal direction, $U$ is the surface confinement potential, and $\mathcal{H}_T$ is the Hamiltonian of the transverse modes. 

For a cylindrical nanowire, the rotational symmetry about the longitudinal axis results in angular momentum $\left(\ell \right)$ subbands. This is because 
\begin{equation}
[\mathcal{H}_T,\hat{L}_z] = 0
\end{equation}
where $\hat{L}_z$ is the angular momentum operator in the z-direction. Hence, $ \ell$  is a good quantum number. The $\ell$ subbands are eigenstates of the $\hat{L}_z$ operator, labelled by their eigenvalue 
\begin{equation}
\hat{L}_z\ket{\ell} = \hbar\ell\ket{\ell}
\label{eq:l}
\end{equation}

We can consider the square cross-section in Fig.~\ref{fig:SNS_3d} as a perturbation to an ideal cylindrical geometry. Each of the transverse subbands in a square cross-section can be written as a superposition of angular momentum eigenstates. In Sec.~\ref{sec:labelling}, we compute the average angular momentum of each transverse subband, and observe only a small difference (see Fig.~\ref{fig:ABSL}) from the unperturbed quantized eigenvalues ($\hbar \ell$). We will thus work within the zeroth order of this perturbation and use the language of angular momentum subbands in our analysis.

The azimuthal motion of the Andreev quasiparticles couples with the applied magnetic field, resulting in a quasiparticle phase pickup. Oscillations in the maximal supercurrent (critical current) with field have been measured by \citeauthor{IQC}, and \citeauthor{Frolov}. Unlike the Fraunhofer interference in wide planar junctions, the field is aligned with the current and the oscillations do not show any periodicity. 



Using Peierls substitution, we include the orbital effect of the vector potential in the phase of the transverse hopping. For a constant magnetic field along the $z$-direction, the vector potential can be written as 
\begin{equation}
\mathbf{A} = B \cdot x\mathbf{\hat{y}}
\label{eq:A}
\end{equation}

Within the tight-binding approximation, the on-site and hopping matrices in the particle-hole Nambu space are given by 
\begin{equation}
\alpha_{N/S}=
\begin{bmatrix}
{h} & \Delta_{N/S}\\
\Delta_{N/S}^* & -{h}^*
\end{bmatrix}
\label{eq:alpha3d}
\end{equation}
where ${h} = 2t_x+2t_z+\abs{t_y}\left[2+(2\pi n_x\Phi_a)^2\right]-\mu$
\begin{equation}
\beta_{x,z} =
 \begin{bmatrix}
-t_{x,z} & 0\\
0 & t_{x,z}
\end{bmatrix}
\end{equation}
\begin{equation}
\beta_y =
 \begin{bmatrix}
-t_ye^{i2\pi n_x\Phi_a} & 0\\
0 & t_ye^{-i2\pi n_x\Phi_a} 
\end{bmatrix}
\label{eq:betaxz}
\end{equation}
where $\Phi_a$ is the flux quanta per unit cell of the nanowire cross-section, and $n_x$ is the lattice site index in the $x-$direction. This factor alters the on-site energy ($\alpha_{N/S}$) and contributes a phase to the hopping term corresponding to the gauge chosen for the vector potential (Eq.~\ref{eq:A}).

\subsection{Andreev bound states in a magnetic field}
\begin{figure*}[!htb]
     \begin{center}
        \subfigure[Subgap density of states vs $\phi$ at $\Phi=0.01$]{%
 \label{fig:ABS1Flux}         
    \includegraphics[width=0.45\textwidth]{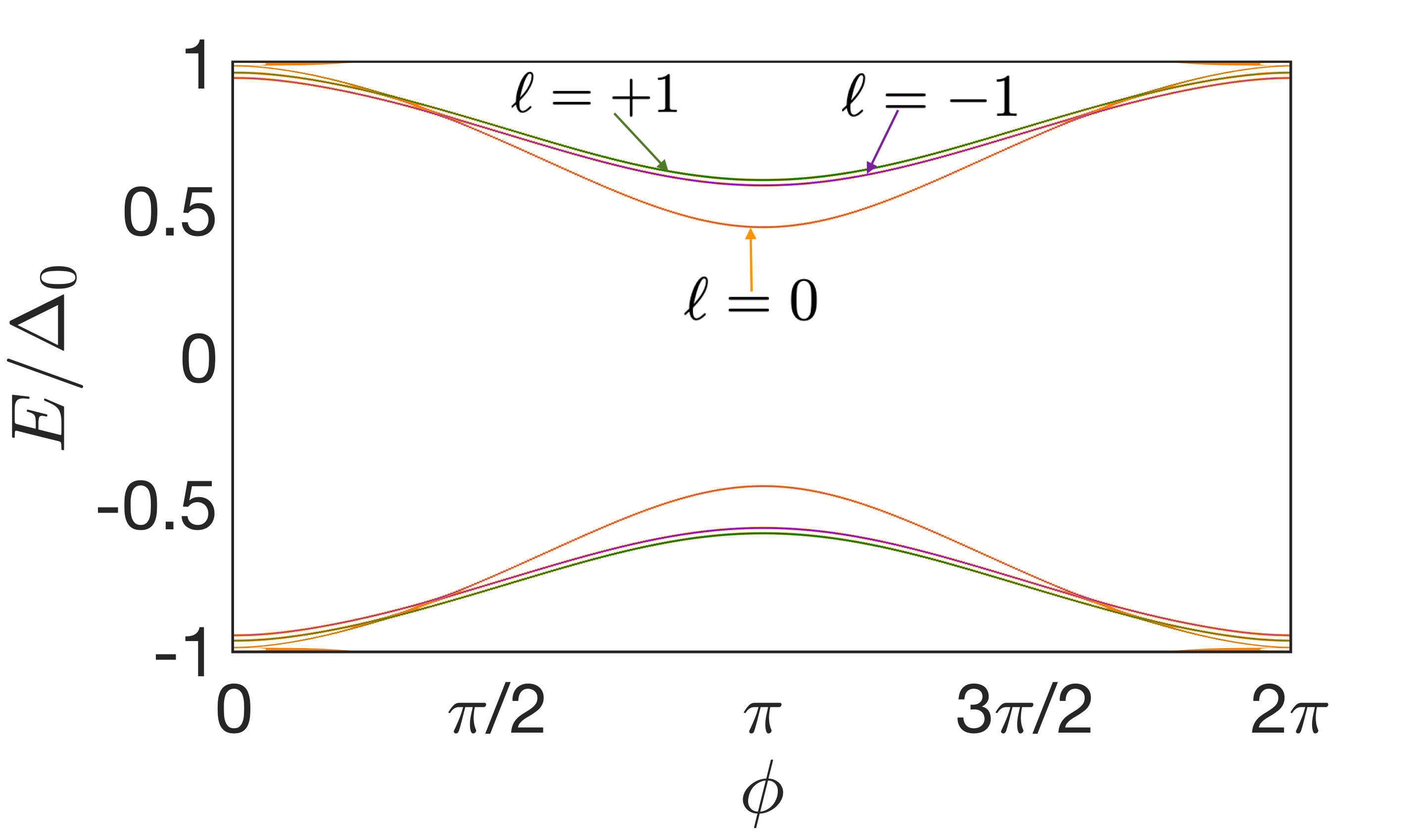}
        }%
        \subfigure[Subgap density of states vs $\phi$ at $\Phi=0.03$]{%
            \label{fig:ABS3Flux}
            \includegraphics[width=0.45\textwidth]{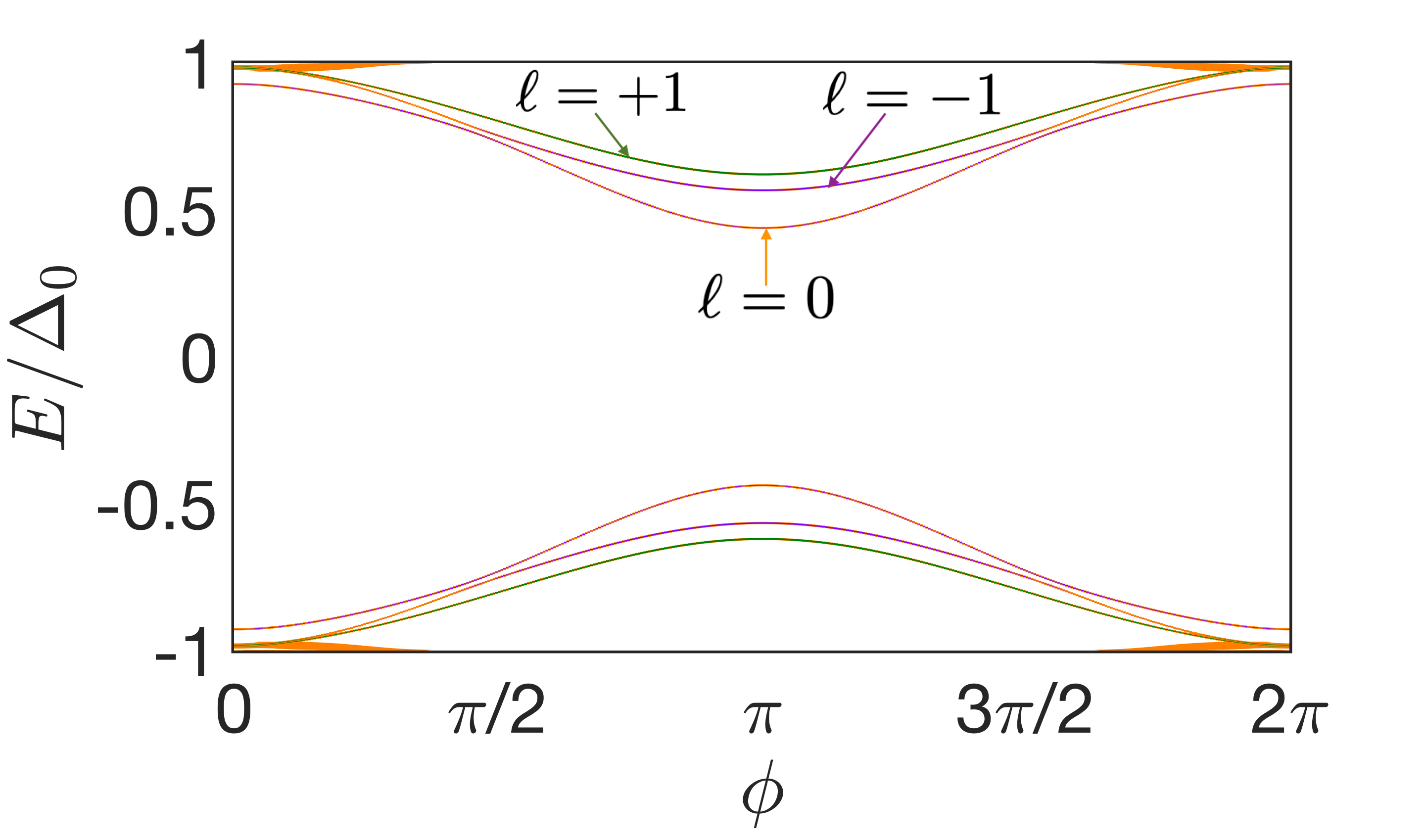}
        }\\%
  \subfigure[$\langle \hat{L}_z \rangle$ at $\Phi=0.01$, $\phi=\pi$]{%
              \label{fig:ABS_label_0Flux}
            \includegraphics[width=0.45\textwidth]{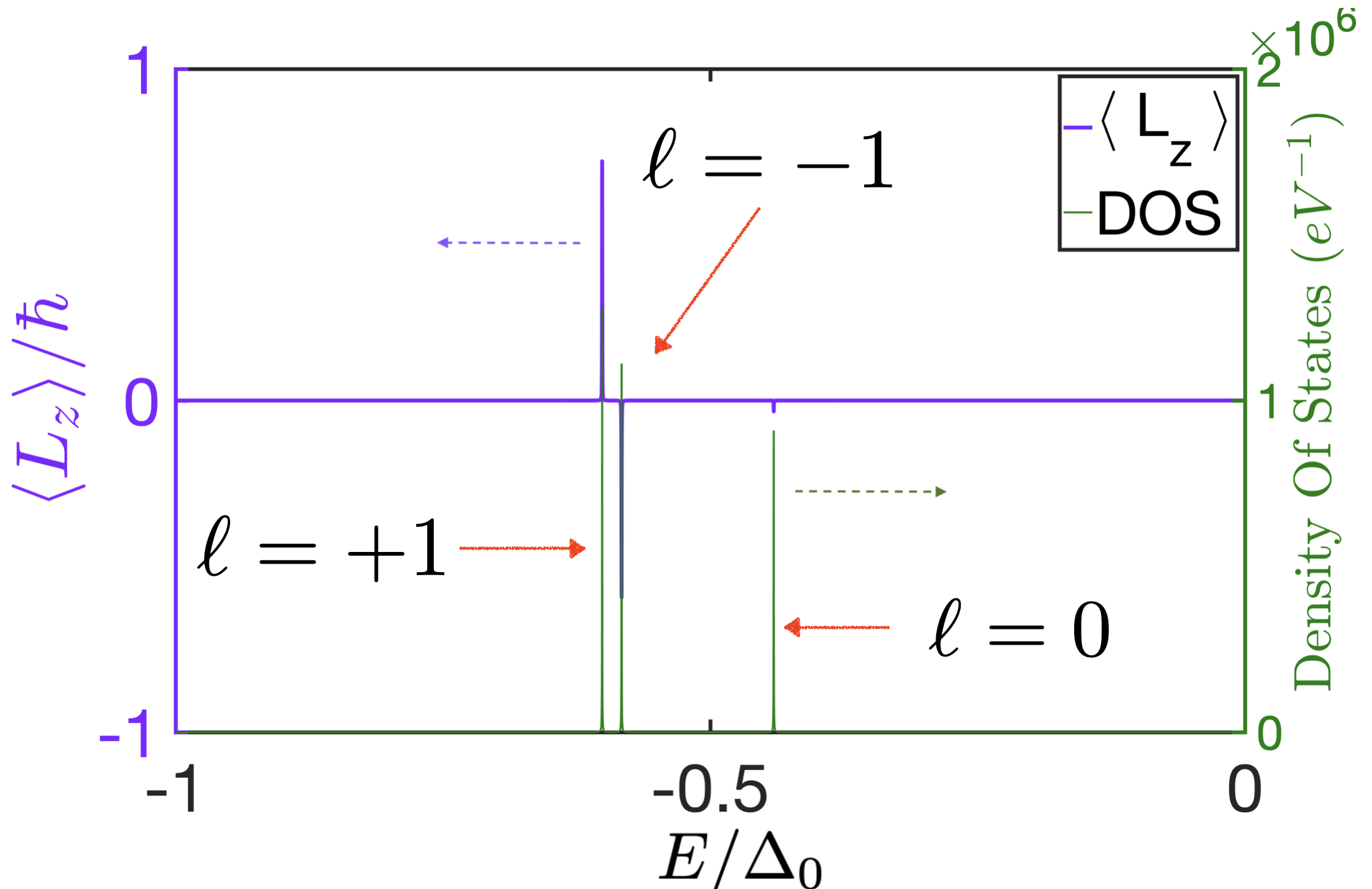}
        }%
        \subfigure[$\langle \hat{L}_z \rangle$ at $\Phi=0.03$, $\phi=\pi$]{%
            \label{fig:ABS_label_01Flux}
            \includegraphics[width=0.45\textwidth]{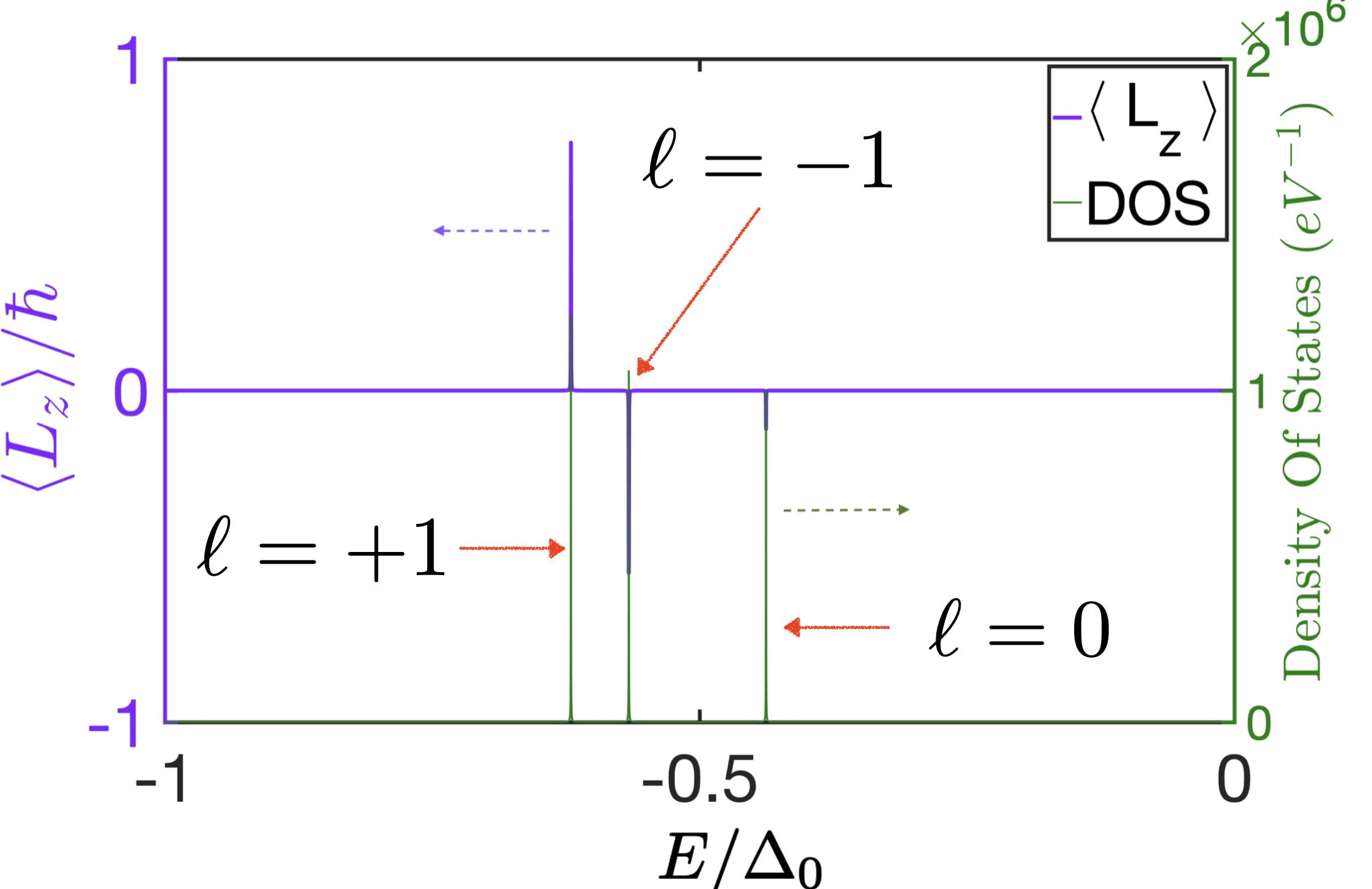}
        }%

    \end{center}
    \caption{%
       The Andreev bound state spectrum for an SNS junction with three occupied subbands is plotted in (a) an axial magnetic field with $\Phi=0.01$ and (b)  $\Phi=0.03$. The subgap density of states (green) and average angular momentum quantum (purple) at $\phi=\pi$ is plotted in (c) $\Phi=0.01$ and (d) $\Phi=0.03$. The nanowire has a square-cross section (Fig.~\ref{fig:SNS_3d}) and hence the transverse subbands do not have quantized angular momenta. Nevertheless, the computed average angular momenta indicate than each subband is primarily composed of a single angular momentum eigenstate, and have been correspondingly labelled.
       The nanowire length $L=8$ nm, chemical potential $\mu=5\Delta_0$ and healing length $\xi_0 = 90$ nm. 
}%
\label{fig:ABSL}
 \end{figure*}

Figures~\ref{fig:ABS1Flux}, and~\ref{fig:ABS3Flux} plot the subgap density of states as obtained from the spectral Green's function (Eq.~\ref{eq:DOS}) for a nanowire with an InAs effective mass $m^*$ = 0.023$m_e$\cite{NAKWASKI19951} ($m_e$ is the bare electron mass), radius R = 30 nm and chemical potential $\mu=5 \Delta_0$. As described in Appendix~\ref{sec:BeyondAA}, the bound states anti-cross at $\phi=\pi$ due to normal reflections at the N/S interfaces. A normalised flux of $\Phi = 0.01$ ($0.03$) is applied in Fig.~\ref{fig:ABS1Flux} (Fig.~\ref{fig:ABS3Flux}), which breaks symmetry in the transverse direction (Eq.~\ref{eq:A}) and lifts the degeneracy of the $\pm \ell$ subbands.
Here, $\Phi=B\cdot S/\Phi_0$, where $S$ is the cross-sectional area, and $\Phi_0 = h/e$ is the flux quantum. 
 

    
    
    The chemical potential is adjusted to populate three subbands. In Fig.~\ref{fig:ABSL} we see that the $\ell=0$ subband remains unaffected, while the $\ell \neq 0 $ subbands split in the presence of an axial flux. The splitting is proportional to the flux and the subband angular momentum. The process used to label the subbands is described in Sec.~\ref{sec:labelling}. 
    
    The characteristic length-scale associated with an occupied subband is called the healing length $(\xi_{\ell})$\cite{Kulik,Bagwell}, and is given by 
    \begin{equation}
    \xi_{\ell} = \frac{\hbar v_{F,\ell}}{2\Delta_0}
    \label{eq:xi}
    \end{equation} 
     $v_{F,\ell}$ is the fermi-velocity and is given by 
        \begin{equation}
    v_{F,\ell} = \sqrt{2\left(\mu - \frac{\hbar^2}{2m^*R^2}\left(\ell^2+\Phi^2 \right)\right)/m^*}
    \label{eq:fermi_vel}
        \end{equation}

        
        We have a ``short junction'' when the nanowire is shorter than the healing length ($L<\xi_{\ell}$). Note that the healing length depends on the angular momentum quantum number, and the classification of the junction as long/short is subband dependent.



We now describe a procedure to label the angular-momentum subbands using the correlation Green's function ($G^n$).

  \subsection{\label{sec:labelling}Average angular momentum  of the transverse subbands}
The angular momentum of the subbands can be computed as the expectation of the $\hat{L}_z$ operator (Eq.~\ref{eq:l}), however, we do not have access to the wavefunctions in a numerical simulation. We do have the correlation Green's function $-iG^< = G^n$ which gives the particle-hole density per-unit energy. Using this, we find the expectation of the $\hat{L}_z$ operator as a function of energy
\begin{equation}
\langle \hat{L}_z \rangle  = \frac{\Tr\left[G^n(E)\cdot \hat{L}_z \right]}{\max_E\{\Tr\left[G^n(E)\right]\}}
\label{eq:ExLz}
\end{equation}
where $\max_E\{\Tr\left[G^n(E)\right]\}$ returns the peak subgap quasiparticle concentration.
In the Cartesian coordinate system, the $\hat{L}_z$ operator is written as 
\begin{equation}
\hat{L}_z = \hat{x}\cdot \hat{p}_y - \hat{y}\cdot \hat{p}_x
\label{eq:Lz}
\end{equation}
The position operators $\hat{x}, \hat{y}$ are diagonal in the tight-binding basis, with each entry a multiple of the lattice constant ($a$). For example, if we consider two points along the $x$ and $y$ axis, and one along $z$, we have the following position operators
\begin{equation}
\mathbb{I}_x \otimes \hat{y} = 
a\left(\begin{matrix}
    1     & 0 & 0 & 0\\
   0 &   2  & 0 & 0\\
   0 & 0 & 1 & 0\\
   0 & 0 & 0 & 2\\
   \end{matrix}\right)
   \hspace{1cm}
   \hat{x}\otimes\mathbb{I}_y = 
a\left(\begin{matrix}
    1     & 0 & 0 & 0\\
   0 &   1  & 0 & 0\\
   0 & 0 & 2 & 0\\
   0 & 0 & 0 & 2\\
   \end{matrix}\right)
   \label{eq:x}
  \end{equation}
  
 The linear momentum operators can be written as 
 \begin{equation}
 \hat{p}_i = -i\frac{ m}{\hbar}[\hat{x}_i,\mathcal{H}_0]
 \label{eq:p}
 \end{equation}
where the subscript $i$ is used to denote $x,y$ basis of the position and momentum operators. Using Eq.~\ref{eq:Lz},~\ref{eq:x},~\ref{eq:p}, we write the $\hat{L}_z$ operator and using Eq.~\ref{eq:ExLz} we find the expectation of $\hat{L}_z$ as a function of energy.

We employ the above procedure to compute the average angular momentum of the Andreev bound states in an SNS junction. The chemical potential is fixed to give us three occupied subbands ($\ell = 0, \pm 1$). 
 The average angular momentum of the subbands at $\Phi=0.01$, $\phi = \pi$ is shown in Fig.~\ref{fig:ABS_label_0Flux}. Next, we increase the axial magnetic flux through the nanowire to $\Phi=0.03$. The $\ell = \pm 1$ subband states further split (Fig.~\ref{fig:ABS3Flux}), and the angular momentum of the subbands is plotted in Fig.~\ref{fig:ABS_label_01Flux} for $\phi=\pi$. The zero angular momentum subband ($\ell=0$) has no azimuthal motion, and hence is unaffected by the axial field. 
A non-zero $\langle L_z \rangle$ for the $\ell = 0$ subband (Figs.~\ref{fig:ABS_label_0Flux},~\ref{fig:ABS_label_01Flux}) results from its hybridisation with the nearby $\ell = -1$ subband. This consequently decreases the $\langle L_z \rangle$ magnitude for the $\ell = -1$ subband w.r.t. $\ell = +1$.
 
From Figs.~\ref{fig:ABS_label_0Flux},\ref{fig:ABS_label_01Flux} we infer that each transverse subband is primarily composed of a single angular momentum eigenstate. This resemblance to the subband structure of an ideal cylindrical nanowire is the basis for Sec.~\ref{sec:1d_eff}, where we build an effective subband model by including of a fixed number of angular momentum subbands. 
In the next section we explain the importance of including only the zero angular momentum subband in the superconducting contacts.
\subsection{Zero angular momentum subband in the Superconductor}
The BdG Hamiltonian in a superconductor is given by 

\begin{equation}
\begin{bmatrix}
\mathcal{H}_0 &\Delta \\
\Delta^* & -\mathcal{H}_0^*
\end{bmatrix}
\begin{bmatrix}
u\\v
\end{bmatrix}
=
E
\begin{bmatrix}
u\\v
\end{bmatrix}
\label{eq:BdGSC}
\end{equation}
 For a cylindrical geometry with an azimuthal vector potential, $\mathcal{H}_0$ is given by
\begin{equation}
\mathcal{H}_0 = -\frac{\hbar^2}{2m^*}\frac{\partial^2}{\partial z^2} + \frac{1}{2m^*}\left(-i\hbar \frac{1}{R}\frac{\partial}{\partial \theta} -eA_{\theta}\right)^2 - \mu
\end{equation}
As discussed, the radial confinement due to the nanowire's cylindrical geometry gives rise to angular momentum subbands labelled by $\ell$.

We will analyze the eigenenergies of this superconductor in the presence and absence of a magnetic field.

\subsubsection{Zero magnetic field, $\mathbf{B = 0, A = 0}$}

Using the ansatz wavefunction $\exp(ik_zz)\exp(i\ell\theta)$, the diagonal elements of the BdG Hamiltonian can be written as
 $$h_{\ell} = \frac{\hbar^2k_z^2}{2m^*} + \frac{\hbar^2\ell^2}{2m^*R^2} - \mu = \frac{\hbar^2k_z^2}{2m^*} - \mu_{\ell}$$
 where $\mu_{\ell}$ is the effective chemical potential 
\begin{equation}\mu_{\ell} = \mu-\frac{\hbar^2\ell^2}{2m^*R^2} \label{eq:mu_eff} \end{equation}
 The BdG Hamiltonian simplifies to 
 \begin{equation}
 \mathcal{H}_{BdG} =
 \begin{bmatrix}
h_{\ell} &\Delta \\
\Delta^* & -h_{\ell}
\end{bmatrix}
 \end{equation}

and its eigenvalues $E$ are given by
\begin{equation}
E=\pm \sqrt{h_{\ell}^2 + \Delta_0^2}
\end{equation}
This is the well-known superconductor dispersion relation, with a gap of $\Delta_0$ on either side of the fermi level.

\subsubsection{Constant axial magnetic field, $\mathbf{B}=B_z\hat{z}$}

In the Coulomb gauge we can write the vector potential for this magnetic field as 
\begin{equation}
    \mathbf{A} = A_{\theta}\hat{\bm{\theta}}
\end{equation}
From Stoke's law
\begin{equation}
    \oint{\mathbf{A}\cdot Rd\bm{\theta}} = \int{\mathbf{B}\cdot d\mathbf{A}}
\end{equation}
Exploiting the symmetry of the cylindrical geometry, the above equation can be simplified to 
\begin{equation}
    A_{\theta} = \frac{\Phi\hbar}{e R}
\end{equation}
 Using the same ansatz $\exp(ik_zz)\exp(i\ell\theta)$, the diagonal elements can be written as
  \begin{equation}
  \zeta^e_{\ell} = \frac{\hbar^2k_z^2}{2m^*} + \frac{\hbar^2\left(\ell-\Phi\right)^2}{2m^*R^2} - \mu = \frac{\hbar^2k_z^2}{2m^*}-\mu_{\ell} - \mathcal{E}_{\ell} 
  \label{eq:zeta_e}
  \end{equation}
  
  \begin{equation}
  \zeta^h_{\ell} = \frac{\hbar^2k_z^2}{2m^*} + \frac{\hbar^2\left(\ell+\Phi\right)^2}{2m^*R^2} - \mu = \frac{\hbar^2k_z^2}{2m^*}-\mu_{\ell} + \mathcal{E}_{\ell} 
   \label{eq:zeta_h}
  \end{equation}

for the electron and hole parts respectively. The effective chemical potential $\mu_{\ell}$ is defined in Eq.~\ref{eq:mu_eff}, and the field-coupling term $\mathcal{E}_{\ell} = \frac{\hbar^2\left(2\ell \Phi\right)}{2m^*R^2}$.
We note that 
\begin{equation}
\zeta^{e (h)}_{\ell} = h_{\ell} \mp \mathcal{E}_{\ell}    
\end{equation}
The BdG Hamiltonian can then be written as
 \begin{equation}
 \mathcal{H}_{BdG} =
 \begin{bmatrix}
h_{\ell} &\Delta \\
\Delta^* & -h_{\ell}
\end{bmatrix} 
- \mathcal{E}_{\ell} \mathbb{I}
 \end{equation}
and the eigenvalues $E$ are given by
\begin{equation}
E=\pm \sqrt{h_{\ell}^2 + \Delta_0^2} - \mathcal{E}_{\ell}
\end{equation}
Thus, we see that a magnetic field ``shifts'' the superconducting gap. It is no longer centred at the fermi level. 

\begin{figure}[!htb]
     \begin{center}
        \subfigure[$\ell \neq 0$ in the contacts]{%
 \label{fig:lneq0}         
    \includegraphics[width=0.45\textwidth]{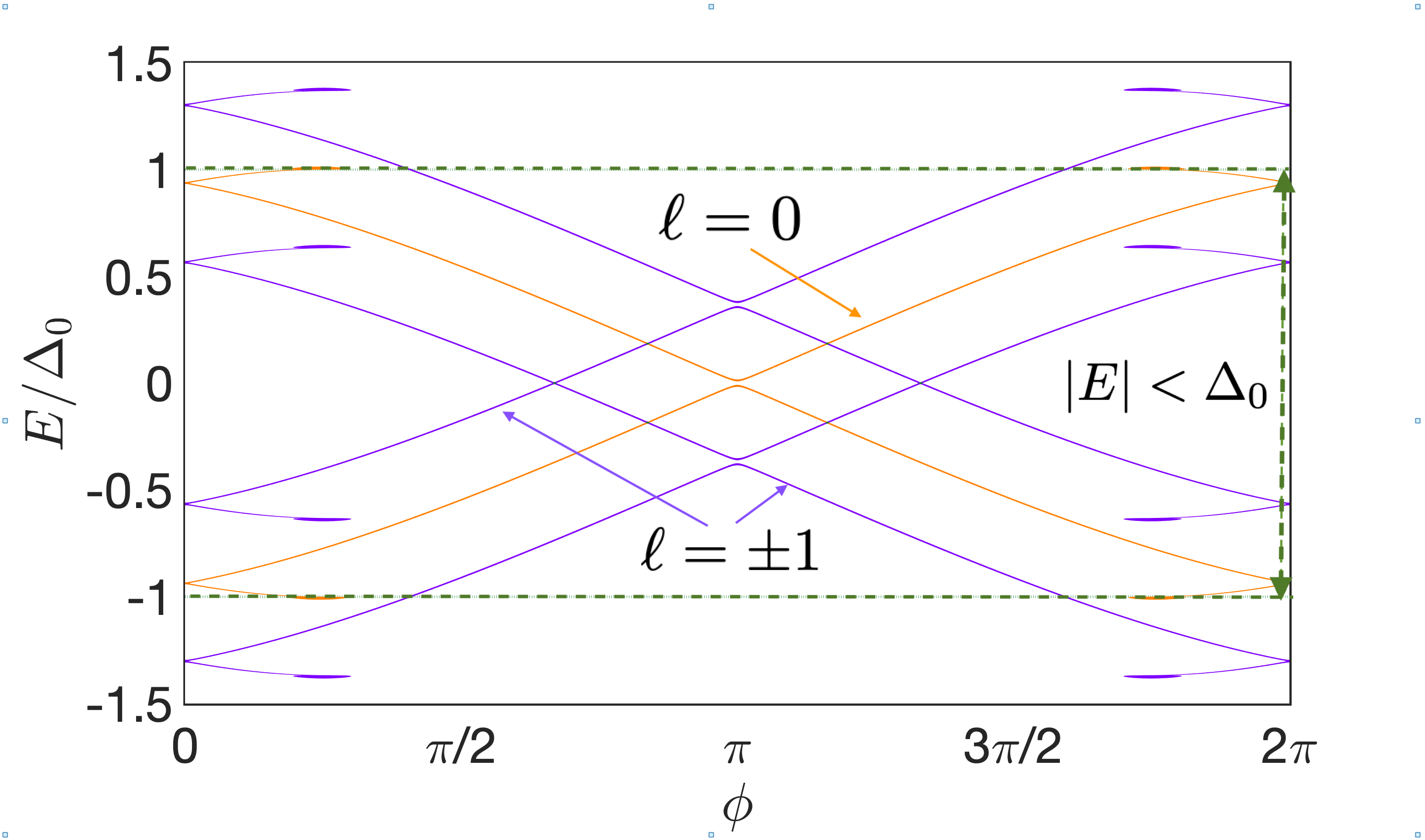}
        }\\%
        \subfigure[$\ell = 0$ in the contacts]{%
            \label{fig:leq0}
            \includegraphics[width=0.45\textwidth]{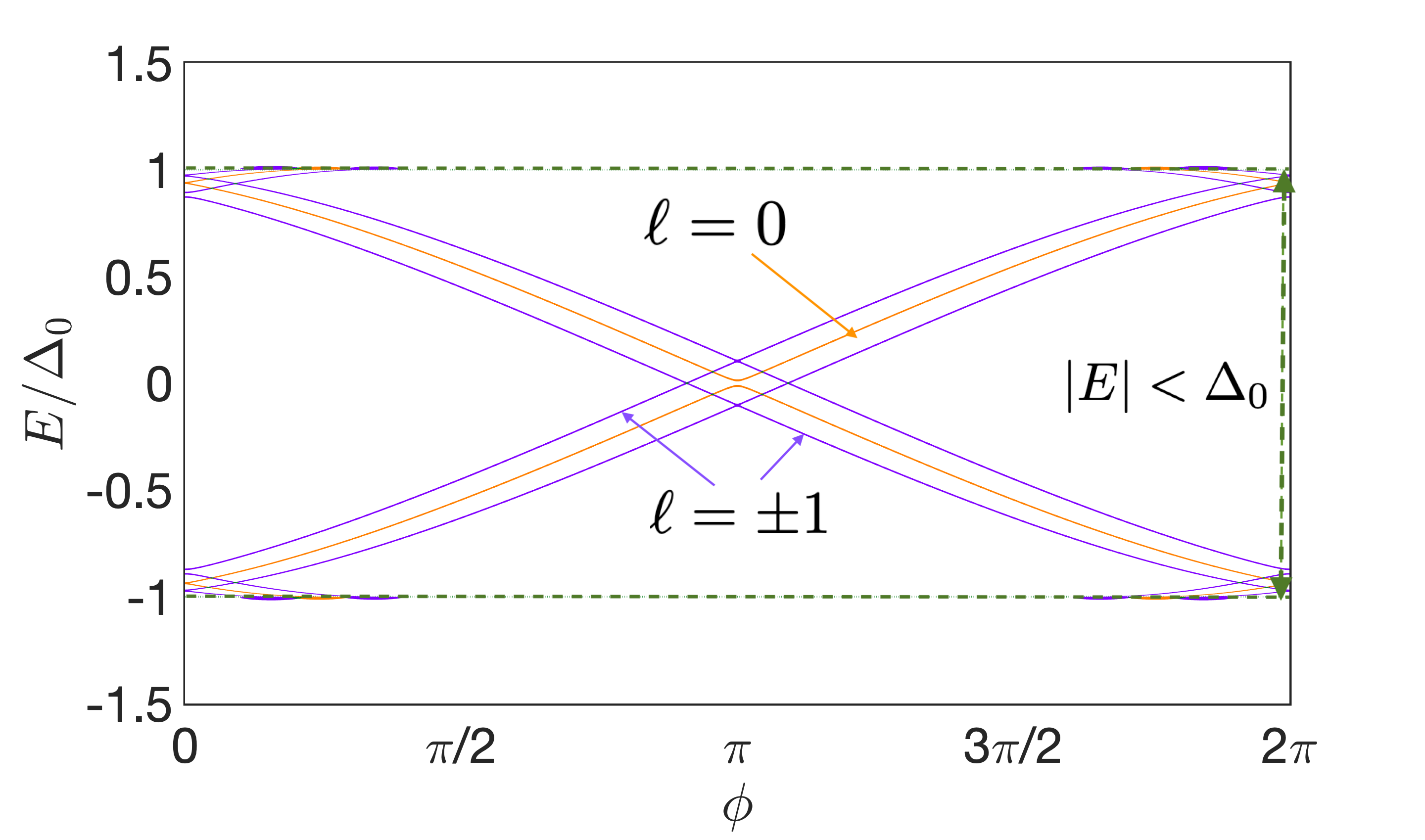}
        }
\end{center}
 \caption{%
The Andreev bound state spectrum in an SNS unction for a nanowire length $L= 160$ nm, chemical potential $\mu=30\Delta_0$ and healing length $\xi_0=222$ nm. The purple curves are the $\ell=\pm 1$ subbands, and the $\ell=0$ state are plotted in orange. The green dotted horizontal lines indicate the sub-gap $(E<\Delta_0)$ region. We observe a shift of the superconducting gap when we have (a) $\ell \neq 0$ in the contacts. In (b), we constrain the contacts to have $\ell = 0$ and confirm that the gap does not shift. The ABS curves of the $\ell = \pm 1$ subbands are phase shifted while the $\ell = 0$ subband is unaffected. The plot linewidths are proportional to the density of states.
 }%
 \label{fig:SupGap}
\end{figure}

While this may be a good model for a superconducting ``nanowire'', experimental setups usually 
involve a superconductor sputtered into quasi-planar contacts\cite{IQC,Frolov} which naturally support only the $\ell=0$ subband. The geometry of the superconducting contacts implies a large $\ell \neq 0$ subband energy, and can safely be assumed to remain unpopulated. This difference in geometry prompts the inclusion of $\ell \neq 0$ subbands in the nanowire, and their exclusion in the superconductor. 

For $\ell=0$, we have $\mathcal{E}_{\ell}=0$, and thus the superconducting gap stays centered at the fermi level. This is illustrated in Fig.~\ref{fig:SupGap} -- when $\ell \neq 0$ subbands are included in the superconductor (Fig.~\ref{fig:lneq0}), the ABS move vertically as a result of the shift in the superconducting gap. However, with only the zero angular momentum subband $\ell=0$ in the superconductor (Fig.~\ref{fig:leq0}), the ABS curves are horizontally phase shifted whilst the superconducting gap remains fixed. This shift is proportional to the applied flux, and the angular momentum of quasiparticles occupying the subband.

From Eqs.~\ref{eq:zeta_e},~\ref{eq:zeta_h}, the effective chemical potential for electron-like (hole-like) particles in the $\ell$ subband in the N-section is given by
\begin{equation}
    \mu_{\ell}^{e(h)} = \mu - \frac{\hbar^2}{2m^*R^2}\left(\ell \mp \Phi\right)^2
    \label{eq:mu_eff2}
\end{equation}
 The electron and hole wavenumbers can then be written as a function of energy $(E)$ 
\begin{equation}
    k^{e(h)}_{\ell}(E) = \frac{\sqrt{2m^*}}{\hbar}\sqrt{\mu^{e(h)}_{\ell} \pm E}
    \label{eq:kekh}
\end{equation}
%
\subsection{\label{sec:1d_eff} The 1-dimensional effective subband model}
As outlined above, it is important to ensure that we only have the $\ell = 0$ subbands in the contacts. We also note from Eqs.~\ref{eq:mu_eff},~\ref{eq:mu_eff2} that we can incorporate the effect of the angular momentum subbands via an effective potential $\mu_{\ell}$, and a field-coupling term $\mathcal{E}_{\ell}$. 

The tight-binding Hamiltonian of the nanowire can be written as
\begin{equation}
 \mathcal{H} = \begin{bmatrix}
    A  & B & 0 & \dots & 0 \\
    B^{\dagger} & A & B & 0 & 0\\
    0 & \ddots & \ddots & \ddots & 0\\
    \vdots & 0 & \ddots & \ddots & B\\
    0 & 0 & 0 & B^{\dagger} & A
    \end{bmatrix}
    \label{eq:H0}
    \end{equation}
    \begin{equation}
A = \left(\alpha_{-l} \oplus \dots \oplus \alpha_{+l} \right)  \text{ and }   B= \mathbb{I}_{N_{\ell}} \otimes \beta      
\label{eq:AB}
\end{equation}
\begin{equation}
\alpha_{l} = 
\begin{bmatrix}
    2t_z - \mu + t_y\left(l-\Phi\right)^2 & 0 \\
    0 & -2t_z + \mu-t_y\left(l+\Phi\right)^2
    \end{bmatrix} 
        \label{eq:alpha_l}
    \end{equation}
    \begin{equation}
    \beta = 
\begin{bmatrix}
    -t_z & 0 \\
    0 & t_z
    \end{bmatrix}   
        \label{eq:beta_l}
\end{equation}
where $N_{\ell}$ is the number of subbands. (For example, $N_{\ell}=3 \implies $ ($\ell={-1,0,1}$)) and $\mathbb{I}_n$ is the $n \times n$ identity matrix.

Meanwhile, the Hamiltonian of the contacts takes a similar form with
\begin{equation}
A = \left(\mathbb{I}_{N_{\ell}} \otimes \alpha_{0} \right)  \text{ and }   B= \mathbb{I}_{N_{\ell}} \otimes \beta 
     \end{equation}

\section{\label{sec:oscillations} Supercurrent Oscillations}
\subsection{Clean junction}
\begin{figure}[!htb]
     \begin{center}
         \subfigure[ $\ell = 0$]
 {\label{fig:CC_1subband}         
    \includegraphics[width=0.45\textwidth,keepaspectratio]{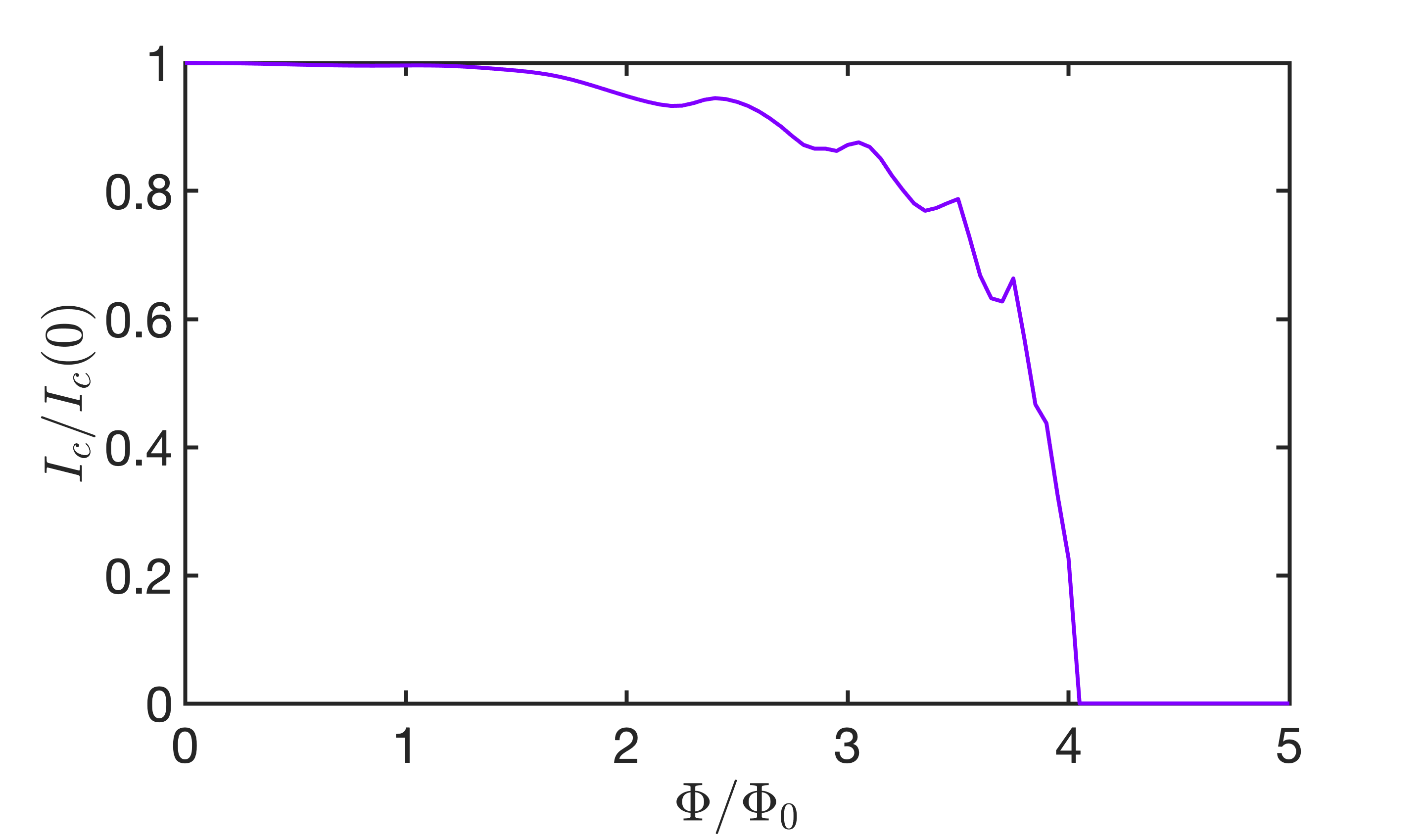}
        }\\%
        \subfigure[ $\ell = 0$]
            {\label{fig:CPR1_B}   
            \includegraphics[width=0.45\textwidth,keepaspectratio]{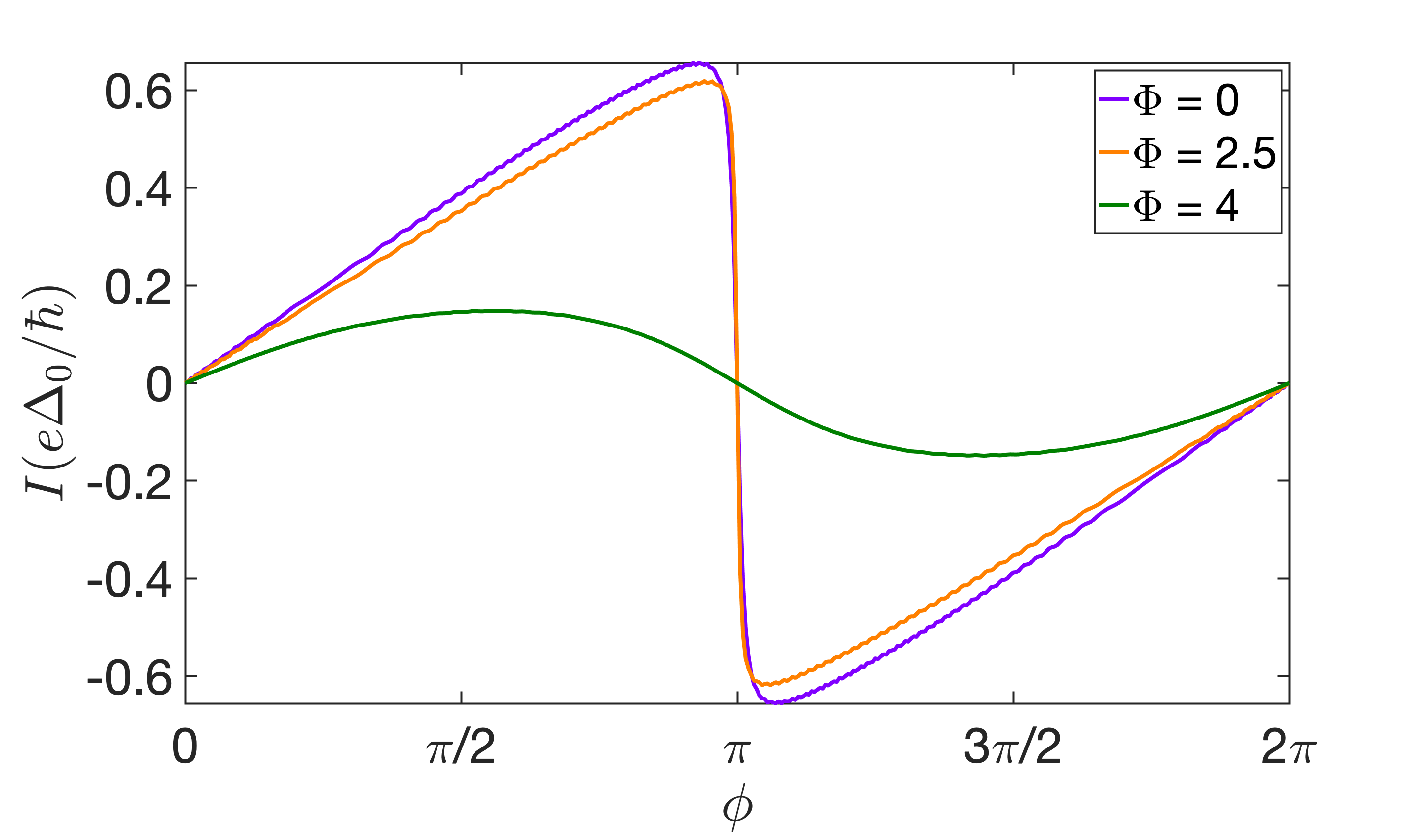}
        }
\end{center}
 \caption[Critical current and current-phase-relation in a clean nanowire with one occupied subband $(\ell=0)$]{(a) The critical current is plotted as a function of the applied magnetic flux for a single occupied subband. (b) The current-phase-relation (CPR) is plotted for $\Phi=0$, $\Phi=2.5,$ and $\Phi=4$. With a single occupied subband, the CPR retains its sawtooth shape, peaking near $\phi=\pi$. In the absence of inter-subband interference, oscillations in the critical current are not observed. The simulations were performed for $L=160$ nm, $\mu=30\Delta_0$, $\xi_0=222$ nm.}%
\end{figure}

We compute the CPR of an SNS junction at finite axial magnetic fields, assuming a shell conduction model, and a nanowire diameter of 60 nm. Temperature is set to $T=100$ mK in all the simulations. In Fig.~\ref{fig:CPR1_B}, we show the CPR as a function of the magnetic flux for a single occupied subband. With only the $\ell=0$ subband populated, there is no phase shift in the ABS, and the CPR retains its saw-tooth shape with a maximum near $\phi=\pi$. The critical current as a function of the flux is plotted in Fig.~\ref{fig:CC_1subband}. The gradual fall in the critical current  can be attributed to the decrease in average quasiparticle momentum with increasing flux, as shown in Eq.~\ref{eq:fermi_vel}. Eventually, at $\Phi = 4.04$ the band depopulates $\left[\min\left(\mu_{\ell}^e,\mu_{\ell}^h\right) = 0\right]$ and the current falls to zero. We observe in Fig.~\ref{fig:CC_1subband} that the critical current does not monotonically decrease to zero, particularly for $\Phi \in [3,4]$. The appearance of these small oscillations is due to the interference with the quasiparticles normally reflected from the N/S interfaces. 
 As discussed in Appendix~\ref{sec:BeyondAA}, the discontinuity in the density of states gives rise to normal reflections. These reflected quasiparticles interfere and result in the non-monotonic decrease of the single subband critical current.
 
\begin{figure*}[!htb]
     \begin{center}
         \subfigure[ $|\ell| = 0,1$]
 {\label{fig:CC_3subband}         
    \includegraphics[width=0.45\textwidth,keepaspectratio]{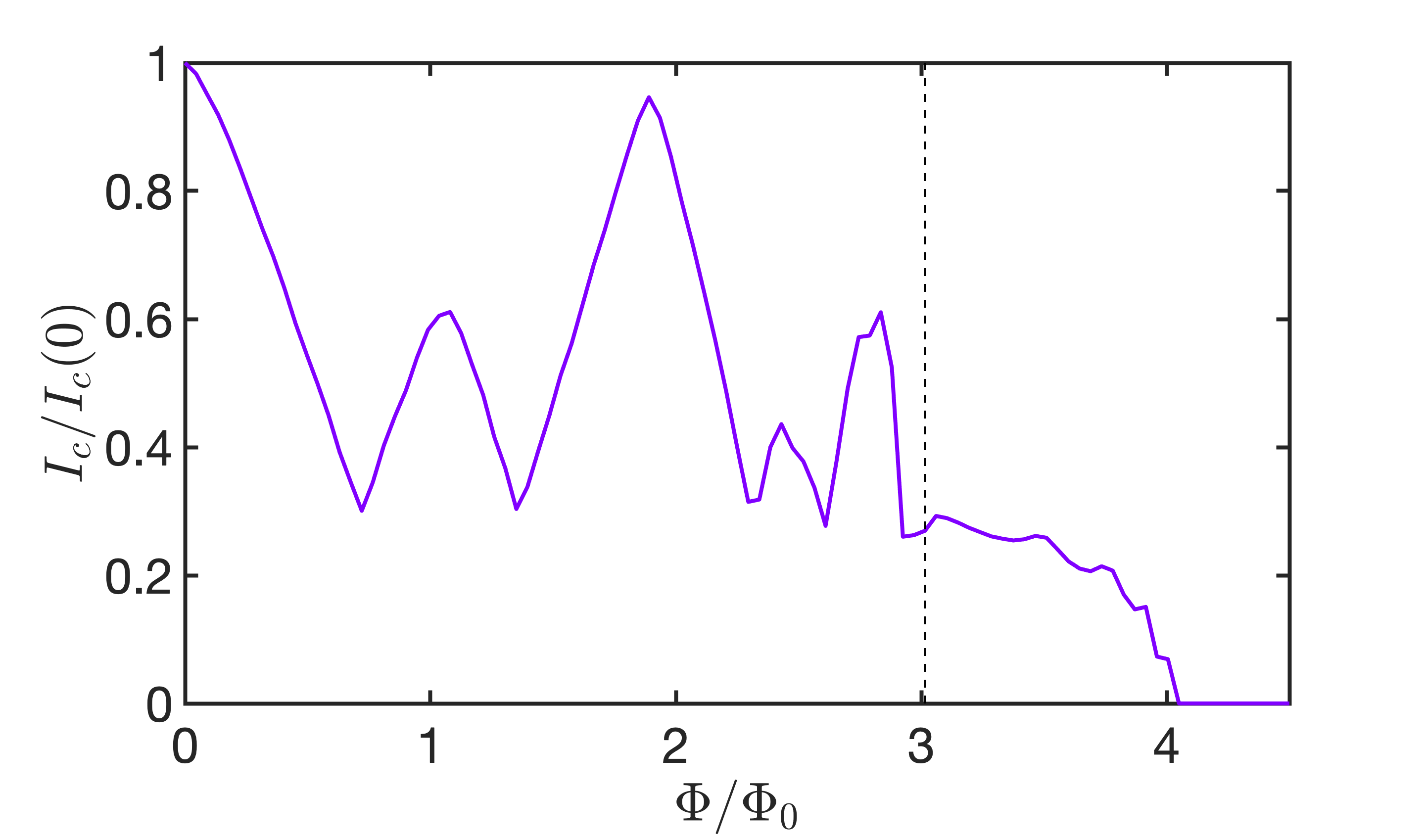}
        }%
        \subfigure[ $|\ell| = 0,1$; $\Phi=0$]
            {\label{fig:CPR3_B}   
            \includegraphics[width=0.45\textwidth,keepaspectratio]{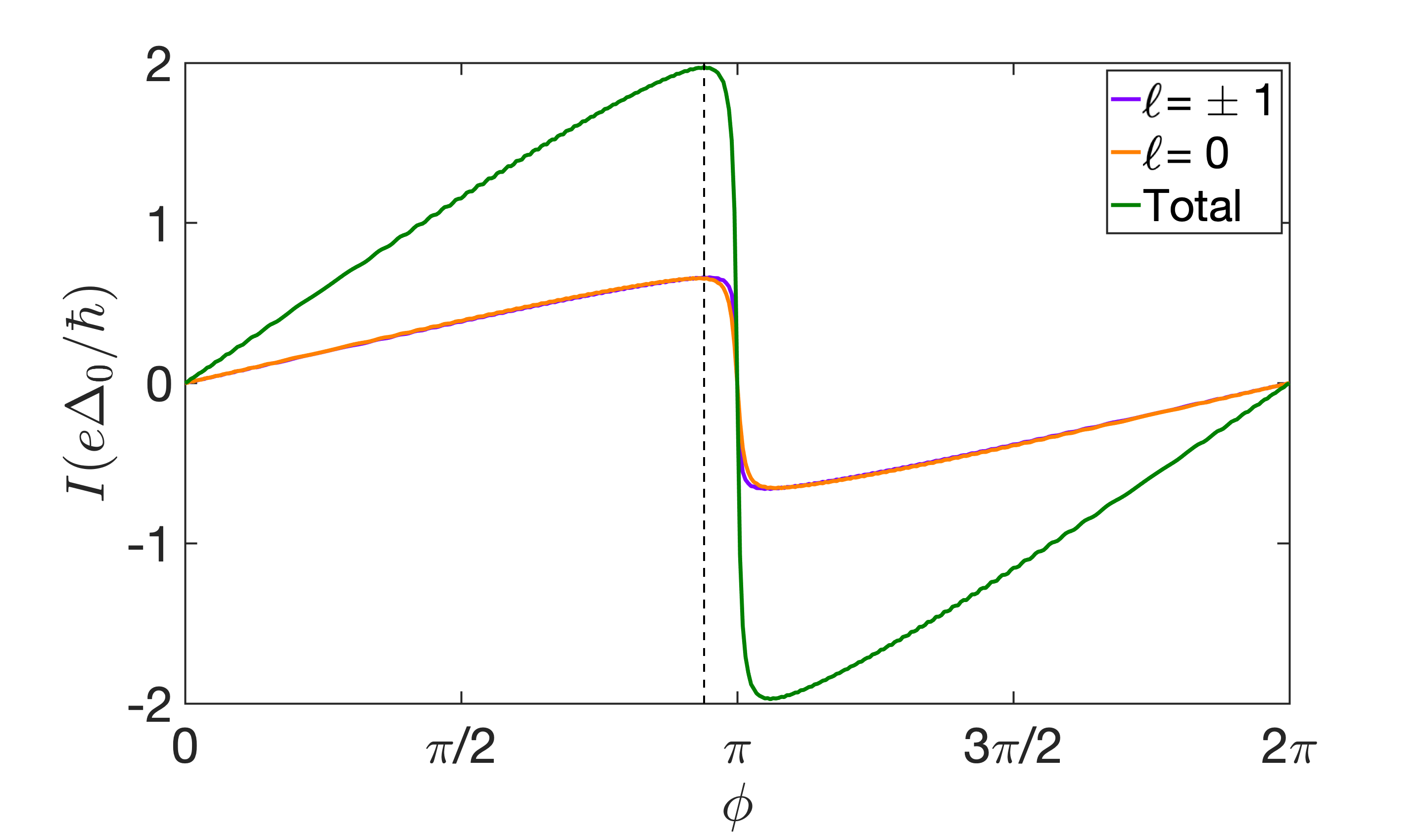}
        }\\%
         \subfigure[ $|\ell| = 0,1$; $\Phi=0.72$]
 {\label{fig:CPR1_min}         
    \includegraphics[width=0.45\textwidth,keepaspectratio]{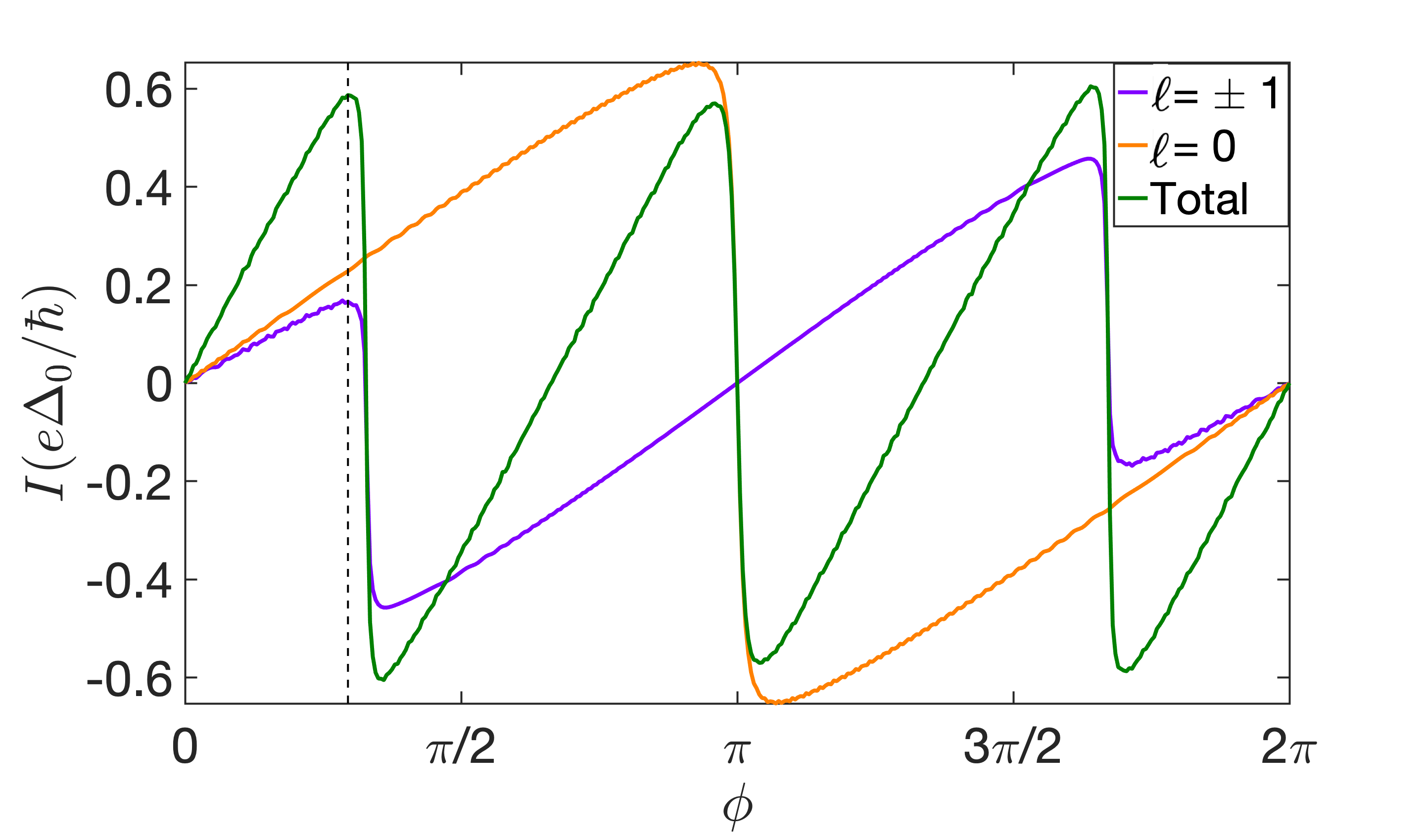}
        }%
        \subfigure[ $|\ell| = 0,1$; $\Phi=1.08$]
            {\label{fig:CPR1_secmin}   
            \includegraphics[width=0.45\textwidth,keepaspectratio]{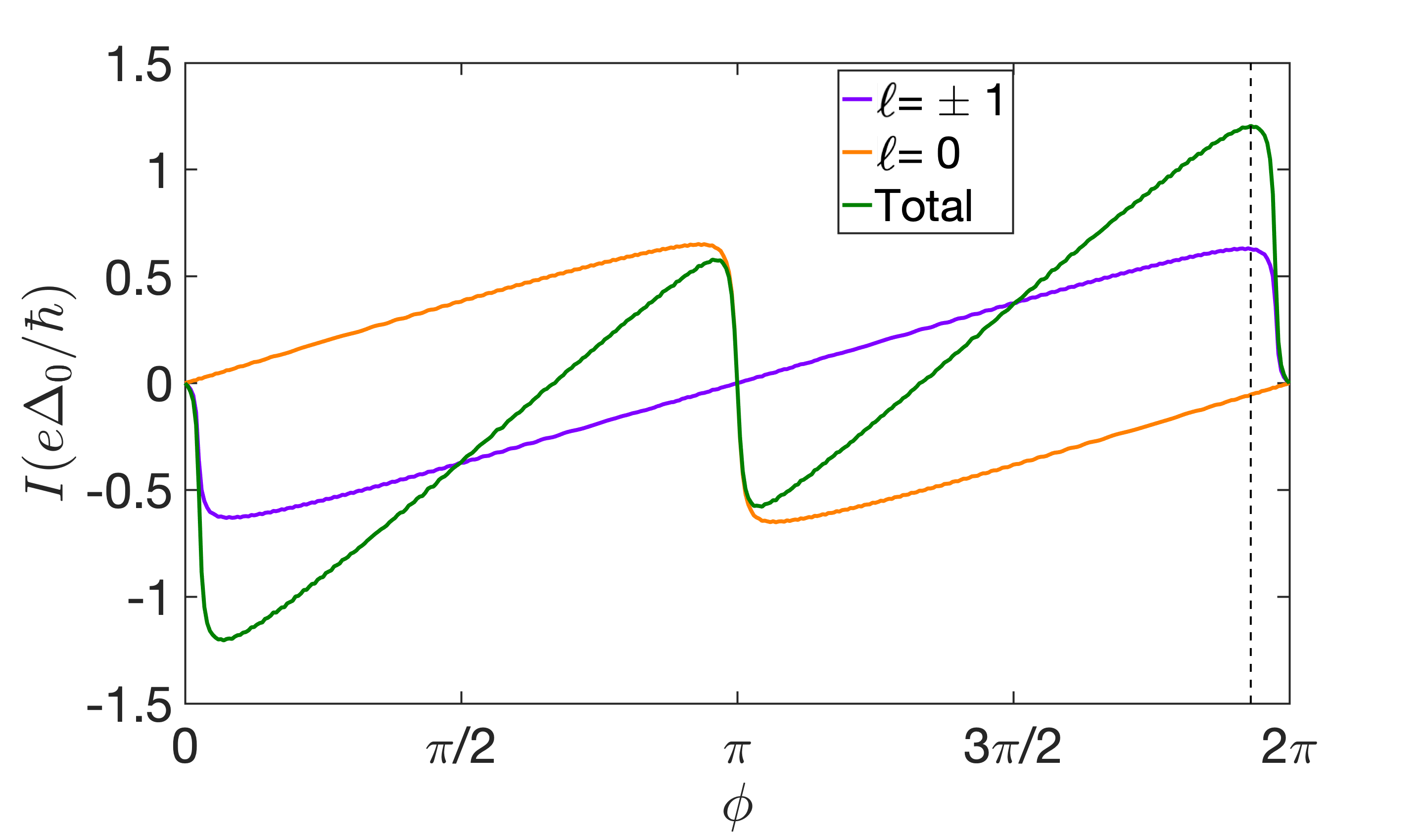}
        }
\end{center}
 \caption[Critical current oscillations and current-phase-relation in a clean nanowire with three occupied subbands $(\ell=0, \pm 1)$]{%
(a) The critical current is plotted as a function of the applied magnetic flux with three occupied subbands $(\ell = 0, \pm 1)$. The black dotted vertical line indicated the flux at which the $\ell=\pm 1$ subbands depopulate. The current-phase-relation (CPR) is plotted for (b) the primary maxima at $\Phi=0$, (c) the minima at $\Phi=0.72$, and (d) secondary maxima at $\Phi=1.08$ (see text). In (b),(c) and (d), the current supported by the $\ell=0$ and $\ell=\pm 1$ subband(s) is plotted in orange and purple respectively. The sum of these is the total CPR, which is plotted in green. The black dotted vertical lines indicate the phase difference corresponding to the critical current.
The $|\ell|=1$ subbands pick up a phase proportional to the difference in the quasiparticle momenta and hence the $|\ell|=1$ subband CPR transforms with the applied field. The simulations were performed for $L=160$ nm, $\mu=30\Delta_0$, $\xi_0=222$ nm.}%
\label{fig:3subband}
\end{figure*}

%
Next, we consider the case when three subbands are occupied ($|\ell| \leq 1$). The magnetic field evolution of the CPR is plotted in Fig.~\ref{fig:3subband}. Since the ABS for the $|\ell| = 1$ subbands are phase shifted in presence of a flux (Fig.~\ref{fig:leq0}), the total current is no longer maximum near $\phi = \pi$. The current in the junction is the sum over the individual subband currents, and consequently, the flux-dependent phase shift results in an interference pattern for the field evolution of the critical current. The phase shift in a subband CPR is proportional to the difference in the electron-hole wavenumbers $(k_{\ell}^e-k_{\ell}^h)$ (Eq.~\ref{eq:kekh}), and the length $L$ of the junction. Hence, the fluxes at which the subband currents constructively interfere need not occur at integer multiples of the flux quantum $\Phi_0=h/e$.

In Fig.~\ref{fig:CC_3subband} we plot the critical current for three occupied subbands as a function of the axial flux. We see several oscillations of the critical current before the $|\ell| = 1$ subbands depopulate at $\Phi = 3.04$.  At zero flux, the CPR of each subband is maximum near $\phi = \pi$ and hence they all add up constructively. As illustrated in Fig.~\ref{fig:CPR3_B}, each subband contributes equally to the critical current.

As the flux is increased, the electron-hole pairs in the $|\ell| = 1$ subbands pickup a phase and the subband CPRs no longer interfere constructively. Consequently, the critical current decreases with flux. At $\Phi=0.72$, the $\ell =0$ and $\ell = \pm 1$ subband CPRs are maximally out-of-phase, resulting in a local minima (node). The subband and total CPR at this node is plotted in Fig.~\ref{fig:CPR1_min}. Beyond $\Phi=0.72$, the critical current switches phase from $\phi<\pi$ to $\phi>\pi$ and the current increases again. This increase persists till $\Phi =1.08$ at which point the current is maximum near $\phi=2\pi$. At this flux, the $|\ell| = 1$ subband current peaks near $\phi=2\pi$ while it is negligible near $\phi=\pi$. Hence, this secondary peak -- which only involves contribution from $|\ell|=1$ subbands -- is approximately a two-third of the primary peak and corresponds to a phase pickup of $\pi$ in the aforementioned subbands. The subband and total CPR for the secondary peak is shown in Fig.~\ref{fig:CPR1_secmin}. As noted earlier, the magnitudes of the primary and secondary peaks progressively diminish due to the decrease in average quasiparticle velocity.  

\begin{figure}[!htb]
     \begin{center}
            {\includegraphics[width=0.5\textwidth]{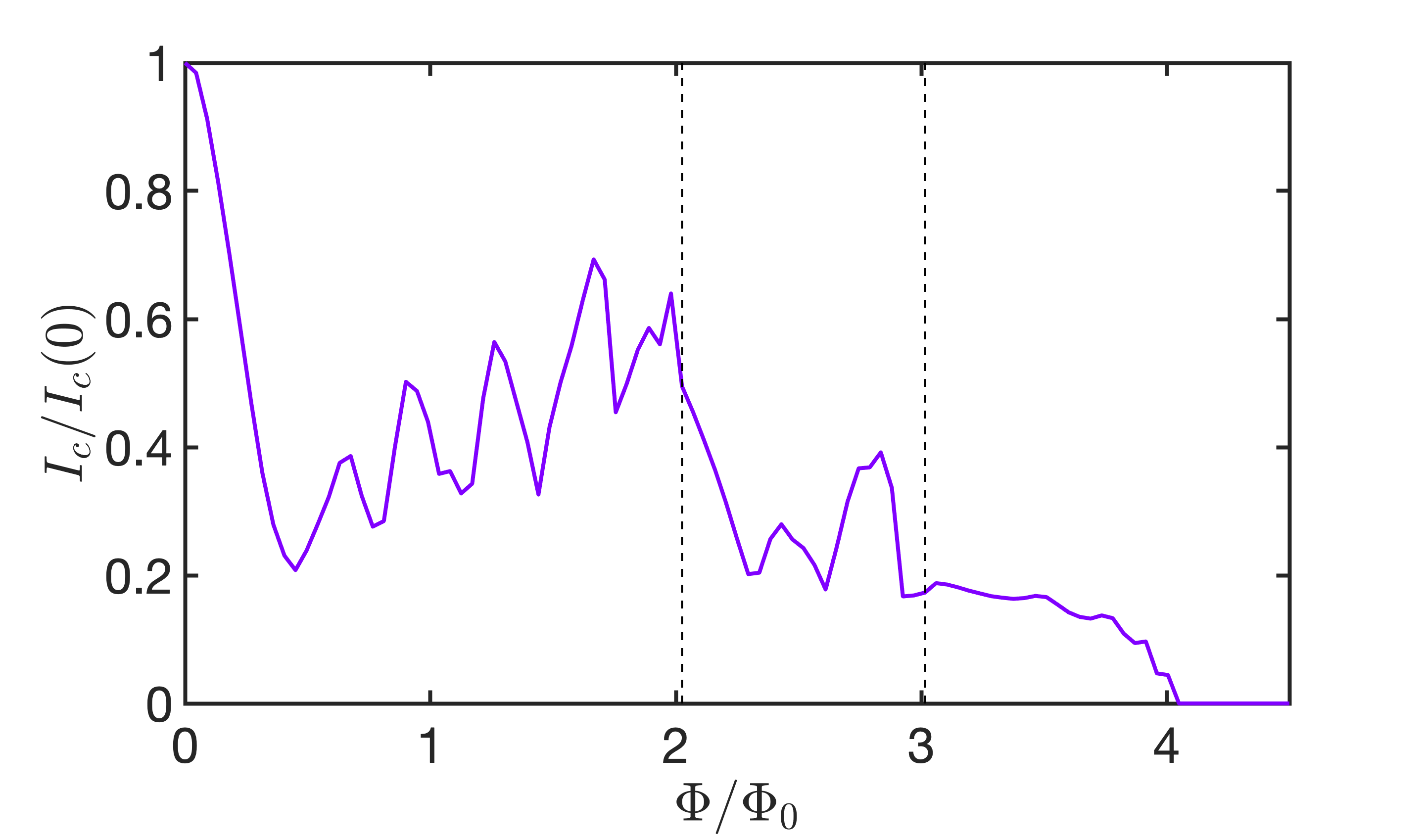}        }
\end{center}
 \caption[Critical current oscillations in a clean nanowire with five occupied subbands.]{%
 The critical current is plotted as a function of the applied magnetic flux with five occupied subbands. The black vertical dotted lines denote the depopulation of the subbands, in a descending order of the angular momentum quantum number. The $|\ell|=1,2$ subbands pick up a phase proportional to the difference in the quasiparticle momenta and hence the critical current oscillates with the applied flux. The simulations were performed for $L=160$ nm, $\mu=30\Delta_0$, $\xi_0=222$ nm.}%
 \label{fig:CC_5subband}
\end{figure}

Finally, we consider the situation when five subbands are occupied ($|\ell| \leq 2$). The critical current is plotted as a function of the magnetic flux in Fig.~\ref{fig:CC_5subband}. Once again, at $\Phi=0$ the subband currents are all in-phase and constructively interfere to give a maximum. In  presence of a magnetic field, the quasiparticles in the $|\ell|=1$ and $|\ell|=2$ subbands pick up different phases and hence, they do not appear to constructively interfere again in presence of a magnetic field to recover the zero field critical current. 

The absence of such oscillations with a single occupied subband (Fig.~\ref{fig:CC_1subband}) confirms the subband supercurrent interference as the causal agent.  
\subsection{Effect of disorder}
In order to simulate experimentally relevant conditions, we include a random uncorrelated onsite disorder potential $u \in [-W,W]$ in the semiconductor. This models phase-coherent scattering events in the junction. We parameterise the disorder by the mean free path ($\lambda_{mf}$), which is estimated from the disorder-averaged normal state conductance ($g$) using the following relation 
\begin{equation}
 g = \frac{2e^2}{h}N_{\ell}\frac{1}{\left(1+L/\lambda_{mf}\right)}
 \label{eq:mfp}
 \end{equation}
  $N_{\ell}$ is the number of subbands and $L$ is the length of the junction.

\begin{figure}[!htb]
     \begin{center}
             \subfigure[$\lambda_{mf} = 30$ nm, $L = 160$ nm]
             {%
                \label{fig:CC_L160_l30}
            \includegraphics[width=0.23\textwidth,keepaspectratio]{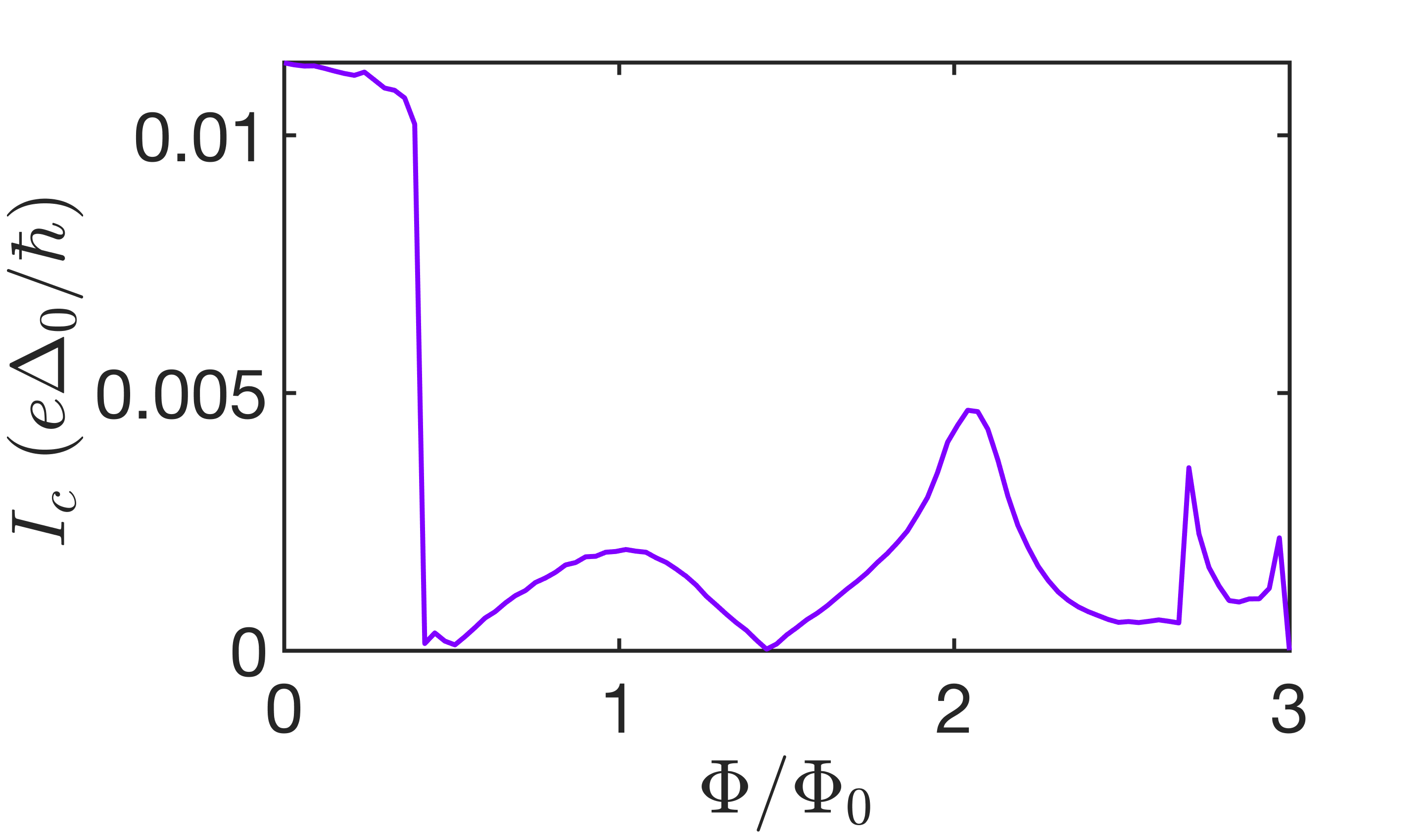}
        }
         \subfigure[$\lambda_{mf} = 30$ nm, $L = 160$ nm]
         {%
            \includegraphics[width=0.23\textwidth,keepaspectratio]{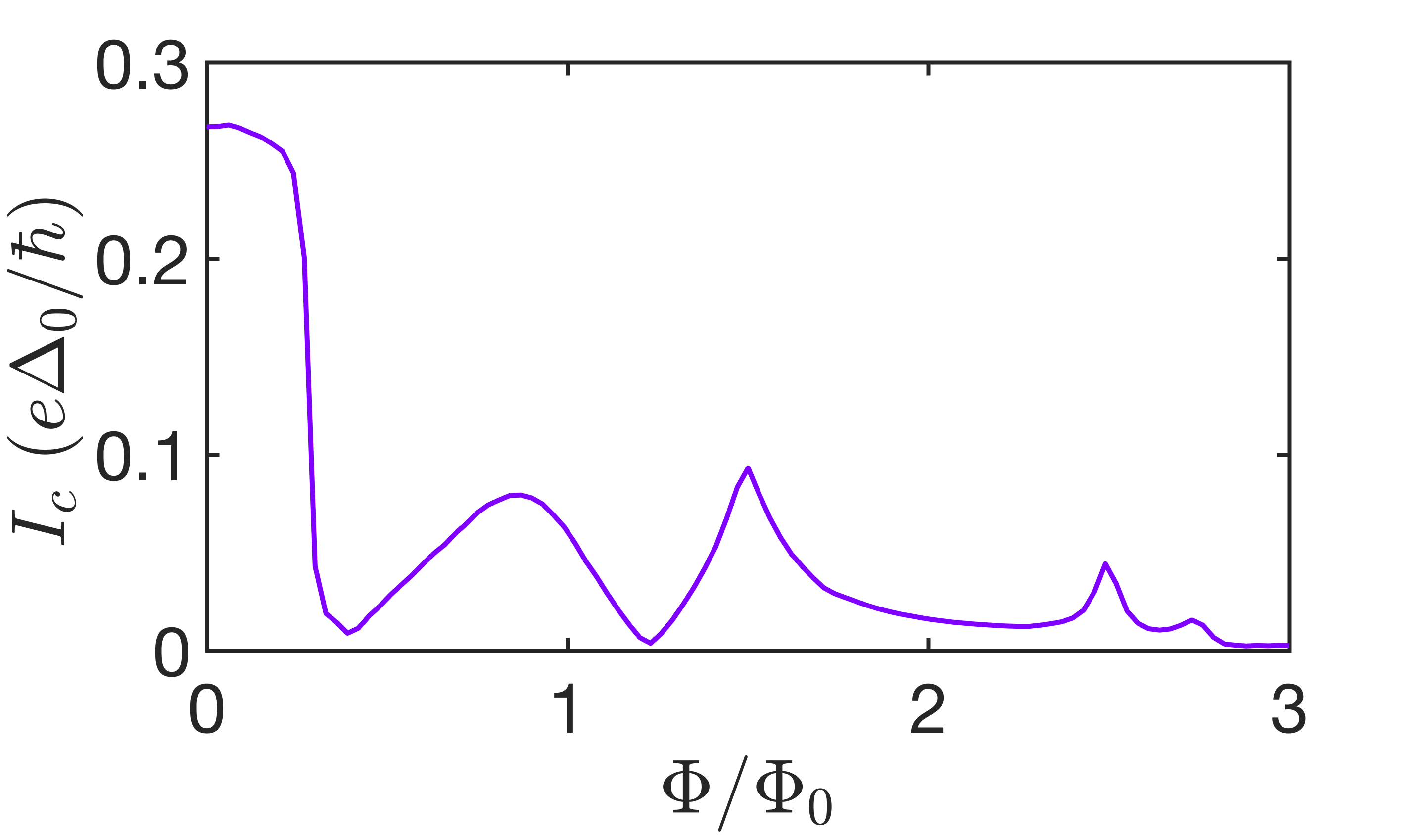}
        }\\%
         \subfigure[$\lambda_{mf} = 80$ nm, $L = 160$ nm]
         {%
            \includegraphics[width=0.23\textwidth,keepaspectratio]{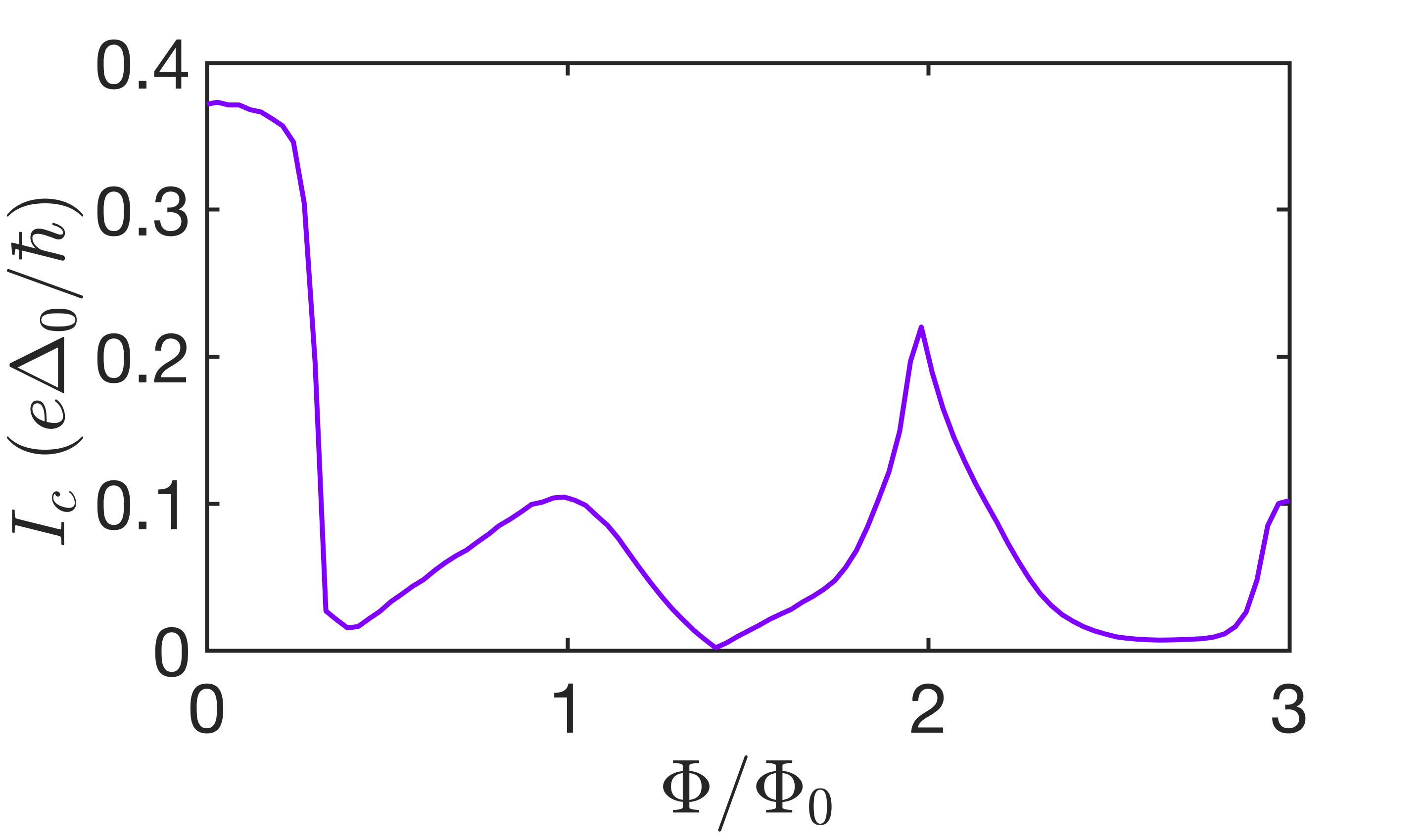}
            }
                 \subfigure[$\lambda_{mf} = 30$ nm, $L = 240$ nm]
                 {%
            \includegraphics[width=0.23\textwidth,keepaspectratio]{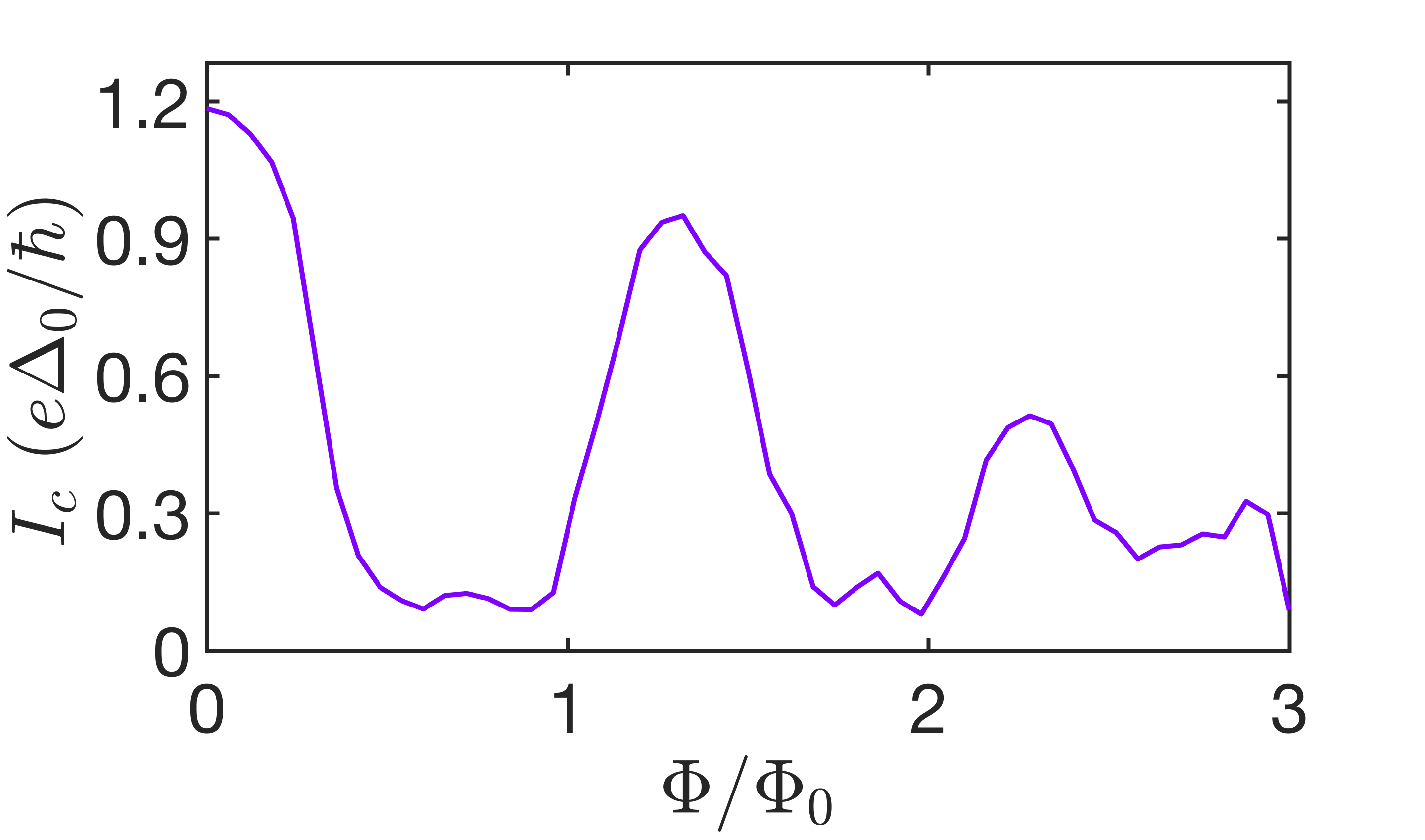}
        }%
        
\end{center}
 \caption{%
 The critical current oscillations for a disordered SNS Josephson junction. The disorder is parameterised by the mean-free path ($\lambda_{mf}$) which is calculated from the normal state disorder-averaged conductance using Eq.~\ref{eq:mfp}. Each sub-figure shows the critical current evolution for a particular realization of the disorder, and is labelled by the mean-free path $\lambda_{mf}$ and nanowire length $L$. The chemical potential $\mu=30\Delta_0$ for all the plots. (a) and (b) plot the oscillations for two different realisations of a random disorder potential resulting in $\lambda_{mf}= 30$ nm.}
  \label{fig:SNSDisorder}
\end{figure}

In Fig.~\ref{fig:SNSDisorder} we plot the critical current oscillations in a nanowire for particular realisations of the disorder. While the initial decay and the oscillations are still present, the secondary maxima are suppressed. In a clean nanowire with a saw-tooth CPR (which peaks near $\phi=\pi$ at zero field), at a magnetic flux $\Phi^*$ the $|\ell|=1$ subbands pickup a phase of $\pi$ and their CPRs peak near $\phi=2\pi$. The $\ell=0$ subband retains its sawtooth CPR with a negligible current near $\phi=2\pi$. As described in the previous subsection, this results in the secondary maximum.
Upon adding disorder to the nanowire, we depart from this saw-tooth CPR, tending towards a sinusoidal CPR which peaks further away from $\phi=\pi$ at zero field. Thus, there exists no $\Phi^*$ at which the $|\ell| = 1$ subband current peaks while the $\ell=0$ subband current is negligible. As a consequence of the sinusoidal CPR, on picking up a phase of $\pi$ the $|\ell| = 1$ subbands destructively interfere with the $\ell = 0$ subband, and this causes the suppression of the secondary maxima.

\begin{figure}[!htbp]
     \begin{center}
    \includegraphics[width=0.5\textwidth]{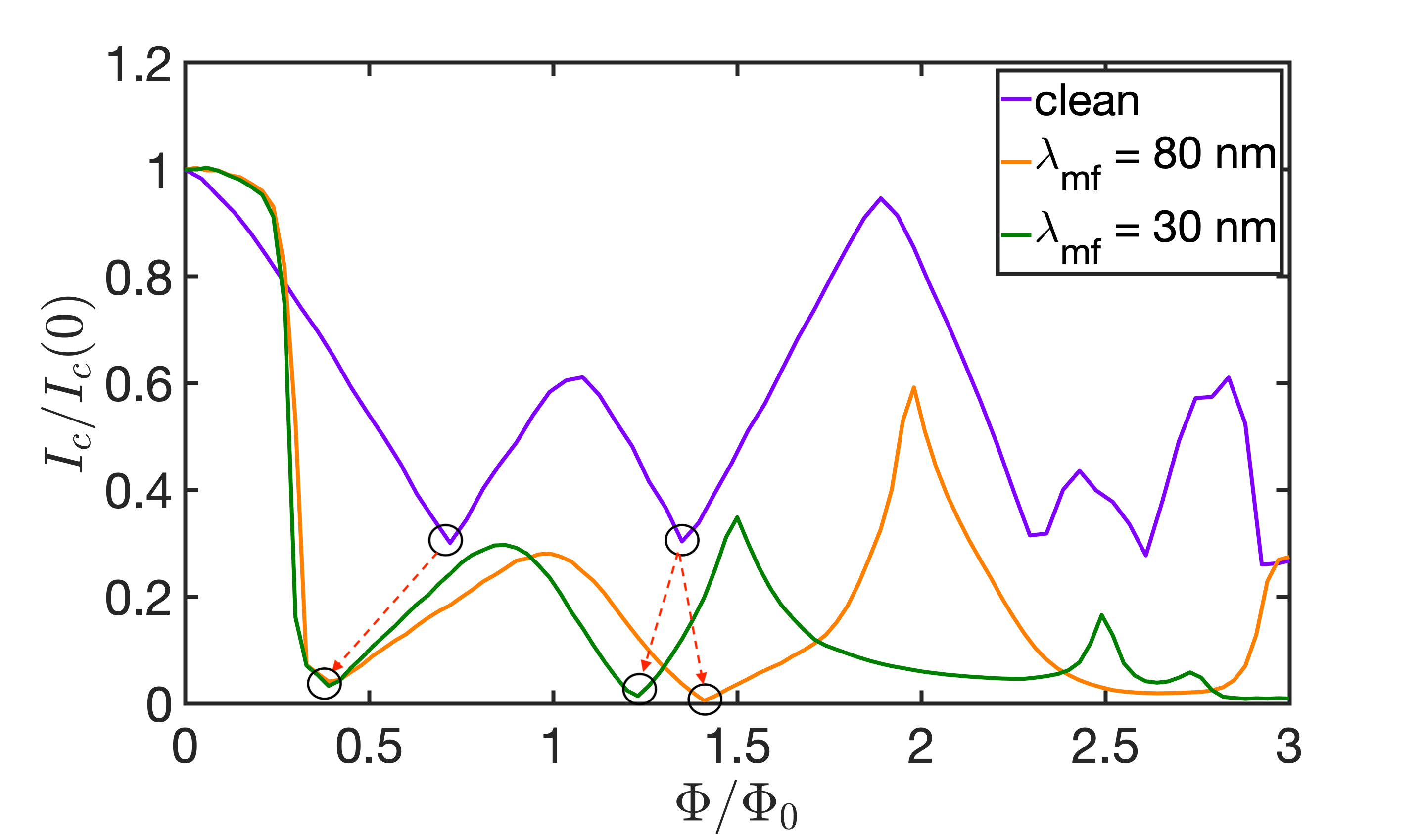}
\end{center}
 \caption{%
 The critical current as a function of the applied magnetic flux for a clean and disordered ($\lambda_{mf} = 30, 80$ nm) junction. Each critical current curve is normalized to its respective zero field value. We observe that the first node in the disordered junction occurs at a lower field as compared to a clean junction.}%
     \label{fig:CleanDis}         
\end{figure}

In the presence of scatterers the effective path traversed by the quasiparticles increases and hence, the subbands destructively interfere at a lower flux. This is shown in Fig.~\ref{fig:CleanDis}, where the first crticial current node in a disordered junction occurs at a lower field as compared to the clean nanowire. 

\begin{figure}[!htb]
     \begin{center}
        \subfigure[Clean Junction]{%
           \label{fig:clean}
            \includegraphics[width=0.45\textwidth]{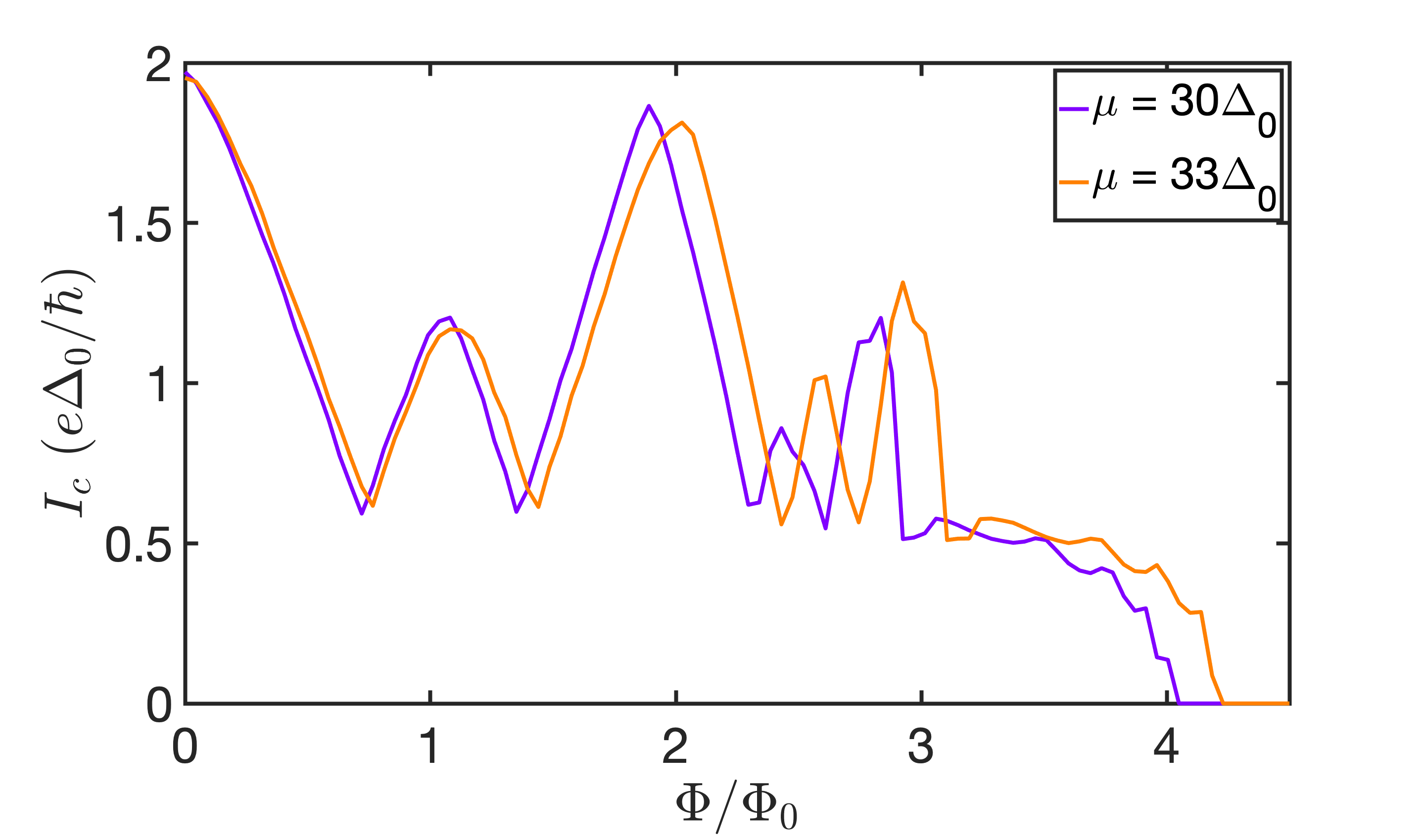}
        }\\%
      \subfigure[$\lambda_{mf} = 30$ nm, $L = 160$ nm ]{%
            \label{fig:dis1}
            \includegraphics[width=0.45\textwidth]{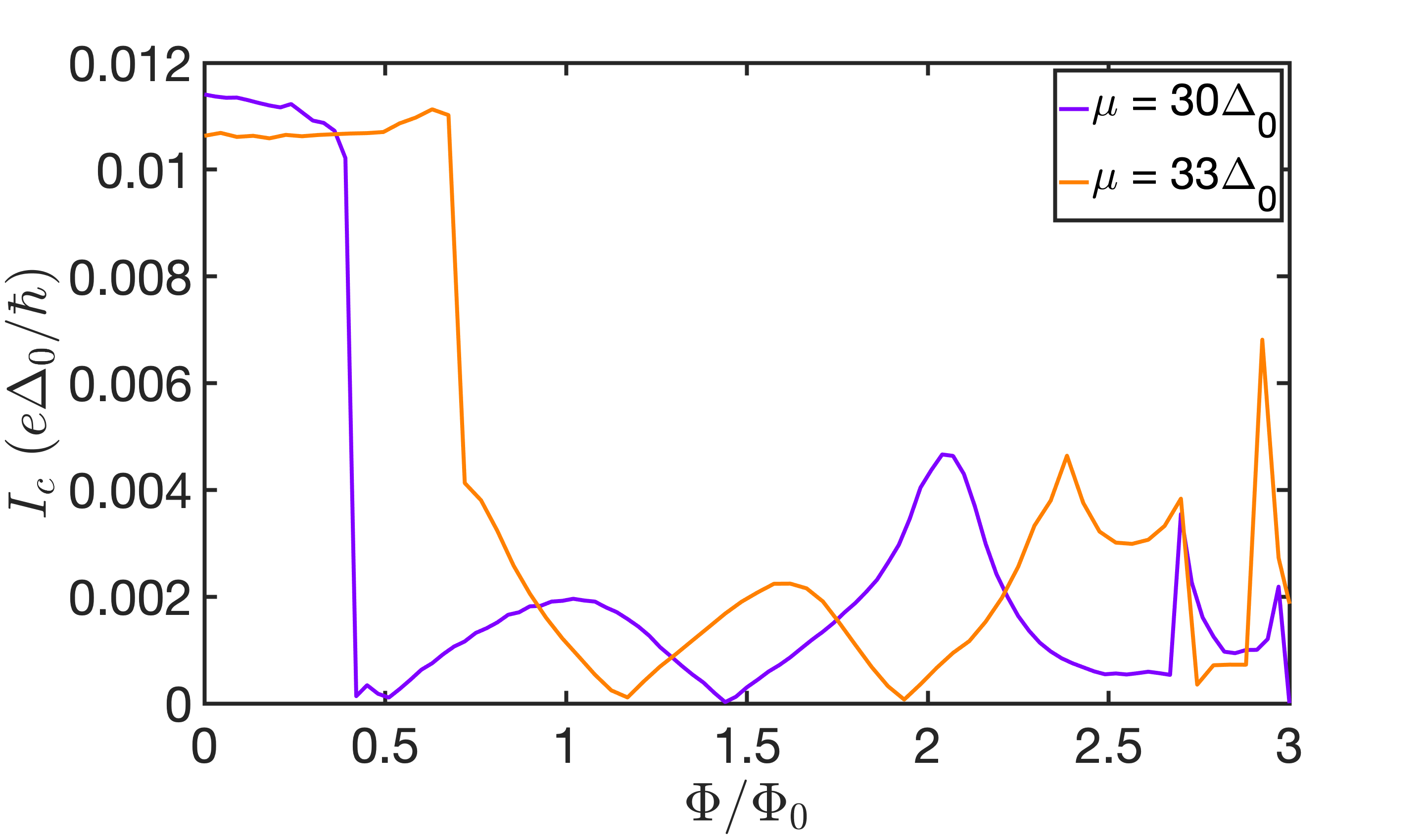}
        }\\%
        \subfigure[$\lambda_{mf} = 30$ nm, $L = 160$ nm]{%
            \label{fig:dis2}
            \includegraphics[width=0.45\textwidth]{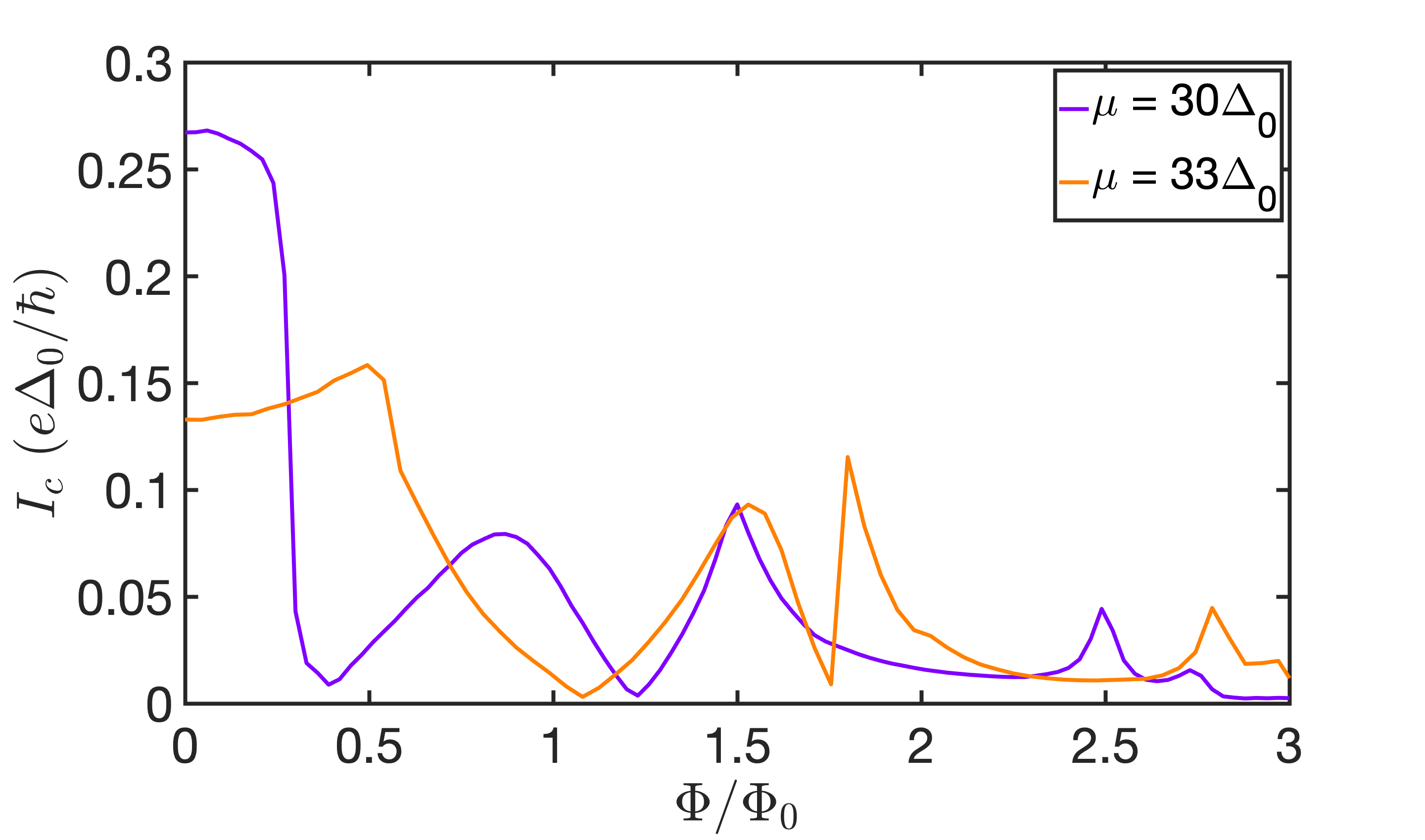}
            }
    \end{center}
    \caption{%
       Small fluctuations in the gate voltage change the chemical potential ($\mu$) between the blue ($\mu = 30\Delta_0$) and red ($\mu=33\Delta_0$) curves. $\Delta_0$ is the superconducting order parameter. We observe small variations in the oscillations on slightly varying the chemical potential in (a) a clean junction. For a disordered junction in (b),(c)  we observe larger fluctuations in the oscillations in response to small variations in the chemical potential. Two instances of the disorder potential are shown.  
     }%
   \label{fig:EffDis1}
\end{figure}

Furthermore, as shown by \citeauthor{Frolov}, the essential effect of disorder can be observed by the dependence of the critical current oscillations on the gate voltage. As shown in Fig.~\ref{fig:clean} for the clean nanowire, small variations in the gate voltage hardly cause any fluctuations in the oscillations. This is because small changes in the chemical potential do not change the number of occupied subbands and only weakly affects the quasiparticle transmission through the junction. However, in a disordered nanowire with a small mean free path, the quasiparticles traverse a longer path in the nanowire and hence, the critical current oscillations are significantly affected by the gate voltage. This is shown in Figs.~\ref{fig:dis1},\ref{fig:dis2} for two disorder realisations.

From Figs.~\ref{fig:SNSDisorder},\ref{fig:EffDis1} we infer that the critical current oscillations are highly sensitive to the gate voltage and the particular realisation of the disorder. Thus, a macroscopic current measurement indirectly gives us information about the microscopic specifics of the junction. However, while our model provides a qualitative understanding of the oscillations, the high sensitivity w.r.t. the microscopic parameters renders a  quantitative description of the experiment highly challenging.


\subsection{\label{sec:deph}Dephasing in the nanowire}
The analysis presented in the previous sections described the phase-coherent flow of quasiparticles in the junction. We now include phase-breaking processes \cite{Golizadeh,datta_2005} that may arise from any time-dependent potential, such as the interaction of an electron with the surrounding bath of phonons, other electrons, or fast-fluctuating charge noise caused by traps in defects. 
Although it is non-trivial to identify the dominant source of dephasing, we can adopt a phenomenological model by introducing a B\"{u}ttiker probe for the lattice background\cite{datta_2005}.  

We subsume these processes within the NEGF formalism by including a self-energy term for the lattice background $\Sigma^r_s$, proportional to the Green's function and the emission-absorption coefficients. This calls for a self-consistent computation of the Green's function and the bath self-energy \cite{Golizadeh}
\begin{align}
    G^r(E) &= \left(E\mathbb{I} + i \eta-\mathcal{H}-\Sigma^r_1-\Sigma^r_2 - \Sigma^r_s \right)^{-1}\\
    \Sigma^r_s &= D \times G^r(E)
\end{align}
where $\times$ denotes element by element multiplication. The elements of the matrix $D$ represent the correlation of the time-dependent interaction potential between pairs of lattice points. Adopting a homogeneous model, we assume uniform, elastic, and spatially uncorrelated interactions resulting in a diagonal $D$,
\begin{equation}
    D_{i,j}=D_0\delta_{i,j}
\end{equation}
for every pair of coordinates $z_i$ and $z_j$ in the nanowire. This model discards the off-diagonal elements of the Green's function, and hence relaxes both the phase and momentum of the quasiparticles in the nanowire.

\begin{figure}[!htb]
     \begin{center}
         \subfigure[]{%
           \label{fig:13a}
            \includegraphics[width=0.45\textwidth]{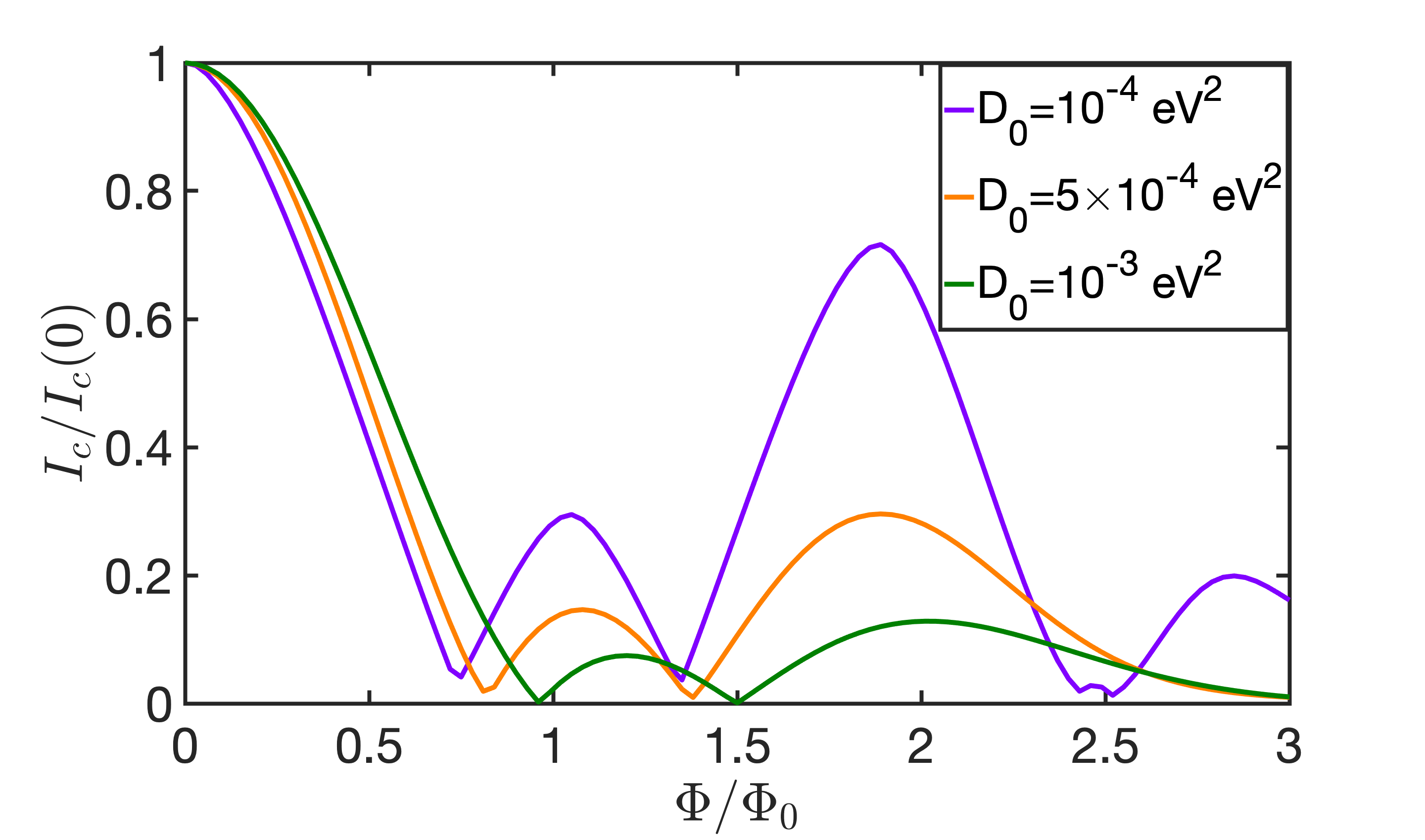}
         }\\%
       \subfigure[]{%
            \label{fig:13b}
            \includegraphics[width=0.45\textwidth]{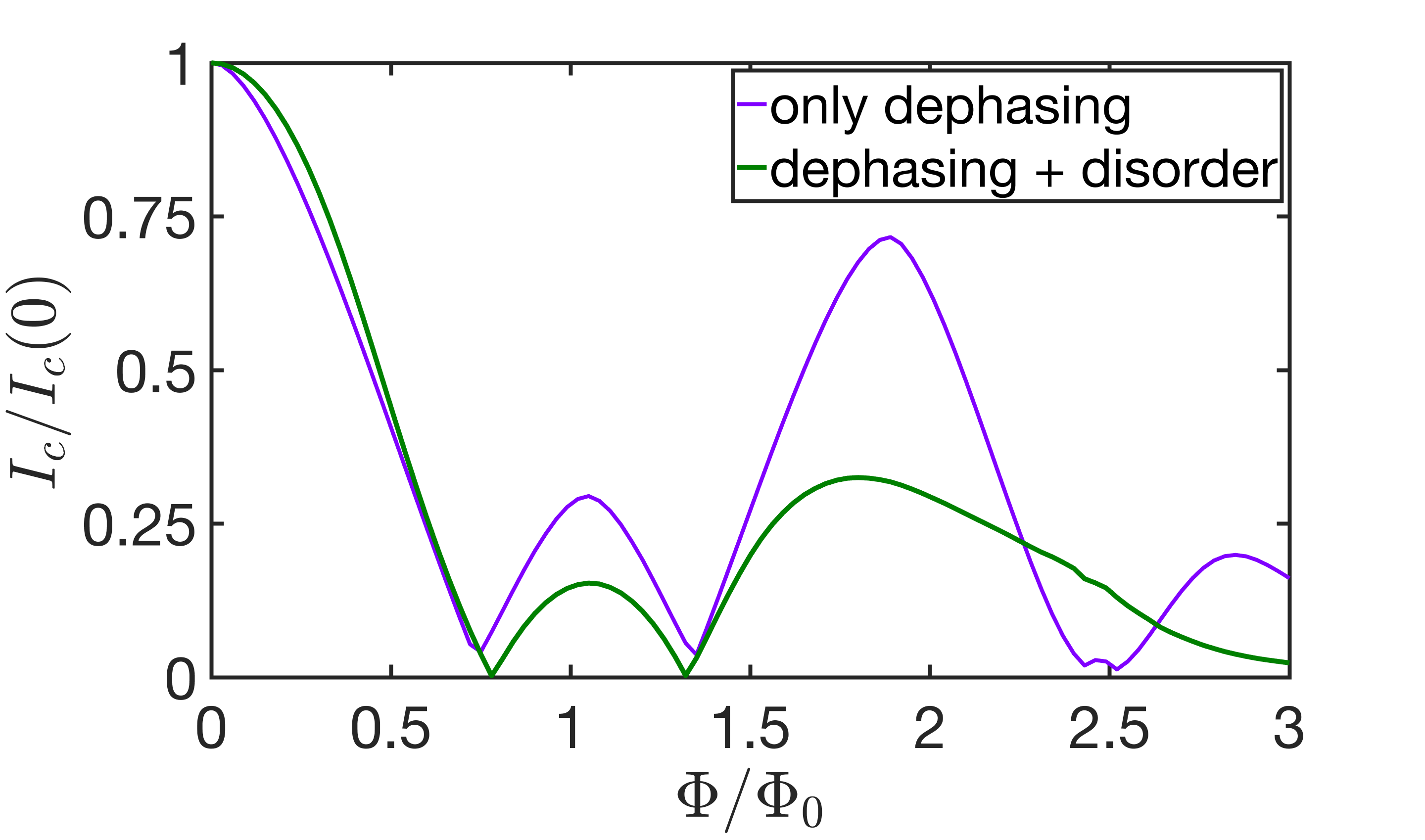}
         }
    \end{center}
    \caption{%
      Critical current oscillations with phase-breaking scattering processes in the nanowire. The dephasing interaction is parameterised by the coupling strengths $D_0$ listed in Table~\ref{sophisticatedtable1}. (a) The nanowire is free of disorder and only phase-breaking processes are involved. (b) The green curve corresponds to a nanowire with a random onsite potential distribution resulting in a mean free path $\lambda_{mf}=30$ nm, in addition to the phase-breaking processes with $D_0=1\times 10^{-4}$ $eV^2$. This is plotted in comparison with the corresponding disorder-free nanowire in (a) (purple curve). Each critical current curve is normalized to its respective zero field value. The simulations were performed for $L=160$ nm, $\mu=30\Delta_0$.
     }%
   \label{fig:PBProcesses}
\end{figure}

The magnitude of $D_0$ limits the phase relaxation length of the junction. Molecular beam epitaxy(MBE) and metalorganic vapour phase epitaxy grown InAs nanowires typically have a phase relaxation length on the order of a few-hundred nanometers $l_{\varphi} \sim 100-300$ nm\cite{Blomers2,Estevez_2010,KouwenhovenInAs}. In our model, $l_{\varphi}$ can be estimated from the statistical properties of universal conductance fluctuations (UCF)\cite{Blomers2,Estevez_2010,KouwenhovenInAs} The details of this calculation is presented in Appendix~\ref{sec:lphi}, and the results are tabulated here in Table~\ref{sophisticatedtable1}
\begin{table}
 \centering
\begin{tabular}{ |c||c| } 
 \hline
$D_0$ (eV$^2$)  &  $l_{\varphi}$ (nm)  \\
 \hline
$1\times10^{-4}$ & 247\\
$5\times10^{-4}$ & 157\\
$1\times10^{-3}$  & 105\\
 \hline
\end{tabular}
\caption{\label{sophisticatedtable1}Phase coherence length as a function of dephasing strength $D_0$}
\end{table}
The critical current oscillations in the presence of elastic dephasing interactions, as listed in Table~\ref{sophisticatedtable1}, are shown in Fig.~\ref{fig:13a}. The green curve in Fig.~\ref{fig:13b} includes a random disorder potential profile in addition to the phase-breaking processes. With dephasing in the nanowire, the excess path traversed due to the disorder potentials does not result in a proportionate phase pick-up. Consequently, in contrast to Fig.~\ref{fig:CleanDis}, the inclusion of disorder does not result in a significant shift in the critical current nodes.

{
One of the main effects of dephasing is in pinning down the critical current nodes so that they're less sensitive to disorder. A phase-coherent simulation would overestimate the disorder-induced quasiparticle phase pickup and consequently, the computed critical current oscillations are sensitive and strongly dependent on the microscopic disorder profile (see Fig.~\ref{fig:SNSDisorder}). However, measured oscillations from \citeauthor{Frolov} exhibit only a gradual variation in the oscillations with changing disorder realizations -- well modeled by the inclusion of dephasing in the nanowire.
Further, dephasing is necessary to reproduce the observed reduction of the relative peak height for subsequent primary maxima. Phase-breaking processes restrict the coherent life time of the quasiparticles in the nanowire, which hinders a constructive interference at higher fields. This is well illustrated in Fig.~\ref{fig:13a}, where the relative peak height of the second primary maximum (third peak) decreases with an increase in dephasing strength.

 Magnetoconductance calculations by \citeauthor{Aritra} reveal the suppression of higher harmonics of the transmission characteristics in a nanowire with dephasing. This is manifest in experiments as an increasing inter-node spacing\cite{Frolov,IQC} with respect to the axial field. As illustrated in Fig.~\ref{fig:PBProcesses}, this feature is captured by the inclusion of phase-breaking processes in our model. 
 }
The oscillations in Fig.~\ref{fig:PBProcesses} are in good qualitative agreement with the experiments by \citeauthor{IQC} and \citeauthor{Frolov}. Thus, we infer that phase-breaking processes play a non-negligible role in III-V semiconductor nanowire Josephson junctions.

\section{Conclusion}
In this paper we employed the Keldysh Non-Equilibrium Green's Function formalism to model quantum transport in semiconductor nanowire Josephson junctions. In our analysis we used a three-dimensional discrete lattice model described by the Bogoliubov-de Gennes Hamiltonian in the tight-binding approximation, and computed the Andreev bound state spectrum and current-phase relations. We went beyond the Andreev approximation limit and investigated the avoided level crossing in the ABS spectrum. Our results confirm the measured critical current oscillations to arise from the subband supercurrent interference in presence of an axial magnetic field. The phase picked up by the quasiparticles depend on the difference of their wavenumbers, the length of the junction and the angular momentum quantum number. Thus, the oscillations do not show any periodicity in the flux quantum through the nanowire cross-section.  We included phase-coherent scattering to model a disordered junction and investigated its effect on the critical current oscillations. We observed that the oscillations in the disordered junction are highly sensitive to the realisation of the random disorder potential, and on small fluctuations of the gate voltage. This high sensitivity makes a quantitative description of the experiment a challenging task. Nevertheless, a macroscopic current measurement conveys valuable information about the microscopic profile of the junction. We include elastic dephasing in the nanowire by modelling weak phase-breaking interactions. A good qualitative match of our results with the experiment is observed, and this underscores the role played by phase-breaking processes in III-V nanowire Josephson junctions.

The relevance of these results is emphasized by outlining the points of comparison with experimental data from \citeauthor{IQC} and \citeauthor{Frolov}. The data exhibits a strong suppression of the switching current at magnetic fields on the order of $100-500\text{ mT}$. Subsequently, oscillations with aperiodic nodes are observed in the field dependence of the switching current. As shown in Fig.~\ref{fig:PBProcesses}, our simulations capture the characteristic features of this evolution. Furthermore, as observed in the experiments, the simulated oscillations display a strong gate-tunability and are not uniquely determined by the junction geometry. Finally, a phase-coherence length in the range $100-300$ nm of MBE\cite{Blomers2} and metalorganic vapor phase epitaxy\cite{Estevez_2010,KouwenhovenInAs} grown InAs nanowire samples corroborates the dephasing in our model, as introduced by a uniform, spatially uncorrelated dephasing parameter $(D_0)$.

\indent {\it{Acknowledgements: }} The authors BM and PS would like to thank Prof.\ Supriyo Datta, Prof.\ Kantimay Das Gupta, Abhishek Sharma and Aritra Lahiri for useful discussions throughout this work. This work is an outcome of the Research and Development work undertaken in the project under the Visvesvaraya PhD Scheme of Ministry of Electronics and Information Technology, Government of India, being implemented by Digital India Corporation (formerly Media Lab Asia). This work was also supported by the Science and Engineering Research Board  (SERB) of the Government of India under Grant number EMR/2017/002853. 
KG and JB acknowledge funding from the Natural Sciences and Engineering Research Council of Canada (NSERC).
\appendix
\section{\label{sec:app} Non Equilibrium Green's Function Formalism (NEGF)}
The retarded Green's function in the energy domain is given by
\begin{equation}
G^r(E) =  \left(E\mathbb{I} + i \eta-\mathcal{H}-\Sigma_1^r-\Sigma_2^r\right)^{-1}
\end{equation}
where $\mathcal{H}$ is the Hamiltonian, $\Sigma_{1,2}^r$ are the retarded self-energies of the semi-infinite contacts, and $\eta$ is an infinitesimal real constant. The advanced Green's function is the Hermitian conjugate of the retarded Green's function ($G^a=G^{r\dagger}$).
The Hamiltonian $\mathcal{H}$ is written in the tight-binding approximation   Eq.~\ref{eq:Hamiltonian}. The surface Green's functions ($g_{s}$) at each contact are recursively evaluated 
\begin{align}
   {g_{sL}}(E) &= \left[\left(E+i\eta\right)\mathbb{I} - \alpha_L - \beta^\dagger {g_{sL}}(E)\beta\right]_{\eta \rightarrow 0}^{-1} \\
   {g_{sR}}(E) &= \left[\left(E+i\eta\right)\mathbb{I} - \alpha_R - \beta {g_{sR}}(E)\beta^{\dagger}\right]_{\eta \rightarrow 0}^{-1}
\end{align}
where the subscript $L,R$ labels the contact. Using this we compute the self-energy

\begin{equation}
\Sigma^r_1 = 
\left(\begin{array}{@{}c|c@{}}
\sigma_1
  & 0 \\
\hline
  0 &
 \bigzero
\end{array}\right)
\text{ , }
\Sigma^r_2 = 
\left(\begin{array}{@{}c|c@{}}
\bigzero
  & 0 \\
\hline
  0 &
 \sigma_2
\end{array}\right)
\end{equation}
where $\sigma_1 = \beta^{\dagger} g_{sL} \beta$, and $\sigma_2 = \beta g_{sR} \beta^{\dagger}$

The anti-Hermitian part of the self-energy is responsible for the finite life-time of the quasiparticles in the junction and broadens the energy levels. This broadening matrix is denoted by $\Gamma_i$.

The Fermi functions in the particle-hole Nambu space is given by
\begin{equation}
 F_i = 
\begin{bmatrix}
f(E,\mu+eV) &0\\
0 & f(E,-\mu-eV)
\end{bmatrix}   
\end{equation}
where $f(E,\mu) = 1/\left(\exp\left[(E-\mu)/kT\right]+1\right)$ is the fermi function, and $V$ is the bias applied to the contact.

The lesser self-energy, or the inscattering matrix can be computed from the broadening matrix and fermi function as 
\begin{equation}
    -i \Sigma^< = \Sigma^{in} = \Gamma_1F_1 + \Gamma_2F_2
\end{equation}
The lesser Green's Function is then computed
\begin{equation}
    -iG^< = G^{n} = G^r\Sigma^{in}G^a
\end{equation}
Next, we construct the current operator 
\begin{align}
I_{op} &= \frac{ie}{h}\left(\mathcal{H}G^{n} - G^{n}\mathcal{H}\right)\\
	  &= \frac{ie}{h}\left(G^r\Sigma_1^{in} - \Sigma_1^{in}G^a-\Sigma_1^rG^n+G^n\Sigma_1^a\right)
\end{align}
Electrons and holes travelling in the same direction carry opposite currents and hence, the current is given by the difference of the partial trace of the current operator over the electron and hole sub-spaces. 
\begin{equation}
    J(E) = \Tr_e\left(I_{op}\right) - \Tr_h\left(I_{op}\right) 
\end{equation}
This can be also be written as
\begin{equation}
    J(E) = \Tr(I_{op}\tau_z)
\end{equation}
where $\tau_z$ is the Pauli operator in the particle-hole Nambu space.
The total current is then evaluated by integrating the current-energy density 
\begin{equation}
    I(\phi) =  \int_{-\infty}^{\infty} J(E) dE
\end{equation}
There's a small technical caveat to keep in mind when using the NEGF current operator. \\


 Using the equations for the retarded and advanced Green's functions,
$$\Sigma^a-\Sigma^r+2i\eta = i\left(\Gamma+2\eta \right)$$
where  $\Gamma = i(\Sigma^r-\Sigma^a)$\\

Pre-multiplying by $G^r$ and post-multiplying by $G^a$, 
\begin{align}
 &G^r\left({G^r}^{-1}-{G^a}^{-1}\right)G^a = iG^r\left(\Gamma+2\eta \right)G^a\\
  &\implies (G^a-G^r) = iG^r\left(\Gamma+2\eta \right)G^a\\
 &\implies A = G^r\left(\Gamma+2\eta \right)G^a
  \end{align}
 Multiplying the fermi-function, 
   \begin{align}
 Af = G^n = G^r\left(\Gamma f+2\eta f\right)G^a = G^r\left(\Sigma^{in}+2f\eta\right)G^a
 \end{align}
 Thus,
\begin{equation}
-iG^< =G^n = Af = G^r\left(\Sigma^{in}+{2f \eta}\right)G^a
\end{equation}

However, if we don't consider the term proportional to the infinitesimal $\eta$ we end up with
\begin{equation}
-iG^< = G^n = Af = G^r\left(\Sigma^{in}\right)G^a
\label{eq:wrong}
\end{equation}
$G^<$ from Eq.~\ref{eq:wrong} misses a term in the current proportional to ${\left(G^rG^a\right)}$, the trace of which increases with the number of bound-states. This leads to erroneous results for longer nanowires, which have a larger number of Andreev Bound States.  The NEGF current-operator for contact $i$ is given by -- \begin{multline}
    J_i(E) = \frac{2e}{h}f(E)\text{Tr}\bigg[\text{real}\big( G^<(E)\Sigma_i^a(E)\\+G^r(E)\Sigma_i^<(E) \big)\tau_z\bigg]
    \end{multline}
Substituting $G^n = Af$ and $\Sigma^{in} = \Gamma f$, the current operator can be simplified to 
\begin{equation}
    J_i(E) = \frac{2e}{h}f(E)\text{Tr}\left[\text{real}\left( G^a(E)\Sigma_i^a(E)-G^r(E)\Sigma_i^r(E) \right)\tau_z\right]
\end{equation}
\section{Andreev bound state spectrum : Beyond Andreev approximation\label{sec:BeyondAA}}
 Andreev reflections across an N/S interface were first analyzed by \citeauthor{BTK} in 1982, and have been prevalent in the literature ever since. Almost always, these results are derived under the {Andreev approximation}\cite{Andreev,Andreev2,BeenakkerReview,Ashida}. This approximation deals with a regime where the chemical potential of the nanowire is much larger than the superconducting order parameter of the leads ($\mu \gg \Delta_0$). In this appendix, we analyze the implications of working in a regime where the Andreev approximation is not valid. Specifically, we consider the implications of being outside the Andreev approximation regime on the ABS spectrum in a clean 1-dimensional SNS junction.

The density of states in a superconductor is gapped by an energy $\Delta_0$ on either side of the fermi level. There is no gap in the normal state spectrum. Thus, quasiparticles in the subgap region face an energy barrier at the interface. This is an energy barrier between states with the same momentum, arising due to a difference in the order parameter ($0, \Delta_0$) across the interface. This has nothing do with an impurity or any non-ideality of the junction. However, it plays a role very similar to any impurity-induced barrier $U$ at the interface, i.e.\ it gives rise to \textit{normal reflections} at the interfaces, which cannot be neglected when $\mu \gg \Delta_0$ is not valid, even for a clean junction.

\begin{figure}[!htbp]
            \centering
            \includegraphics[width=0.5\textwidth,keepaspectratio]{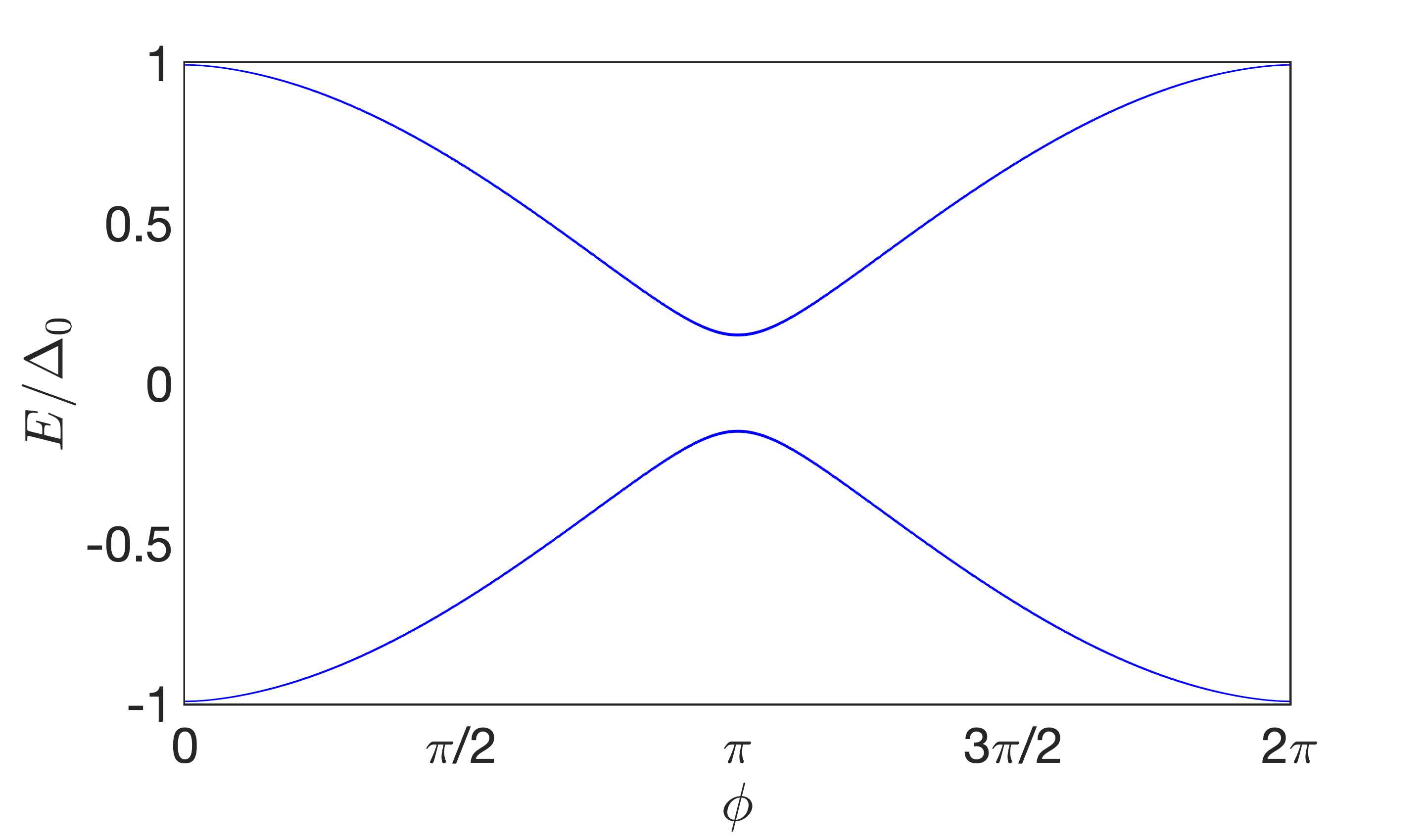}
\caption{Andreev Bound state spectrum as a function of the superconductor phase difference for a clean 1-dimensional SNS junction. The junction is tuned out of the Andreev approximation regime with $\mu = 3\Delta_0$. The avoided level crossing is due to the normal reflections at the interfaces which couple the leftward and rightward moving quasiparticles.}
             	 \label{fig:ABS_noAA}

            \end{figure}

We now examine the spectrum for the chemical potential $\mu$ comparable to $\Delta_0$. Figure.~\ref{fig:ABS_noAA} plots the ABS spectrum for $\mu = 0.5\Delta_0$.

We observe
an avoided level crossing when $\mu \not\gg \Delta_0$, as shown in Fig.~\ref{fig:ABS_noAA}. This anti-crossing can be attributed to the normal reflections which become significant in this parameter space. When these normal reflections are negligible, we have independent rightward and leftward moving excitations in the nanowire, resulting in a crossing at $\phi=\pi$. However, once normal reflections become important, as is the case outside the Andreev approximation, the excitations moving in opposite directions get coupled to each other. This interaction between them, brought about by the normal reflections at the N/S interfaces, results in the anti-crossing.

\begin{figure}[!htbp]
     \begin{center}
        \subfigure[]{%
\label{fig:delta_pi}
    \includegraphics[width=0.45\textwidth]{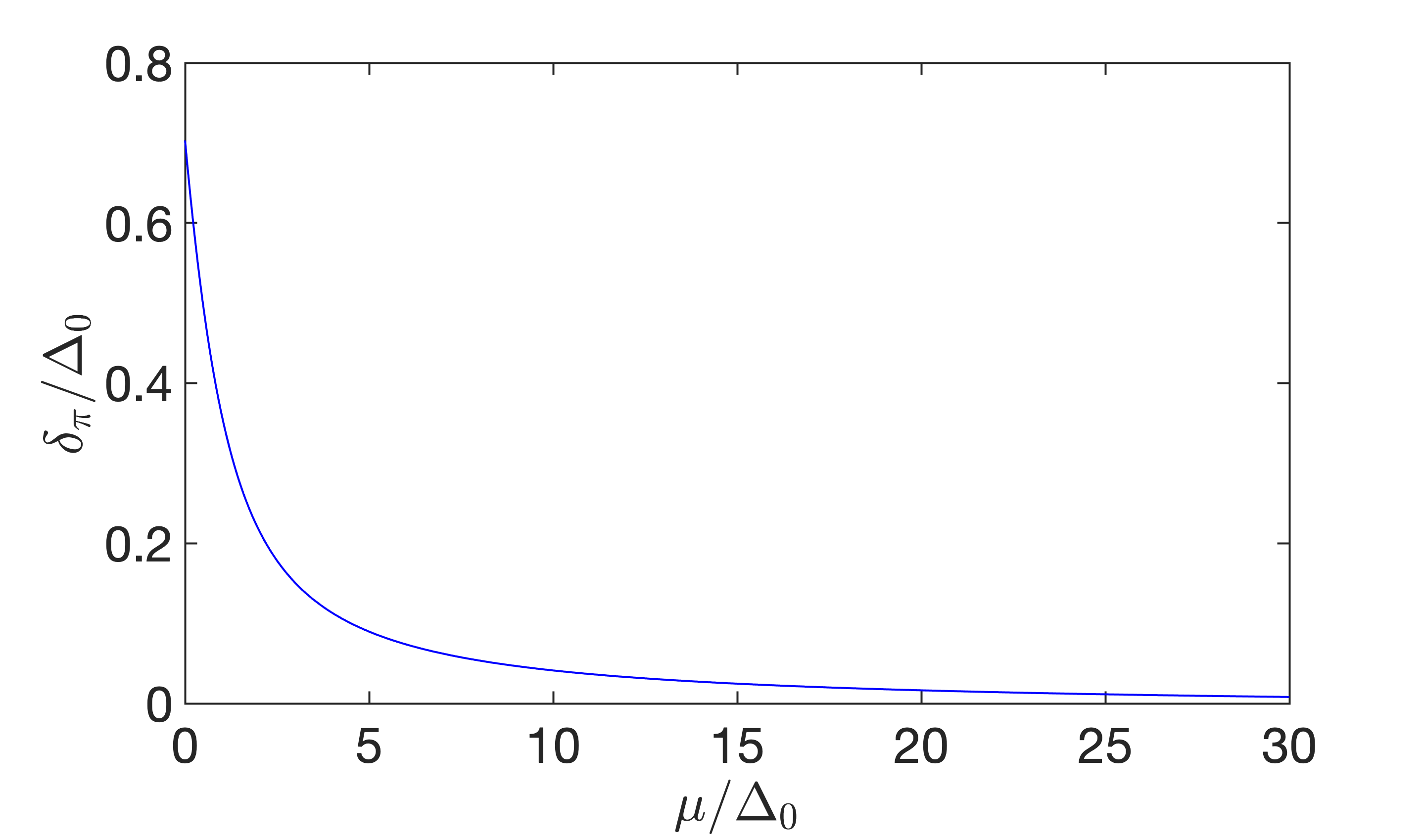}
        }%
        \\
        \subfigure[]{%
 	 \label{fig:delta_pivsL}         \includegraphics[width=0.45\textwidth]{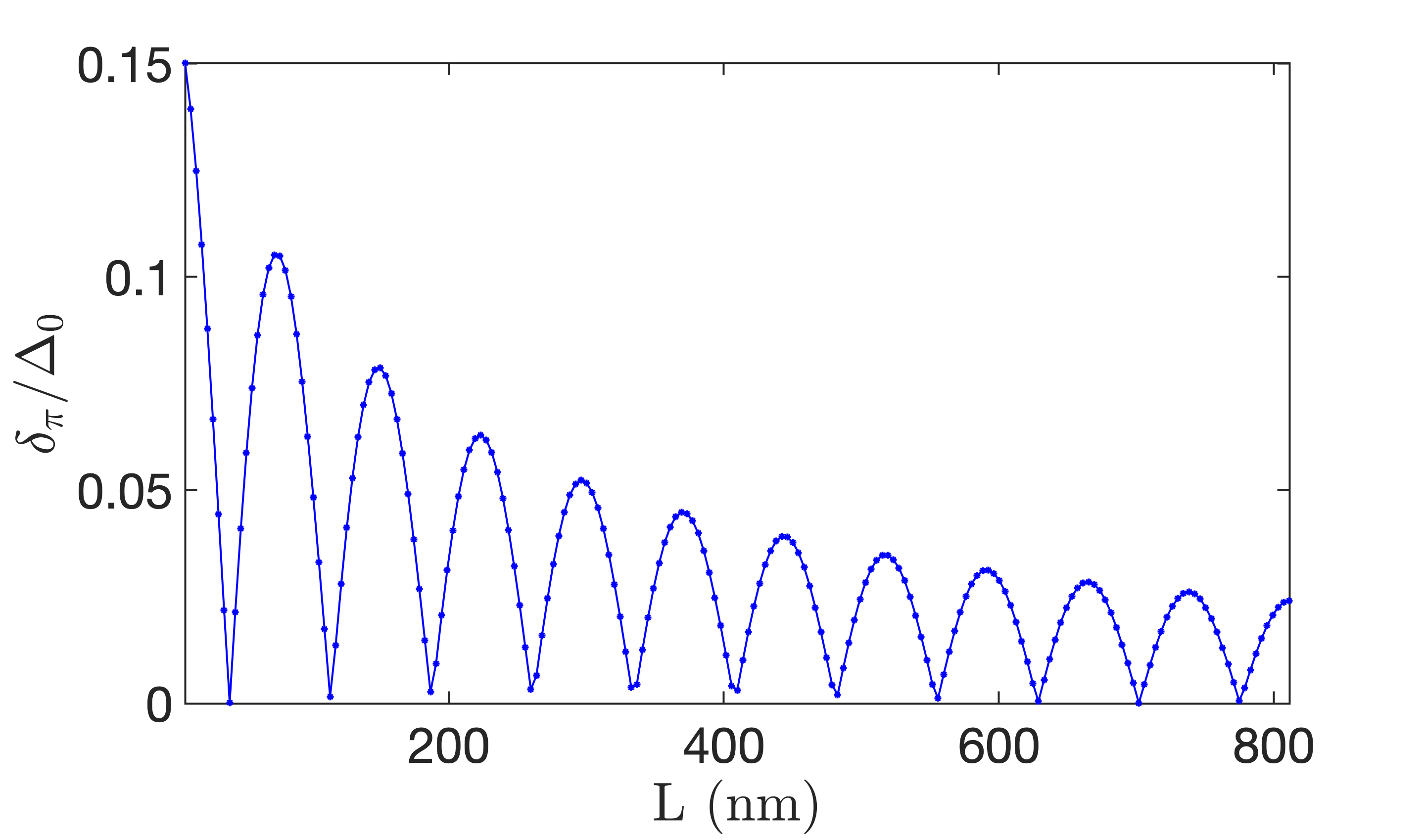}
        }
\end{center}
 \caption{%
  (a) The Andreev bound state energy gap at a superconductor phase difference $\phi=\pi$ is plotted as a function of the ratio of the chemical potential and the order parameter $\mu/\Delta_0$. The states will cross at $\phi=\pi$ as $\mu/\Delta_0 \rightarrow \infty$ (b) The variation in the Andreev bound state energy gap for a fixed ratio of chemical potential to order parameter $\mu/\Delta_0$ at the superconductor phase difference $\phi=\pi$ as a function of the nanowire length L. These oscillations arise from the interference of the normally reflected quasiparticles at the two N/S interfaces.
}%
\end{figure}

In Fig.~\ref{fig:delta_pi}, we plot the ABS energy gap ($\delta_{\pi}$) at $\phi=\pi$  as a function of $\mu/\Delta_0$. It is evident from this plot that $\delta_{\pi}$ decreases as $\mu/\Delta_0$ increases. Thus, in the Andreev approximation regime ($\mu/\Delta_0 \rightarrow \infty$) the bound states cross ($\delta_{\pi} \rightarrow 0$) at $\phi=\pi$. When $\Delta_0$ is kept constant and $\mu$ is varied, the gap varies as $(\mu/\Delta_0)^{-1}$ for $\mu > \Delta_0$. In Appendix~\ref{sec:app2} we verify this dependence analytically by taking a scattering theory approach. Figure~\ref{fig:delta_pivsL} plots the variation of the gap at $\phi=\pi$ with the nanowire length. These oscillations result from the interference of the waves reflected at the two S/N interfaces. 

\section{\label{sec:app2} The Andreev bound state quantisation condition outside the Andreev Approximation Regime}

In this appendix we derive an expression for the Andreev bound state energy in a SNS junction, when the phase difference between the two superconductors is tuned to $\pi$. We show this to be \emph{non-zero} for a finite $\mu/\Delta$ while approaching zero in the limit $\mu/\Delta \rightarrow \infty$, where $\mu$ is the chemical potential of the entire device and $\Delta$ is the induced superconducting gap.

\subsection{Setting up the problem}

We follow the notation from \citet{Kulik}. Consider first the transport of an electron across an N/S interface. It is well known that for a clean interface, an incident electron is perfectly Andreev reflected as a hole. However, there is also a component of the order $\Delta/\mu$ which is reflected as a normal electron. This is mentioned in \citeauthor{Kulik}'s work and can also be derived from first principles scattering theory at the N/S interface. We usually encounter cases when perfect Andreev reflection is assumed with no normal reflection\cite{Andreev,Andreev2,BeenakkerReview,Ashida}, but this is only true to the zeroth order in $\Delta/\mu$.

This normal reflection component will modify the Andreev bound state quantization condition\cite{IQC1,Kulik}
\begin{equation}
    \gamma^2 e^{i(k_0 - k_1)d} e^{i\chi} = 1
\end{equation} where $\frac{1}{\gamma} = \frac{E}{\Delta} + i \sqrt{1-\frac{E^2}{\Delta^2}}$, $k_0$ and $k_1$ are the electron and hole wave vectors and $\chi= \chi_2-\chi_1$ is the phase difference between the superconducting order parameter across the junction.

Consider an SNS junction with a normal region length $d$ extended over $|z|<d/2$, and superconducting contacts defined over $|z|>d/2$. Writing the two-component wave functions

\begin{equation}
\Psi=
\begin{cases}
    \!\begin{aligned}
    A e^{i k_0 z} \mqty(1 \\ 0) + B e^{i k_1 z} \mqty(0 \\ 1) + B' \frac{\Delta}{\mu} e^{-i k_0 z}& \mqty(1 \\ 0)  \\ &  |z| < d/2 
    \end{aligned}\\
     \!\begin{aligned}
   C e^{i \lambda_+ (z - d/2)} \mqty(e^{i \chi_2} \\ \gamma)+ C' \frac{\Delta}{\mu} e^{-i \lambda_- (z - d/2)} &\mqty(e^{i \chi_2} \\ \gamma^*) \\ &  z > d/2
    \end{aligned}\\
     \!\begin{aligned}
    D e^{i \lambda_- (z + d/2)} \mqty(\gamma \\ e^{-i \chi_1} ) + D' \frac{\Delta}{\mu} e^{-i \lambda_+ (z + d/2)} &\mqty(\gamma^* \\ e^{-i \chi_1} ) \\&   z < -d/2
    \end{aligned}\\
\end{cases}
\end{equation}

The terms proportional to $B'$, $C'$ and $D'$ are ignored when perfect Andreev reflection is assumed. We have included a $\Delta/\mu$ coefficient to emphasize a first order expansion beyond the perfect Andreev reflection scenario.

Equating the coefficients at $z = \pm d/2$, we get:

\begin{align}
A e^{ik_0d/2} + B' \frac{\Delta}{\mu} e^{-ik_0d/2} &= C e^{i \chi_2} + C' \frac{\Delta}{\mu} e^{i \chi_2} \\
B e^{ik_1d/2} &= C \gamma + C'\frac{\Delta}{\mu} \gamma^* \\
A e^{-ik_0d/2} + B' \frac{\Delta}{\mu} e^{ik_0d/2} &= D \gamma + D'\frac{\Delta}{\mu} \gamma^*  \\
B e^{-ik_1d/2} &= D e^{-i \chi_1} + D' \frac{\Delta}{\mu} e^{-i \chi_1}  
\end{align}

Solving for $C$ and $D$ (to be compared with Eq~2.13 from \citeauthor{Kulik}), we get:
\begin{multline}
    C = A e^{ik_0d/2} e^{-i \chi_2} (1 + \frac{B'}{A} \frac{\Delta}{\mu} e^{-ik_0d} - \frac{C'}{A} \frac{\Delta}{\mu} e^{i \chi_2} e^{-ik_0d/2} ) \\ = \frac{B e^{ik_1 d/2}}{\gamma} (1 - \frac{C' \gamma^*}{B} \frac{\Delta}{\mu} e^{-ik_1 d/2})
    \end{multline}
    
\begin{multline}
D = \frac{A e^{-ik_0d/2}}{\gamma} (1 + \frac{B'}{A} \frac{\Delta}{\mu} e^{ik_0d} - \frac{D'}{A} \frac{\Delta}{\mu}  \gamma^* e^{ik_0d/2} ) \\ = B e^{-ik_1 d/2}  e^{i \chi_1}\gamma (1 - \frac{D'}{B} \frac{\Delta}{\mu} e^{ik_1 d/2} e^{-i \chi_1}) 
\end{multline}


We now divide the above equations and keep terms to first order in $\Delta/\mu$,
\begin{widetext}
\begin{equation}
\label{eqn:new-kulik}
\gamma^2 e^{i(k_0 - k_1)d} e^{i\chi} = 1 + \frac{\Delta}{\mu} \left(2 i \frac{B'}{A} \sin{k_0 d} + \frac{D'}{B} e^{i k_1 d/2} e^{-i\chi_1} + \frac{C'}{A} e^{-i k_0 d/2} e^{i\chi_2} - \frac{C' \gamma*}{B} e^{-i k_1 d/2} - \frac{D' \gamma*}{A} e^{i k_0 d/2} \right)   
\end{equation}
\end{widetext}

We will now focus on the qualitative behaviour of the solutions of the above equation. In order to do so, we simplify the above equation into a more tractable form:

\begin{equation}
\gamma^2 e^{i(k_0 - k_1)d} e^{i\chi} = 1 + \frac{\Delta}{\mu} \epsilon + i \frac{\Delta}{\mu} \eta 
\end{equation}
where the exact form of $\epsilon$ and $\eta$ can be derived from equation~\ref{eqn:new-kulik}. 
\subsection{Expression for the energy at $\chi = \pi$}
Let us tune the phase difference $\chi = \pi$. Defining $\phi = \cos^{-1}({E/\Delta})$, we can write $\gamma = e^{-i \phi}$. Using this relation, the quantization condition can be simplified to,
\begin{equation}
e^{i[(k_0 - k_1)d - 2\phi + \pi]} = 1 + \frac{\Delta}{\mu} \epsilon + i \frac{\Delta}{\mu} \eta 
\end{equation}
Equating the real and imaginary parts, 
\begin{align}
\cos{[(k_0 - k_1)d - 2\phi + \pi]} &= 1 + \frac{\Delta}{\mu} \epsilon\\
\sin{[(k_0 - k_1)d - 2\phi + \pi]} &= \frac{\Delta}{\mu} \eta \label{eq:im} 
\end{align}
Since $\Delta/\mu \ll 1$, under the small angle approximation Eq.~\ref{eq:im} can be simplified to

\begin{align}
(k_0 - k_1)d - 2 \cos^{-1}({E/\Delta}) + \pi &= \frac{\Delta}{\mu} \eta
\end{align}
For $E/\Delta \ll 1$
\begin{equation}
    \!\begin{aligned}
    (k_0 - k_1)d &+  E/\Delta \approx \frac{\Delta}{\mu} \eta\\
    & \text{from the relation} \left(\cos^{-1}(x) \approx \frac{\pi}{2} - x\right)
    \end{aligned}
\end{equation}
This can be further simplified using $\left(\text{$\frac{\hbar^2 k_{0/1}^2}{2 m} = \mu \pm E$}\right)$,
\begin{equation}
\frac{k_F d}{\mu} E + E/\Delta \approx \frac{\Delta}{\mu} \eta
\end{equation}
\begin{align}
E \approx \frac{\Delta^2 \eta}{\mu + k_F d \Delta} &\approx \eta \Delta\frac{\Delta}{\mu} \left(1 - \frac{k_F d \Delta}{\mu}\right)
\end{align}
This is a finite energy for non-zero $\eta$ and goes to zero in the limit $\Delta/\mu \rightarrow 0$. 

\subsection{Comparison with numerics}
Ignoring the term in the bracket from the final expression for E, we can simplify it to $E \approx \frac{\Delta^2}{\mu}$. We decided to observe the power law dependence of the gap using the numerical simulations. We considered two cases:

\begin{enumerate}
\item Fix $\mu$ and vary $\Delta$: we expect to see a behavior 
\begin{equation}
    E \sim \mathcal{O}\left((\mu/\Delta)^{-2}\right) 
\end{equation}

\item Fix $\Delta$ and vary $\mu$: we expect to see a behavior
\begin{equation}
    E \sim \mathcal{O}\left((\mu/\Delta)^{-1}\right)
\end{equation}
\end{enumerate}

\begin{figure}[!htb]
     \begin{center}
  \subfigure[$\mu$ is kept constant, $\Delta$ is varied]{%
              \label{fig:df}
            \includegraphics[width=0.45\textwidth]{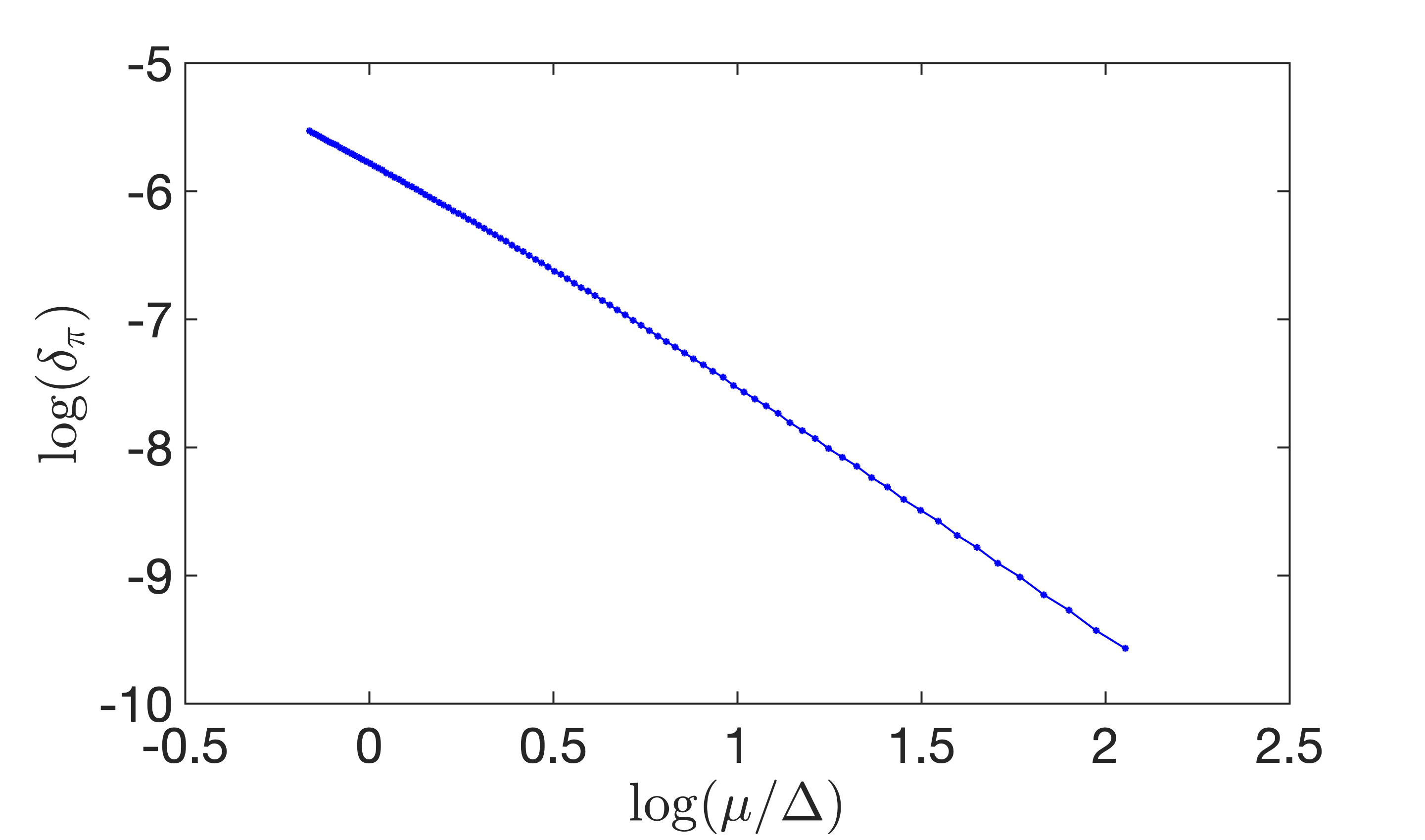}
        }\\%
        \subfigure[$\Delta$ is kept constant, $\mu$ is varied]{%
            \label{fig:muf}
            \includegraphics[width=0.45\textwidth]{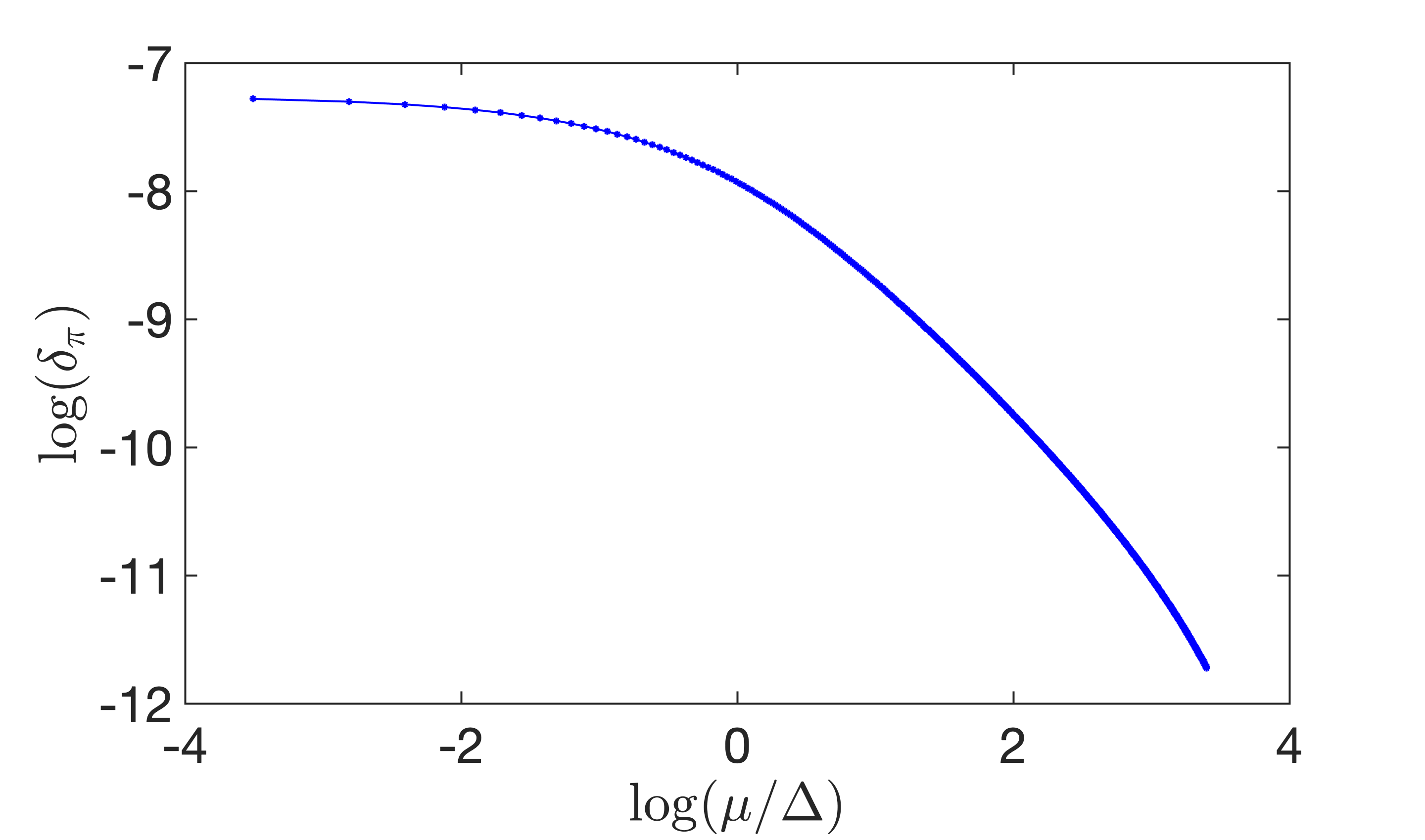}
        }%

    \end{center}
    \caption{%
       The dependence of the ABS energy at $\chi=\pi$ as a function of $\mu/\Delta$ is investigated. $\delta_\pi$ is the gap in the ABS spectrum at $\chi = \pi$, and is thus twice the absolute value of the ABS energy.
       (a) $\mu$ kept constant and the $\Delta$  is varied.  We have $\mu/\Delta$ on the x-axis (in log scale) showing the expected $1/x^2$ scaling behaviour. (b) $\mu$ is varied and $\Delta$ is kept constant. The expected $1/x$ scaling is observed for $\mu>\Delta$.}%
       \label{fig:appC}
 \end{figure}

Figure~\ref{fig:appC} confirms this dependence of the ABS energy on $\mu/\Delta$ for the aforementioned cases.

 \section{\label{sec:lphi} Phase relaxation length estimation}

The dephasing in the nanowire can be parameterised by the phase relaxation length $l_{\varphi}$, which is a length scale over which the phase of the quasiparticles randomise. As explained in Sec.~\ref{sec:deph}, phase-breaking processes are included via a self-energy for the lattice background $\Sigma_s = D \times G^r$; $D = D_0\mathbb{I}$. In this appendix, we estimate $l_{\varphi}$ as a function of the dephasing strength $D_0$.   

\subsection{Estimation from phase coherence lifetime}
The phase relaxation length can be computed from the phase coherence lifetime $\tau_{\varphi}$
\begin{equation}
    l_{\varphi} = 
     \begin{cases} 
      v_F\tau_{\varphi} & \text{(ballistic)} \\
      (\mathcal{D}\tau_{\varphi})^{1/2} & \text{(diffusive)}
   \end{cases}
    \label{eq:lphi}
\end{equation}
where $v_F$ is the Fermi-velocity, and $\mathcal{D}$ is the diffusion constant. 
The anti-hermitian part of the lattice background self-energy limits the phase-coherent lifetime of the quasiparticles, and sets an energy scale for the problem. The phase coherence lifetime can thus be estimated as 
\begin{equation}
    \frac{h}{\tau_{\varphi}(E)} = \Gamma_{s,d}(E)
\end{equation}
where $\Gamma_{s,d}$ is a diagonal 
element of $\Gamma_s = i\left(\Sigma_s-\Sigma_s^{\dagger}\right)$ is the broadening function corresponding to the lattice background. Note that $\Gamma_s$, and hence the $l_{\varphi}$ so estimated from Eq.~\ref{eq:lphi} is a function of energy. The phase relaxation length is then reported as an average over the energy grid,
$l_{\varphi} = \langle l_{\varphi}(E)\rangle_E$. Under the assumption of ballistic transport, we estimate an upper bound on the phase relaxation length $l_{\varphi} \underset{\sim}{<} 150$ nm for $D_0 = 0.001$ eV$^2$, and for $D_0 = 5\times10^{-4}$ eV$^2$, $l_{\varphi} \underset{\sim}{<} 300$ nm. The Fermi-velocity decreases with field, and hence we observe a gradual monotonic degradation in the phase relaxation length. 
\begin{figure}[!htb]
        \subfigure[$D_0=0.001$ eV$^2$]{%
           \label{fig:lphi_D1e-3}
            \includegraphics[width=0.45\textwidth,keepaspectratio]{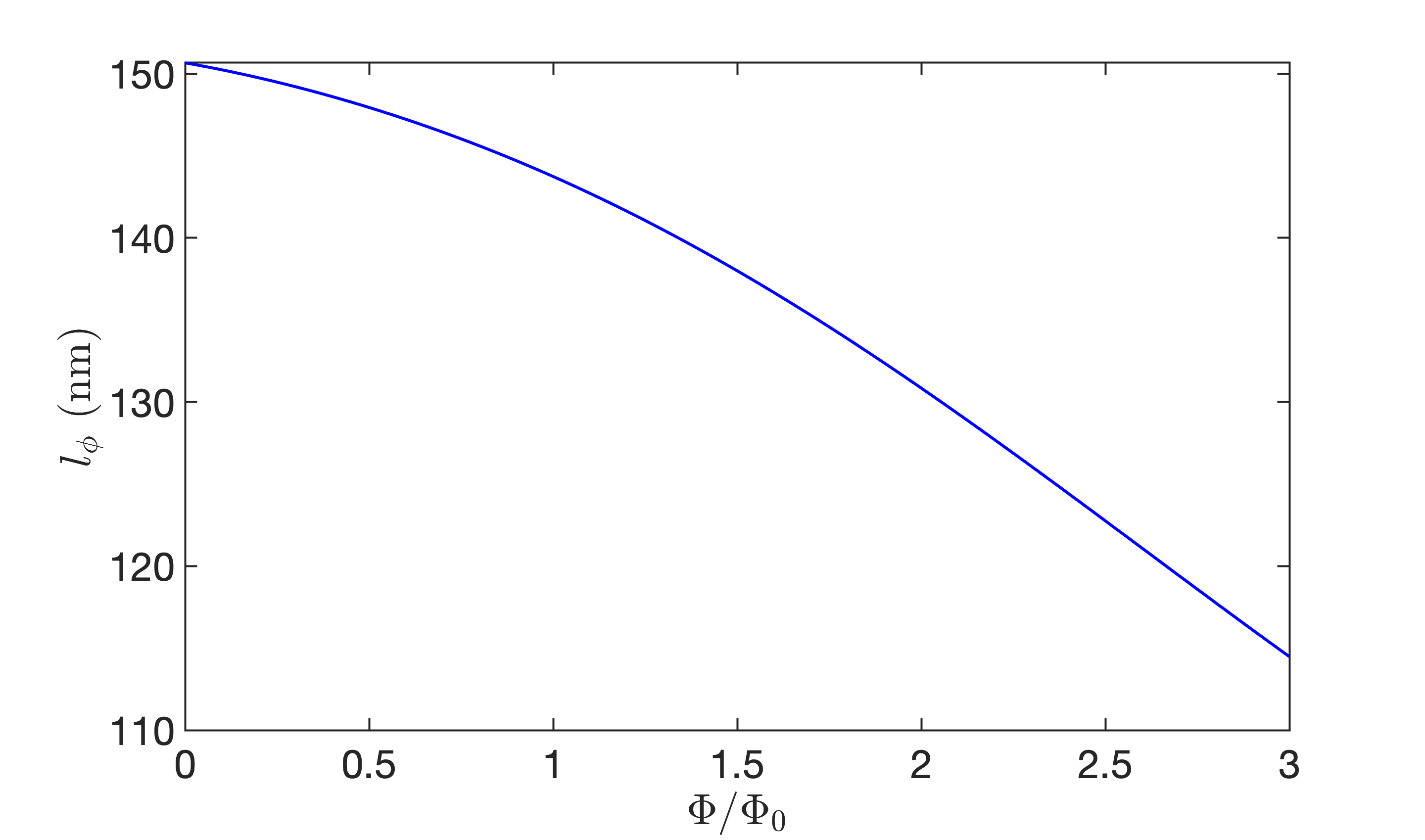}
        }\\%
      \subfigure[$D_0=0.0005$ eV$^2$]{%
            \label{fig:lphi_D5e-4}
            \includegraphics[width=0.45\textwidth,keepaspectratio]{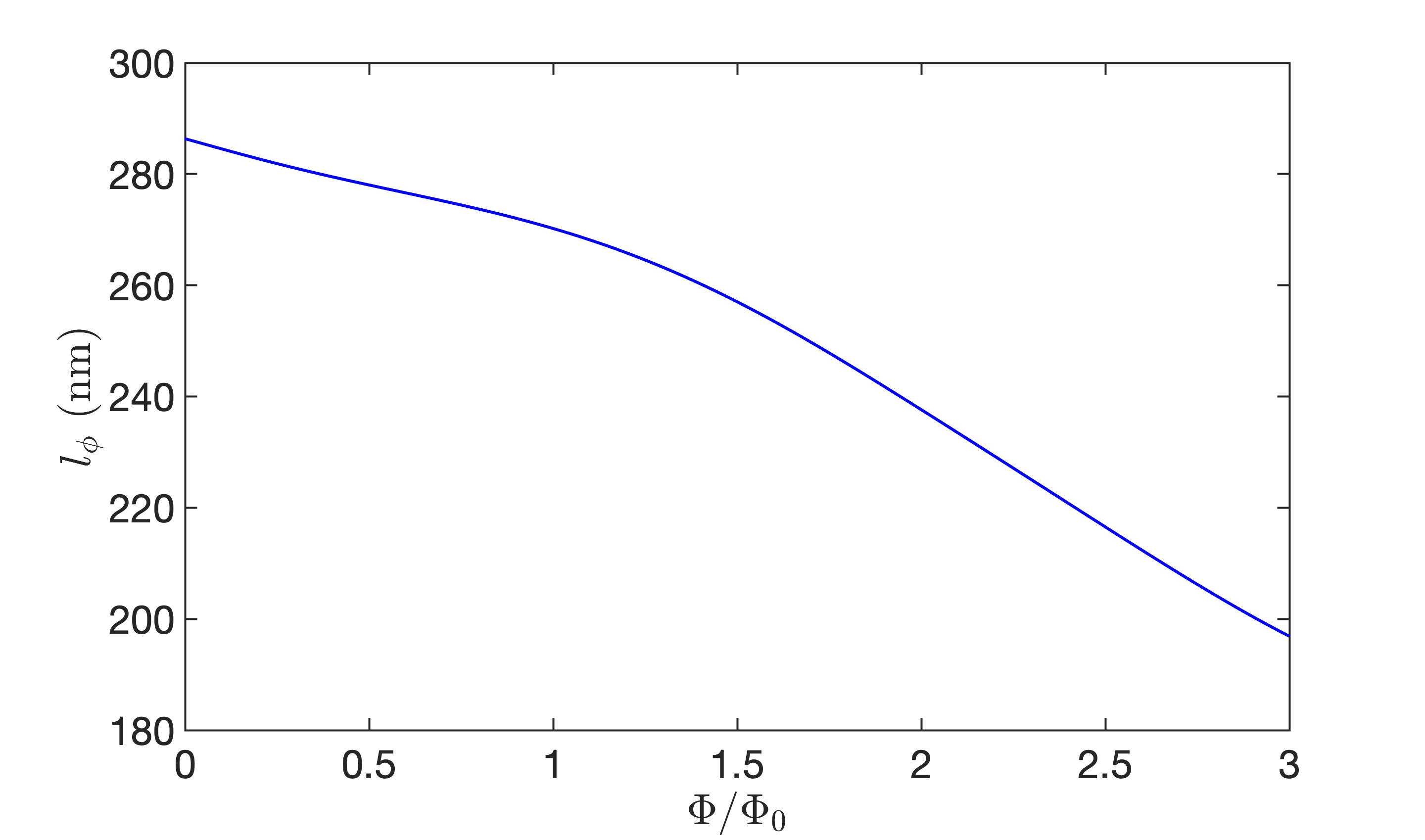}
        }
    \caption[Phase relaxation length $l_{\varphi}$ from the phase coherence lifetime]{%
      Phase relaxation length $l_{\varphi}$, computed from Eq.~\ref{eq:lphi} (ballistic case) for a dephasing strength (a) $D_0=0.001$ eV$^2$, and (b) $D_0=0.0005$ eV$^2$. The estimated $l_{\varphi} \sim 150$ nm, and $l_{\varphi} \sim 300$ nm for (a) and (b) respectively.
      The monotonic decrease in $l_{\varphi}$ is due to the reduction in Fermi velocity with an applied field. The simulations were performed for $L=160$ nm, $\mu=30\Delta_0$.
     }%
   \label{fig:PBProcesses_app}
\end{figure}

\subsection{Estimation from statistical properties of UCF}
 For a nanowire length comparable to $l_{\varphi}$, the normal-state conductance fluctuates with an amplitude of the order of $e^2/h$ in presence of a magnetic field. These aperiodic universal conductance fluctuations (UCF) measured in a magnetic field perpendicular to the nanowire axis can be analysed to extract information on phase coherent transport. The UCF originates from electron phase shifts resulting from the penetration of magnetic flux through closed electron trajectories.  The conductance shows strong fluctuations for low dephasing strengths, while they are smeared out at higher coupling strengths. 

The magnetoconductance fluctuation is denoted by $\delta G$
\begin{equation}
\delta G = G - \langle G \rangle_B
\end{equation}
where the average $\langle .\rangle$ is taken over the magnetic field $B$. The average fluctuation amplitude about the mean conductance is quantified by the root-mean-square $\text{rms}(\delta G)_B = \sqrt{\text{var}{(\delta G)}_B}$. The rms$(\delta G)_B$ decreases monotonically with $D_0$. The phase relaxation length $l_{\varphi}$ can be estimated from the analysis of the autocorrelation function $F$ of $\delta G$. The half-width half-maximum (HWHM) of $F$
corresponds to the correlation field $B_c$, which is a measure of a field range over which the phases of the interference path become uncorrelated.  
\begin{align}
F(\Delta B) &= \langle \delta G(B+\Delta B) \delta G \rangle_B \\
F(B_c) &= \frac{1}{2}F(0)
\end{align}

Assuming the phase relaxation length $(l_{\varphi})$ to be greater than the nanowire diameter $d$, we can extract $l_{\varphi}$ directly from the correlation field\cite{Blomers2,Estevez_2010,Beenakker_lphi}
\begin{equation}
l_{\varphi} = \gamma \frac{h}{e}\frac{1}{B_cd}
\end{equation}
where $\gamma$ is a dimensionless prefactor depending on the transport regime. We work in the the dirty metal limit with $\gamma = 0.95$\cite{Blomers2}. 
\subsubsection{Results}
\begin{figure}[!htb]
        \subfigure[$\delta G$]{%
           \label{fig:magnetoG_fluc}
            \includegraphics[width=0.45\textwidth,keepaspectratio]{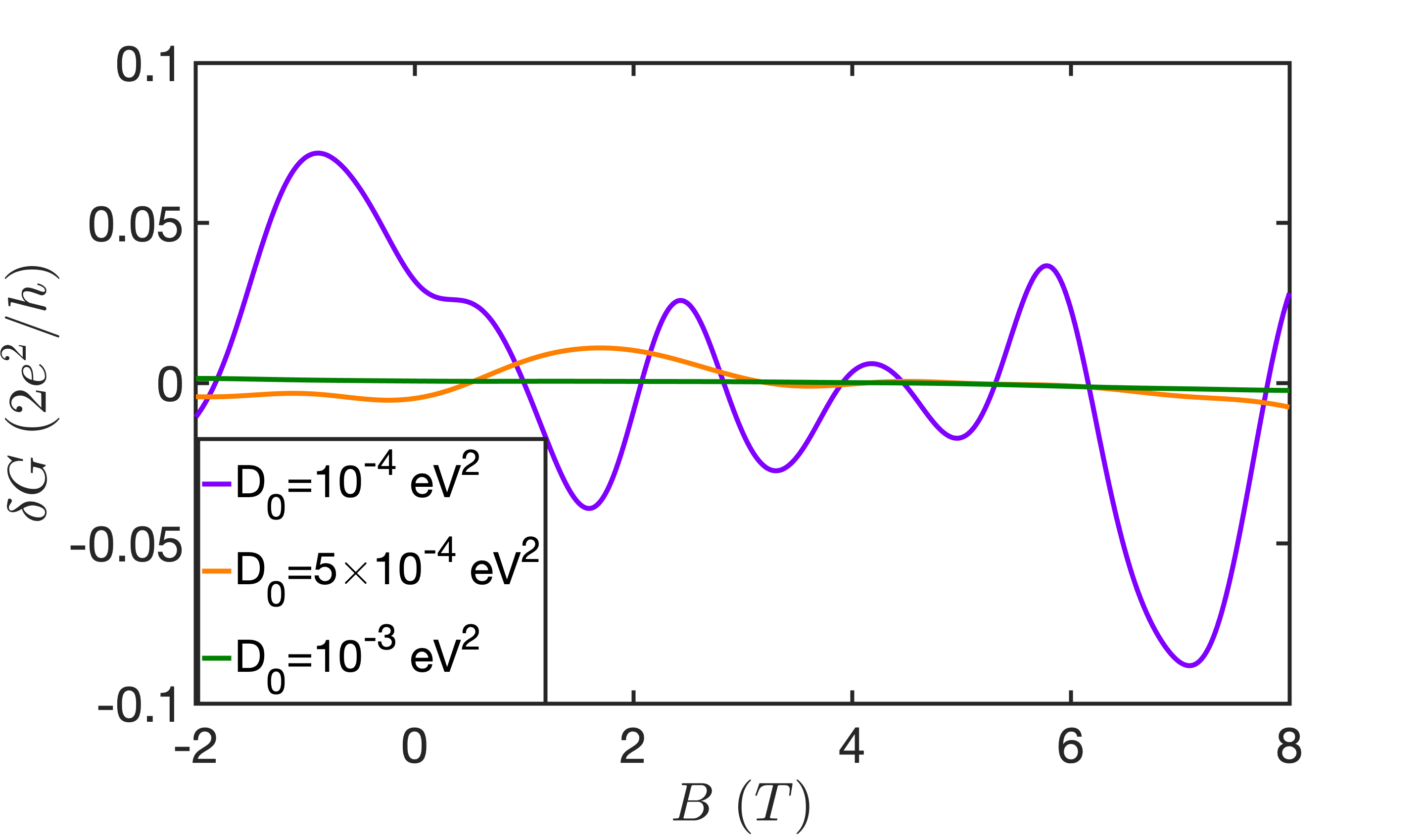}
        }\\%
      \subfigure[$F/F(0)$]{%
            \label{fig:autoCorr}
            \includegraphics[width=0.45\textwidth,keepaspectratio]{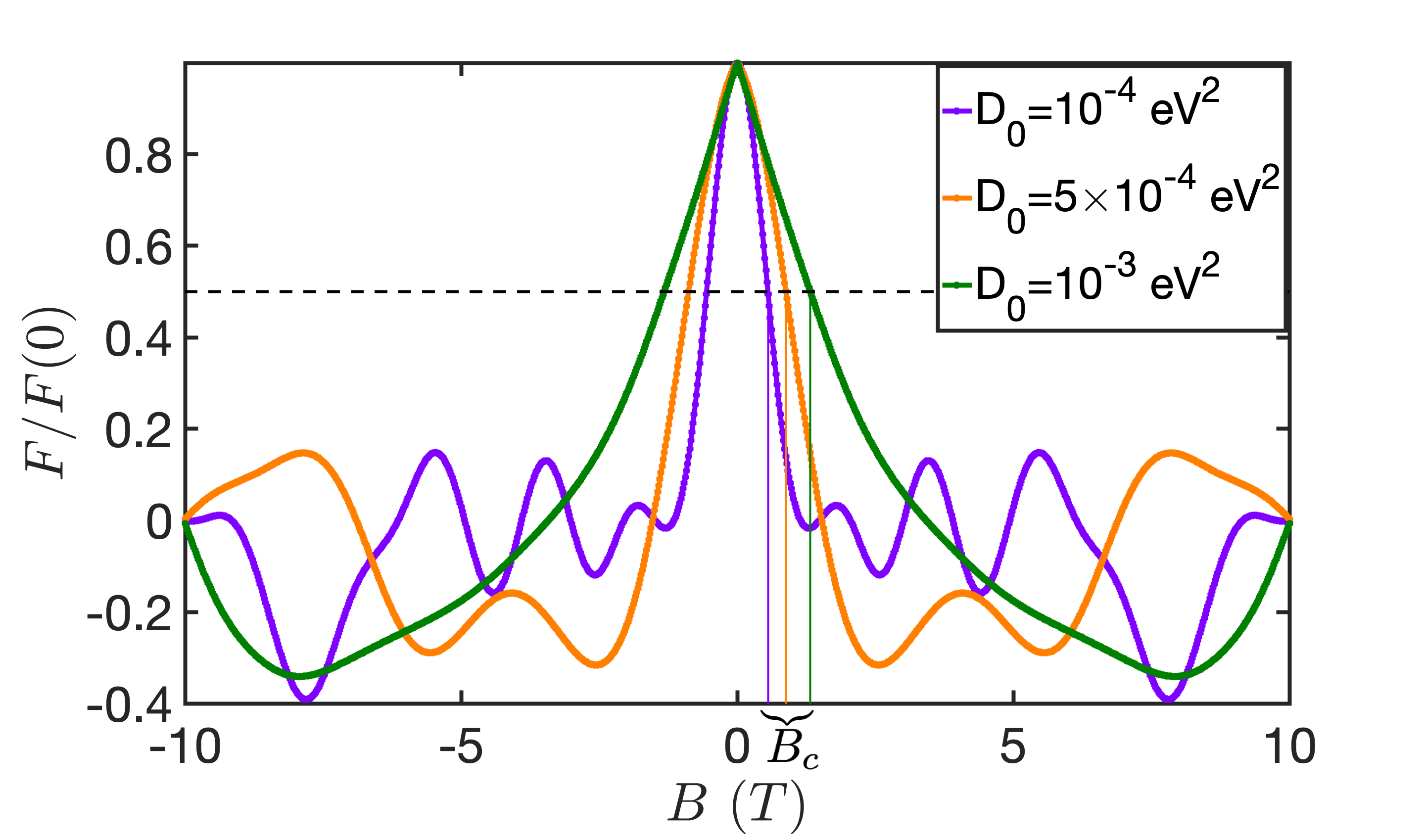}
        }
    \caption[Phase relaxation length $l_{\varphi}$ from statistical properties of UCF]{%
    (a) Magnetoconductance fluctuations $\delta G$ in units of $2e^2/h$ for a dephasing strength $D_0=1\times 10^{-4}$ eV$^2$, $D_0=5\times 10^{-4}$ eV$^2$ and $1\times 10^{-3}$ eV$^2$. The fluctuations arise from field induced electron phase shifts, and hence reduce with dephasing. (b) Autocorrelation of $\delta G$ for the various dephasing coupling strengths. Each curve has been normalised to its respective maximum. The black dotted horizontal line indicates the half-maximum of $F$. The correlation field $B_c$ corresponds to this half-maximum of $F$, and is denoted by a vertical line for each curve. 
     }%
   \label{fig:PBrkProcesses}
\end{figure}

This simulation involves normal-state low-bias transport in presence of a magnetic field oriented in a direction perpendicular to the nanowire axis. The nanowire length $L = 200$ nm, and diameter $d = 30$ nm. To model diffusive transport, an onsite random potential in the range $W \in [-1.5t, 1.5t]$ is introduced at each point in the nanowire, where $t$ is the tight-binding hopping parameter. This corresponds to a mean-free path $\lambda_{mf} \approx 18$ nm. The magnetoconductance fluctuations are plotted in Fig.~\ref{fig:magnetoG_fluc}. The normalised autocorrelation $F/F(0)$ of $\delta G$ is shown in Fig.~\ref{fig:autoCorr}, and the extracted parameters are listed in Table~\ref{sophisticatedtable}. 

% ...
\begin{table}[!htb]
\begin{tabular}{ |c|c|c |c| } 
 \hline
$D_0$ (eV$^2$)  & rms($G$) $(2e^2/h)$ & $B_c$ $(T)$ &  $l_{\varphi}$ (nm)  \\  
 \hline
$1\times10^{-4}$ & 0.038 & 0.56 & 247\\
$5\times10^{-4}$ & 0.0048 & 0.88 & 157\\
$1\times10^{-3}$ & 0.00098 & 1.32 & 105\\
 \hline
\end{tabular}
\caption{\label{sophisticatedtable}Phase coherence length as a function of dephasing strengths $D_0$}
\end{table}

\end{document}